\documentclass[prd,aps,preprintnumbers,twocolumn,showpacs,nofootinbib,superscriptaddress,notitlepage,floatfix]{revtex4-1}
\usepackage{amsthm,amsmath,amssymb}
\usepackage{xcolor}
\usepackage{graphicx}
\usepackage{subcaption}
\usepackage{multirow}
\usepackage[colorlinks, citecolor=blue,anchorcolor=blue,menucolor=blue, linkcolor=blue,filecolor=blue,runcolor=blue,urlcolor=blue,frenchlinks=true]{hyperref}
\usepackage{multirow}
\usepackage{caption}
\usepackage{diffcoeff} 

\begin{document}

\title{Toward the First Gluon Parton Distribution from the LaMET
}

\author{William Good}
\email{goodwil9@msu.edu}
\affiliation{Department of Physics and Astronomy, Michigan State University, East Lansing, MI 48824}
\affiliation{Department of Computational Mathematics, Science and Engineering, Michigan State University, East Lansing, MI 48824}

\author{Kinza Hasan}
\affiliation{Department of Physics and Astronomy, Michigan State University, East Lansing, MI 48824}

\author{Huey-Wen Lin}
\affiliation{Department of Physics and Astronomy, Michigan State University, East Lansing, MI 48824}

\preprint{MSUHEP-24-011}

\begin{abstract}
We present progress towards the first unpolarized gluon quasi-PDF from lattice QCD using high-statistics measurements for hadrons at two valence pion masses $M_\pi \approx 310$ and $690$~MeV computed on an $a \approx 0.12$~fm ensemble with $2+1+1$-flavors of HISQ generated by the MILC collaboration.
In this study, we consider two gluon operators for which the hybrid-ratio renormalization matching kernels have been recently derived and a third operator that has been used in prior pseudo-PDF studies of the gluon PDFs.
We compare the matrix elements for each operator for both the nucleon and pion, at both pion masses, and using two gauge-smearing techniques.
Focusing on the more phenomenologically studied nucleon gluon PDF, we compare the ratio and hybrid-ratio renormalized matrix elements at both pion masses and both smearings to those reconstructed from the nucleon gluon PDF from the CT18 global analysis.
We identify the best choice of operator to study the gluon PDF and present the first gluon quasi-PDF under some caveats.
Additionally, we explore the recent idea of Coulomb gauge fixing to improve signal at large Wilson-line displacement and find it could be a major help in improving the signal in the gluon matrix elements, using the perturbative calculation to confirm our results.
This work helps identify the best operator for studying the gluon quasi-PDF, shows higher hadron boost momentum is needed to implement hybrid-ratio renormalization reliably, and
suggests the need to study more diverse set of operators with their corresponding perturbative calculations for hybrid-ratio renormalization to further gluon quasi-PDF study.

\end{abstract}

\maketitle

\section{Introduction}

Parton distribution functions (PDFs) are nonperturbative functions that represent the probability of finding (anti)quarks and gluons within a hadron at a specific fraction of the hadron's total momentum.
These functions act as crucial inputs for many high energy scattering experiments~\cite{Harland-Lang:2014zoa,Dulat:2015mca,Abramowicz:2015mha,Accardi:2016qay,Alekhin:2017kpj,Ball:2017nwa,Hou:2019efy,Bailey:2019yze,Bailey:2020ooq,Ball:2021leu,ATLAS:2021vod}.
The nucleon gluon PDF $g(x)$ is especially important to determine the cross sections in $pp$ collisions, Higgs boson productions, $J/\psi$ photo production and jet production~\cite{CMS:2012nga,Kogler:2018hem,mammeiproposal, Dainese:2019rgk, Amoroso:2022eow}.
In addition to proton structure, there is much interest in elucidating the structure of the pion because of its role in association to chiral-symmetry breaking as the pseudo--Nambu-Goldstone boson of quantum chromodynamics (QCD)~\cite{Roberts:2021nhw,Arrington:2021biu,Aguilar:2019teb}.
The experimental data~\cite{Badier:1983mj,Betev:1985pf,Conway:1989fs,Wijesooriya:2005ir,Aicher:2010cb} is very limited in the pion case, as the pion's short lifetime forbids its use as a scattering target.
The experiments at the future Electron Ion Colliders based in U.S.~\cite{Achenbach:2023pba} and China~\cite{Anderle:2021wcy} along with the proposed COMPASS++ and AMBER facilities~\cite{Adams:2018pwt} will advance our knowledge on gluon PDFs, in the meantime, Lattice QCD serves as a tool enabling us to study gluon PDFs from first principles.

Lattice QCD is a theoretical framework that allows us to calculate nonperturbative QCD quantities with full systematic control.
$x$-dependent calculations for hadron structures in Lattice QCD have multiplied since Large Effective Momentum Theory (LaMET) was proposed in 2013~\cite{Ji:2013dva,Ji:2014gla,Ji:2017rah}.
LaMET, with its application to PDF studies sometimes referred to as the quasi-PDF method, relies on measuring matrix elements nonlocal, bilinear quark/gluon operators in boosted hadron states.
The Fourier transform of these matrix elements are referred to as quasi-PDFs, which can be matched to the lightcone PDFs via a matching procedure which is accurate to powers of inverse parton momentum.
We direct readers to the following reviews on LaMET in Refs.~\cite{ Constantinou:2020hdm,Ji:2020ect,Constantinou:2022yye}.
However, the necessity to have signal out to far separation distances and large momentum has previously forbidden the use of the quasi-PDF method on the gluon PDF from lattice~\cite{Fan:2018dxu}.
The primary method used in LQCD studies of the unpolarized and helicity gluon PDFs~\cite{Fan:2020cpa,Fan:2021bcr,HadStruc:2021wmh,Salas-Chavira:2021wui,Fan:2022kcb,Delmar:2023agv,Good:2023ecp,HadStruc:2022yaw,Khan:2022vot,Karpie:2023nyg} has instead been the pseudo-PDF method~\cite{Radyushkin:2017cyf}, which relies on a short distance factorization and matching the lightcone PDF to the position space matrix elements.
The pseudo-PDF method requires one to use, typically phenomenological-inspired, model forms for the PDF and fit the model parameters.
It is, therefore, desired to obtain the gluon PDF through LaMET to make comparisons between results from the two methodologies.

This paper is organized as follows. We provide some theoretical background in Sec.~\ref{sec:theory}, giving the definitions for the gluon operators, hybrid-ratio renormalization, the quasi-PDF, and matching to the lightcone PDF.
In Sec.~\ref{sec:bMEs}, we explain the numerical setup, define the two-point and three-point correlators, compare the signal for different operators, explain how the bare matrix element are extracted, and present our bare matrix elements for different operators, hadrons, and smearings.
We present the results of our study in Sec.~\ref{sec:result}, including renormalized matrix element comparison between operators and to phenomenological results, a tentative look at the first nucleon gluon quasi-PDF from the data with the best signal, and an early exploration of Coulomb gauge fixing to improve the signal.
The final conclusions and future outlook can be found in Sec.~\ref{sec:conclusion}.

\section{Theoretical Background}\label{sec:theory}

\subsection{Gluon Operators}

Obtaining a lightcone PDF using LaMET starts with the matrix elements of some coordinate-space correlator $O(z)$ having separation in the $z$-direction,
\begin{equation}\label{eq:MEs}
    h^{\text{B}}(z, P_z) = \langle H(P_z) | O(z) | H(P_z) \rangle,
\end{equation}
where $| H(P_z) \rangle$ is the ground state of the hadron $H$ with boost momentum $P_z$.
For the gluon PDF, there is some freedom in the choice of $O(z)$, minding multiplicative renormalizability.
The form of the operators should be~\cite{Zhang:2018diq,Balitsky:2019krf,Wang:2019tgg}
\begin{equation}\label{eq:genGluOp}
    O^{\mu \nu} (z) = F_a^{\mu \gamma}(z)W(z,0)F^{\nu}_{a,\gamma}(0)
\end{equation}
or a combination of such operators, where $F_a^{\mu\alpha} = \partial^\mu A_a^\alpha - \partial^\alpha A_a^\mu - g f_{abc}A_b^\mu A_c^\alpha$ is the gluon field strength tensor, and
\begin{equation}\label{eq:WilsonLine}
    W(z, 0) = \mathcal{P}\exp \left[ -ig\int_0^z \dl{z'} A^z(z')\right]
\end{equation}
is the Wilson line for gauge invariance with $A^z = A^z_a t_a$.
Only some choices of operator indices and summations are known to be multiplicatively renormalizable~\cite{Zhang:2018diq,Balitsky:2019krf}.
We will focus on three operators for the unpolarized gluon PDF
\begin{align}\label{eq:Op1}
O^{(1)}(z) &= F^{z i}(z)W(z,0)F^{z}_{\text{ }i}(0)  \\
\label{eq:Op2}
O^{(2)}(z) &= F^{z \mu}(z)W(z,0)F^{z}_{\text{ }\mu}(0)  \\
\label{eq:Op3}
O^{(3)}(z) &= F^{t i}(z)W(z,0)F^{t}_{\text{ }i}(0) - F^{i j}(z)W(z,0)F_{ij}(0).
\end{align}
Here, the repeated $\mu$ terms denote summation over all Lorentz indices, while $i,j$ means summation over only the transverse indices ($x,y$).
Multiplicative renormalizability at the one-loop level was shown for the first two operators in Ref.~\cite{Zhang:2018diq} and for the last operator in Ref.~\cite{Balitsky:2019krf}.

We choose $O^{(1)}$ and $O^{(2)}$, as these are the only two operators that have hybrid-ratio scheme matching relations derived, and $O^{(3)}$, as it has been shown to produce good signal in the many pseudo-PDF studies~\cite{Fan:2020cpa,Fan:2021bcr,HadStruc:2021wmh,Salas-Chavira:2021wui,Fan:2022kcb,Delmar:2023agv,Good:2023ecp}.

\subsection{Renormalization Procedure}

The hybrid-ratio--renormalized\cite{Ji:2020brr} matrix elements are defined
\begin{equation}\label{eq:hyb_def}
h^R(z,P_z) = \begin{cases}
\frac{h^\text{B}(0,0)}{h^\text{B}(0,P_z)}\frac{h^\text{B}(z,P_z)}{h^\text{B}(z,0)} & z \leq z_s\\
\frac{h^\text{B}(0,0)}{h^\text{B}(0,P_z)}\frac{h^\text{B}(z,P_z)}{h^\text{B}(z_s,0)} \times e^{(\delta m +m_0 )(z-z_s)} &z > z_s
\end{cases},
\end{equation}
where $z_s$ is some scale distance typically chosen to be less than about 0.3~fm, and $\delta m$ and $m_0$ are the mass renormalization and the renormalon ambiguity terms needed to renormalize the linear divergence from the Wilson line self energy.
If $z_s \rightarrow \infty$, we recover the standard ratio-scheme renormalization, which does not take into account the Wilson-line self energy.
The quasi-PDF for the gluon has never been studied directly from lattice data in either renormalization scheme, so we are interested in seeing the hybrid-ratio and ratio-scheme results.
With multiple lattice spacings, $\delta m$ and $m_0$ can be fit independently~\cite{Ji:2020brr,LatticePartonLPC:2021gpi,Zhang:2023bxs}; however, with only a single lattice spacing, it is simpler to fit the sum $\delta m + m_0$ as one term by matching the $P_z = 0$ bare matrix elements to the perturbatively calculated ``Wilson coefficients''.
The Wilson coefficients have only been explicitly calculated for operators $O^{(1)}$ and $O^{(2)}$.
Following Ref.~\cite{Yao:2022vtp}, we write these as
\begin{equation}\label{eq:wil-coef}
    \mathcal{H}^{(i)}\left( 0, \mu^2z^2 \right) = 1 + \frac{\alpha_s}{2\pi}C_A\left(-A^{(i)} L_z + B^{(i)} \right)
\end{equation}
where $L_z = \ln{\left( \frac{4e^{-2\gamma_E}}{\mu^2 z^2}\right)}$ and
\begin{align*}
A^{(1)}&=\frac{11}{6}           &  B^{(1)} &= 4   \\
A^{(2)}&=\frac{11}{6}           &  B^{(2)} &= \frac{14}{3}.
\end{align*}
We fit $\delta m + m_0$ at short distances using the form
\begin{equation}\label{eq:m0-fit}
    (\delta m + m_0)z -I_0 \approx \ln{\left[\mathcal{H}(z,\mu) /h^{\text{B}}(z,0) \right]},
\end{equation}
where $I_0$ is a constant not dependent on $z$.
Typically, one would want to fit using three data points $\{z-a, z, z+a \}$, where $a$ is the lattice spacing, but for coarse lattices, interpolation must be used between data points to get a reasonable fit.

\subsection{The Quasi-PDF and Lightcone Matching}

The Fourier transform of the renormalized matrix elements defines the quasi-PDF, which gives the leading-order behavior of the PDF:
\begin{equation}\label{eq:qPDF_def}
    x\tilde{g}(x,P_z) = \int_{-\infty}^{\infty} \frac{\dl z}{2\pi P_z}e^{ixP_z z}h^\text{R}(z, P_z) .
\end{equation}
It is important to have matrix elements at large enough distances for the integral to converge.
In many cases this is not tractable, since the noise in the lattice matrix elements increases exponentially with distance;
however, based on minimal assumptions on the small-$x$ form of the lightcone PDF, a model involving an exponential decay can be used for a large-distance extrapolation~\cite{Ji:2020brr}.
Obtaining signal at far enough distances to reliably make this extrapolation is still difficult in the gluon case.

The lightcone PDF is then related to the quasi-PDF through a matching relationship:
\begin{multline}\label{eq:matching}
    \tilde{g}(x,P_z) = \int_{-1}^{1}\dl y\, K_{gg}(x,y,\mu/P_z)g(y,\mu) \\+ K_{gq}(x,y,\mu/P_z)q(y,\mu)
    {}+ \mathcal{O}\left(\frac{\Lambda_\text{QCD}^2}{(xP_z)^2},\frac{\Lambda_\text{QCD}^2}{((1-x)P_z)^2}\right)
\end{multline}
where $K_{gg}(x,y,\mu/P_z)$ and $K_{gq}(x,y,\mu/P_z)$ are the perturbatively calculated glue-glue and glue-quark matching kernels, $g(y,\mu)$ and $q(y,\mu)$ are the lightcone gluon and quark PDFs, and $\mu$ is the renormalization scale.
Lightcone PDFs are most often quoted in the modified minimal subtraction ($\overline{\text{MS}}$) scheme.
The kernels handle matching between the lattice schemes and continuum schemes, as well.
The quasi-PDF matching kernels for $O^{(1)}$ and $O^{(2)}$ for ratio and hybrid-ratio renormalization to the $\overline{\text{MS}}$ are derived in Ref.~\cite{Yao:2022vtp}.
Only the pseudo-PDF matching kernels have been explicitly derived in the literature for $O^{(3)}$ in the ratio scheme to $\overline{\text{MS}}$~\cite{Balitsky:2019krf}.
The numerical implementation of the integration in Eq.~\ref{eq:matching} can be written as a matrix-vector multiplication and inverted to find the lightcone PDF from the quasi-PDF.
The perturbative scales $\left(\frac{\Lambda_\text{QCD}^2}{(xP_z)^2},\frac{\Lambda_\text{QCD}^2}{((1-x)P_z)^2}\right)$ suggest that the accuracy of the PDF is limited by the hadron momentum and that the PDF will be more accurate in the mid-$x$ region.

\section{Bare Lattice Matrix Elements}\label{sec:bMEs}


We perform high-statistics calculations on one ensemble with lattice spacing $a \approx 0.12$~fm at two valence pion masses $M_{\pi} \approx 310$ and $690$~MeV generated using 2+1+1 flavors of highly improved staggered quarks (HISQ)~\cite{Follana:2007rc} by the MILC collaboration~\cite{MILC:2013znn} with the lattice volume of $24^3 \times 64$.
Wilson-clover fermions are used in the valence sector and valence quark masses are tuned to reproduce the lightest light and strange masses of the HISQ sea.
The same valance quark parameters are used in the PNDME collaboration~\cite{Gupta:2018qil}.
1,296,640 two point correlator measurements were performed across 1013 configurations to obtain the data presented in this paper.
For the three point correlators we looked at two types of gauge smearings to improve the signal.
We look at data from configurations where with five steps of hypercubic smearing (HYP5) in order to directly compare to previous results from our group~\cite{Fan:2020cpa,Fan:2021bcr,Salas-Chavira:2021wui,Fan:2022kcb,Good:2023ecp}.
We also consider more aggressively smeared lattice where we apply Wilson flow with flow time $T=3 a^2$ (Wilson-3) to the gauge links.

The two-point correlator is defined on the lattice as
\begin{equation}
    C_H^\text{2pt}(P_z;t)=\langle 0|\Gamma\int d^3ye^{-iy_zP_z} \chi(\vec y,t)\chi(\vec 0,0)|0 \rangle
\end{equation}
where $P_z$ is the hadron momentum in the spatial $z$-direction, $t$ is the lattice euclidean time, $\chi(y)$  is the interpolation operator for a specific hadron being analysed and $\Gamma=\frac{1}{2}(1+\frac{\gamma}{4})$ is the projection operator.
The three point correlator is then calculated by combining the gluon loop with the two point correlator.
The three point correlator is defined as
\begin{multline}
     C_H^\text{3pt}(P_z;t_\text{sep},t)=
     \\
     \langle 0|\Gamma\int d^3ye^{-iy_zP_z}\chi(\vec y,t_\text{sep})O(z,t)\chi(\vec 0,0)|0\rangle
\end{multline}
where $t_\text{sep}$ is the source-sink time separation and $t$ is the gluon operator insertion time.

\begin{figure*}
\centering
    \includegraphics[width=0.3\textwidth]{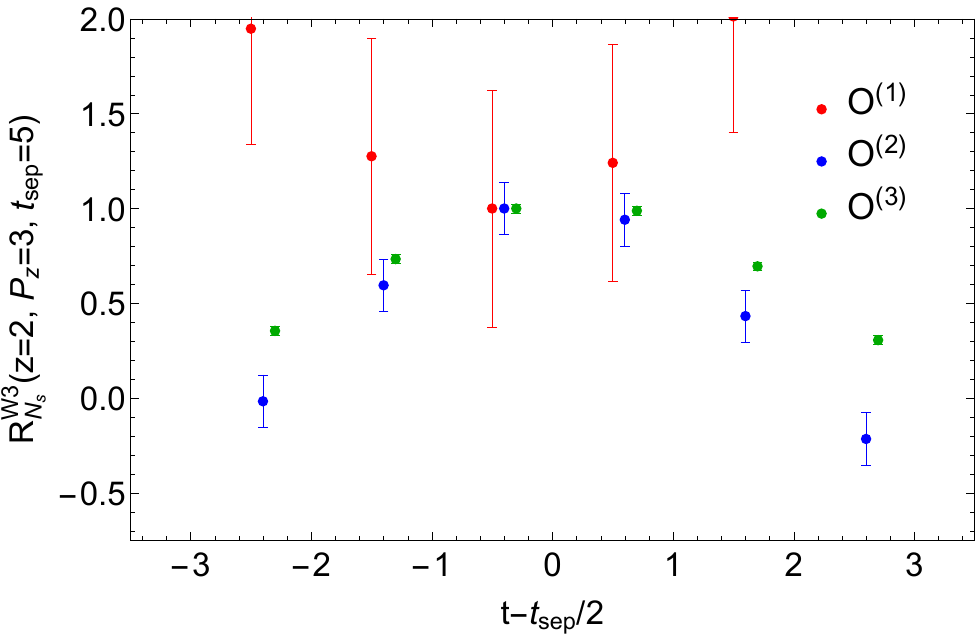}
\centering
    \includegraphics[width=0.3\textwidth]{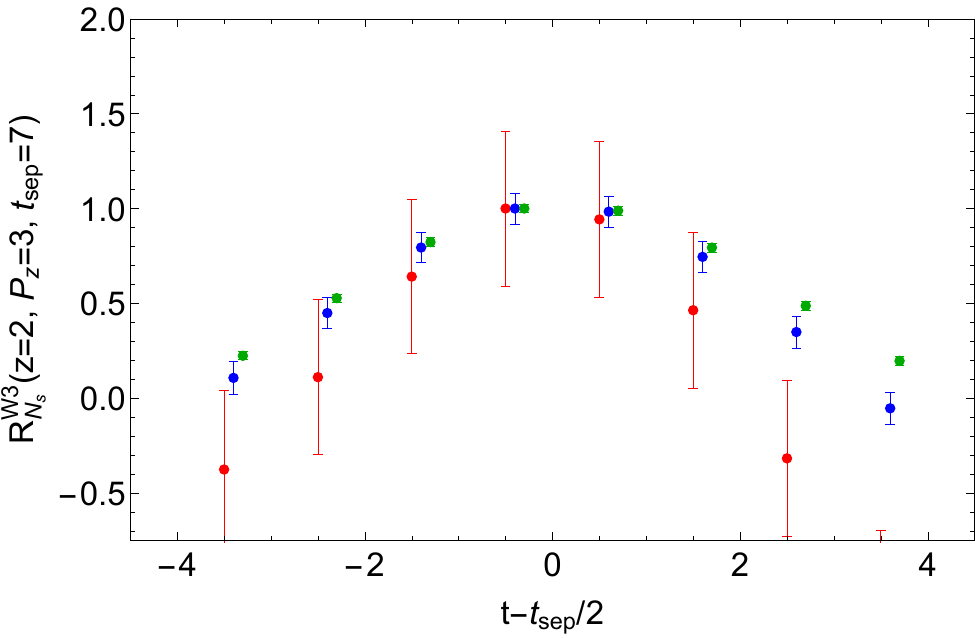}
\centering
    \includegraphics[width=0.3\textwidth]{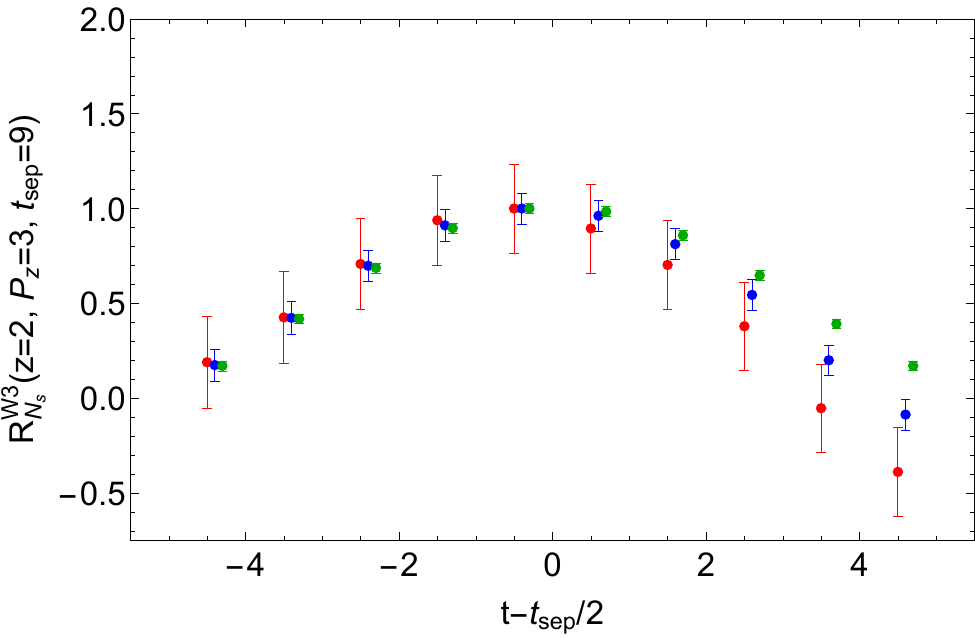}
\centering
    \includegraphics[width=0.3\textwidth]{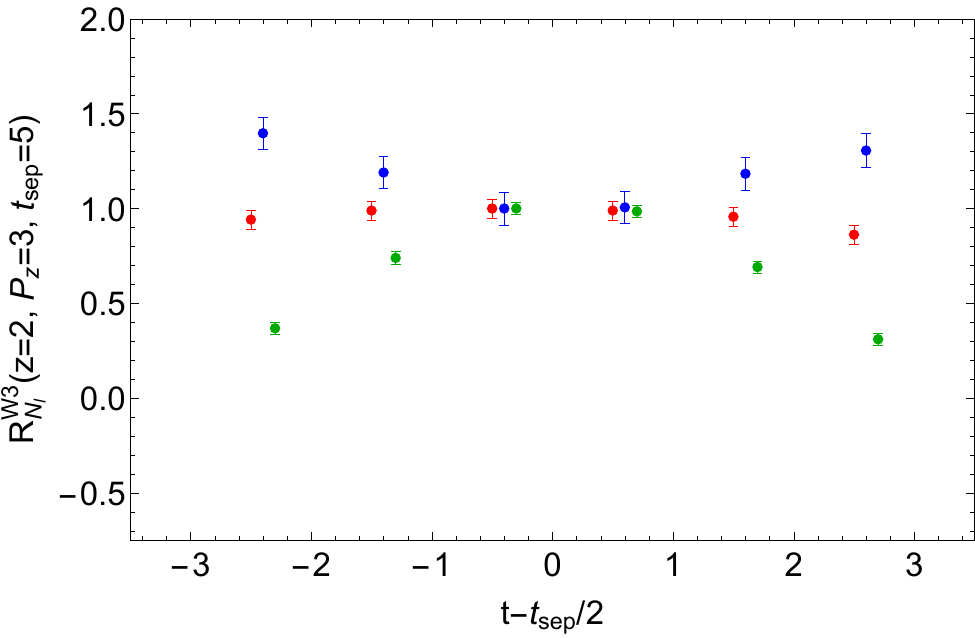}
\centering
    \includegraphics[width=0.3\textwidth]{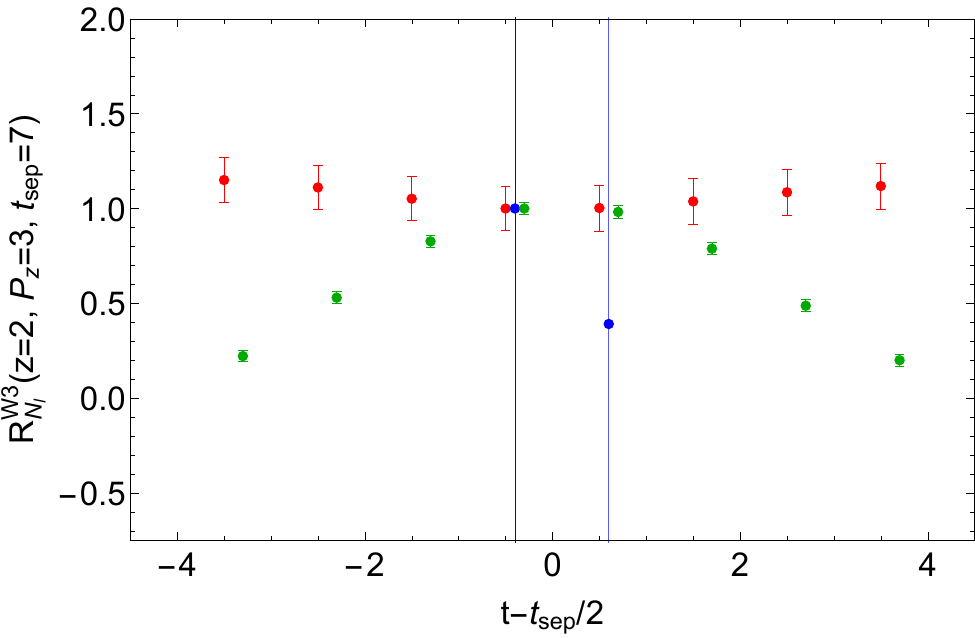}
\centering
    \includegraphics[width=0.3\textwidth]{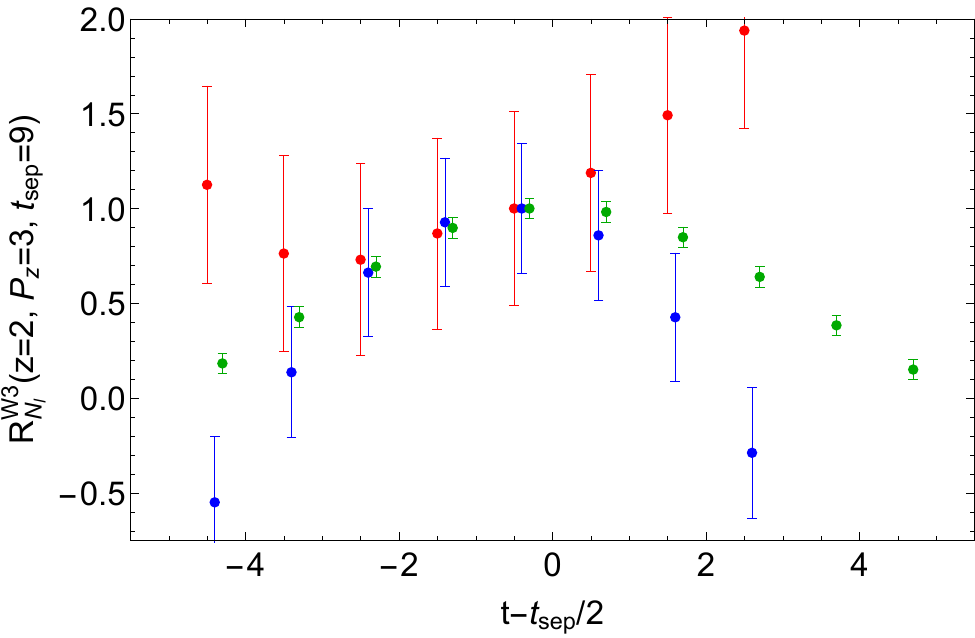}
\centering
    \includegraphics[width=0.3\textwidth]{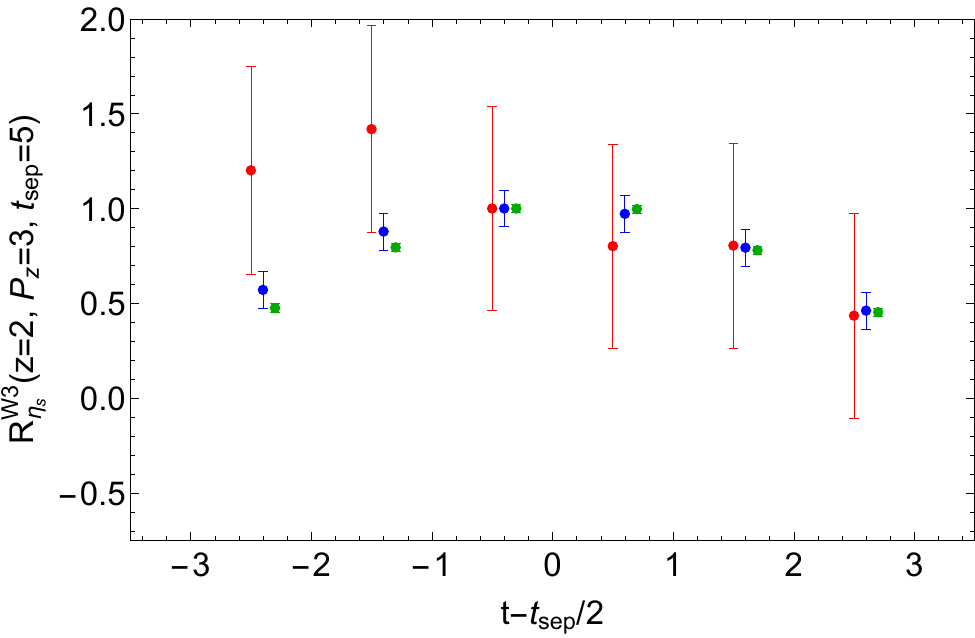}
\centering
    \includegraphics[width=0.3\textwidth]{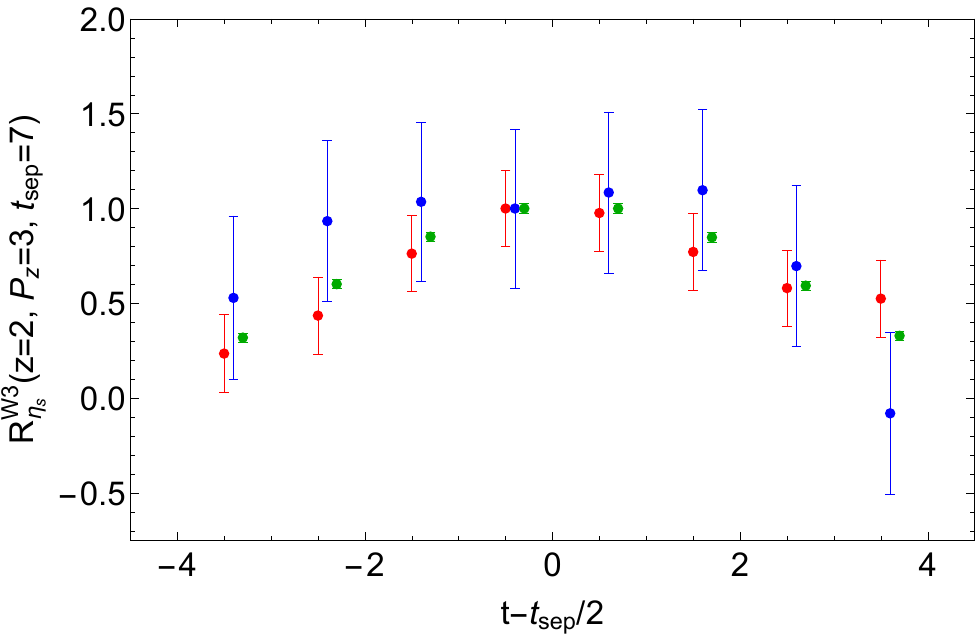}
\centering
    \includegraphics[width=0.3\textwidth]{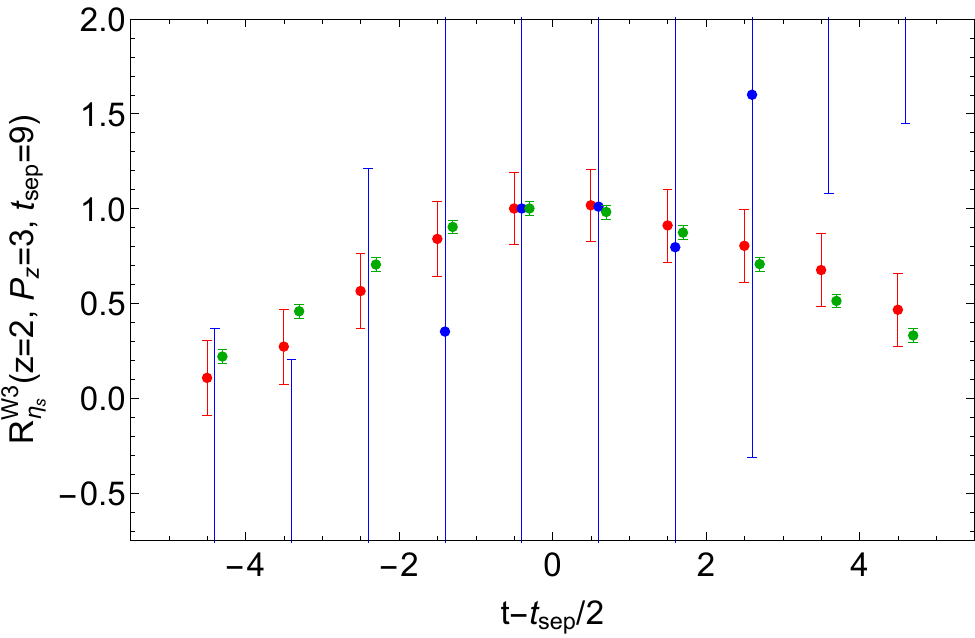}
\centering
    \includegraphics[width=0.3\textwidth]{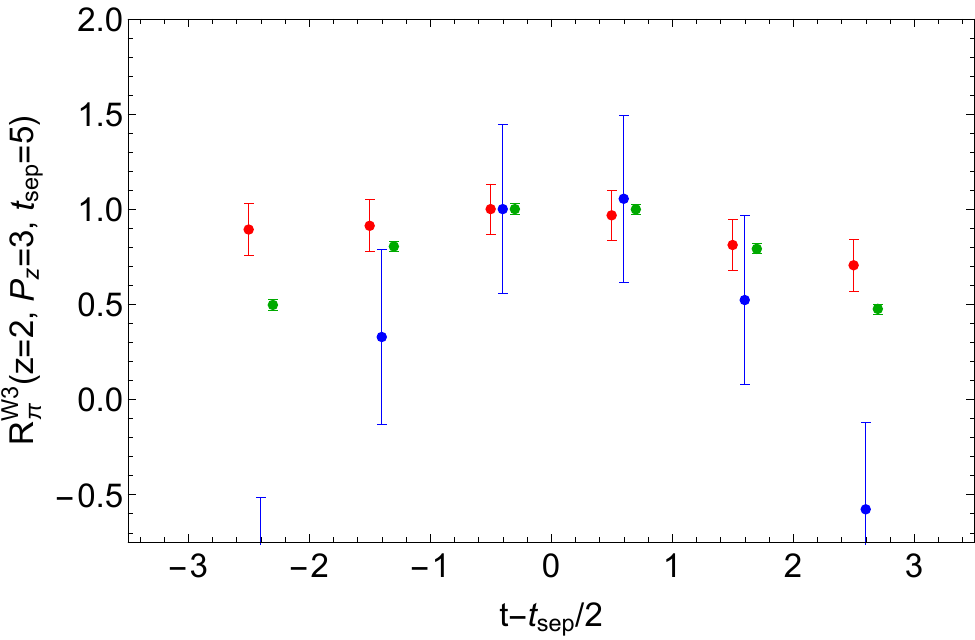}
\centering
    \includegraphics[width=0.3\textwidth]{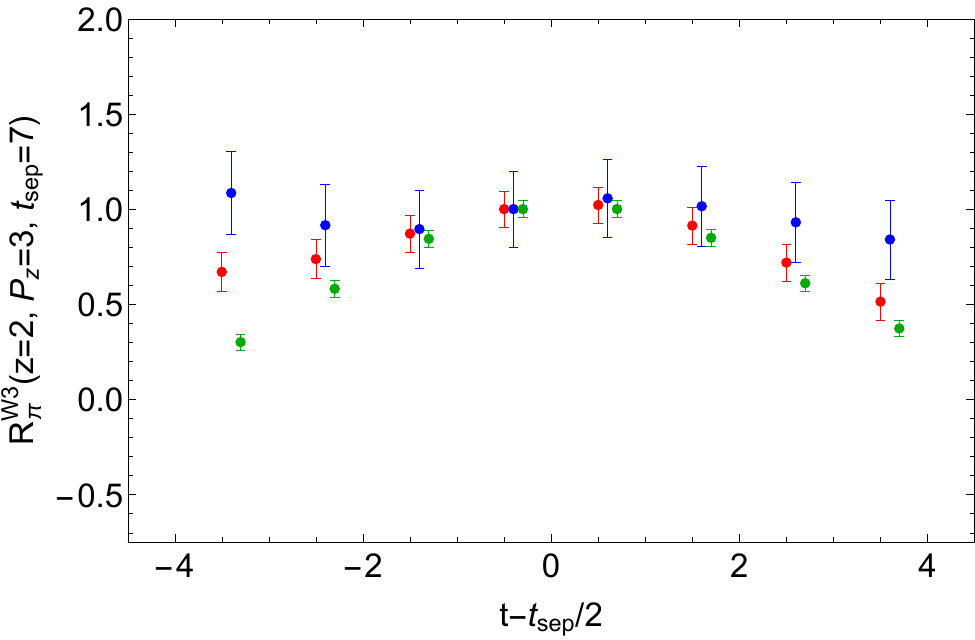}
\centering
    \includegraphics[width=0.3\textwidth]{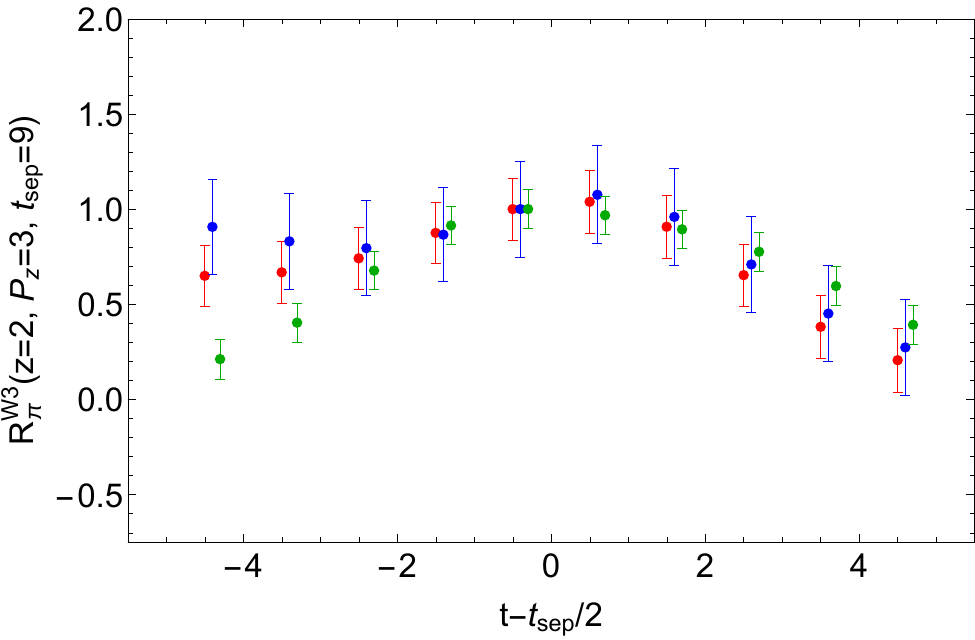}

    \caption{Ratio of the three point to the two point correlators at fixed $t_\text{sep} = 5a,7a,9a$ (left to right columns) for the strange nucleon, light nucleon, $\eta_s$, and $\pi$ with Wilson-3 smearing.
    In each plot, $O^{(1)}$, $O^{(2)}$, and $O^{(3)}$ are each plotted with the mean of left center-most point normalized to 1.
    The second two operators are offset to the right slightly for clarity.}
    \label{fig:W3_ratio_compare}
\end{figure*}

\begin{figure*}
\centering
    \includegraphics[width=0.3\textwidth]{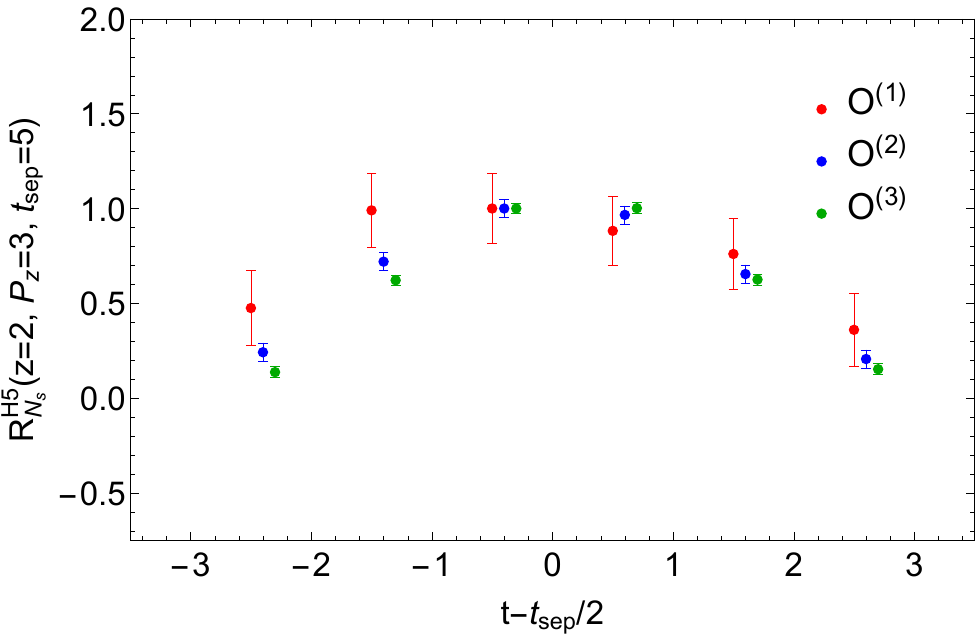}
\centering
    \includegraphics[width=0.3\textwidth]{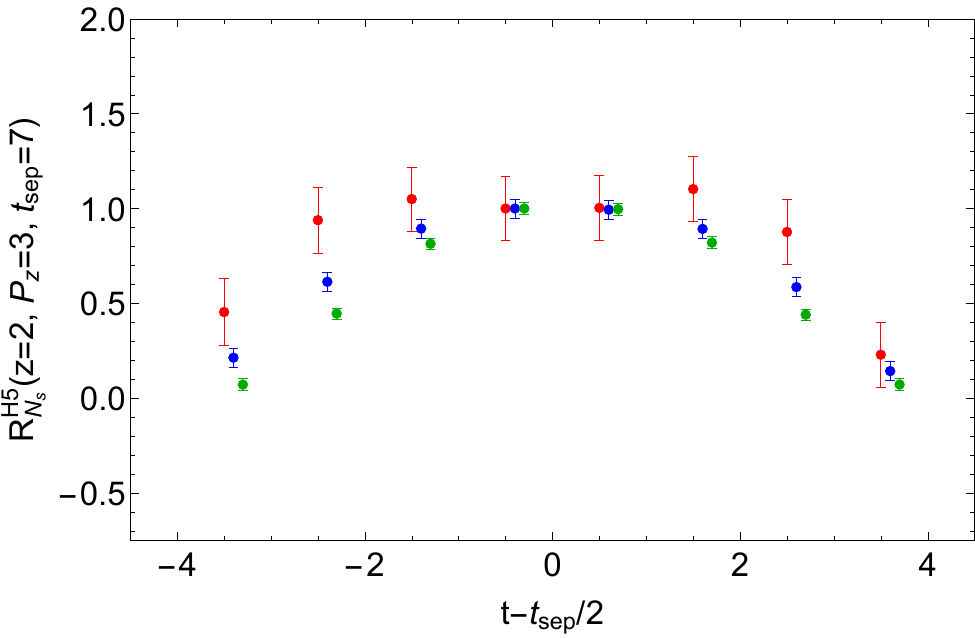}
\centering
    \includegraphics[width=0.3\textwidth]{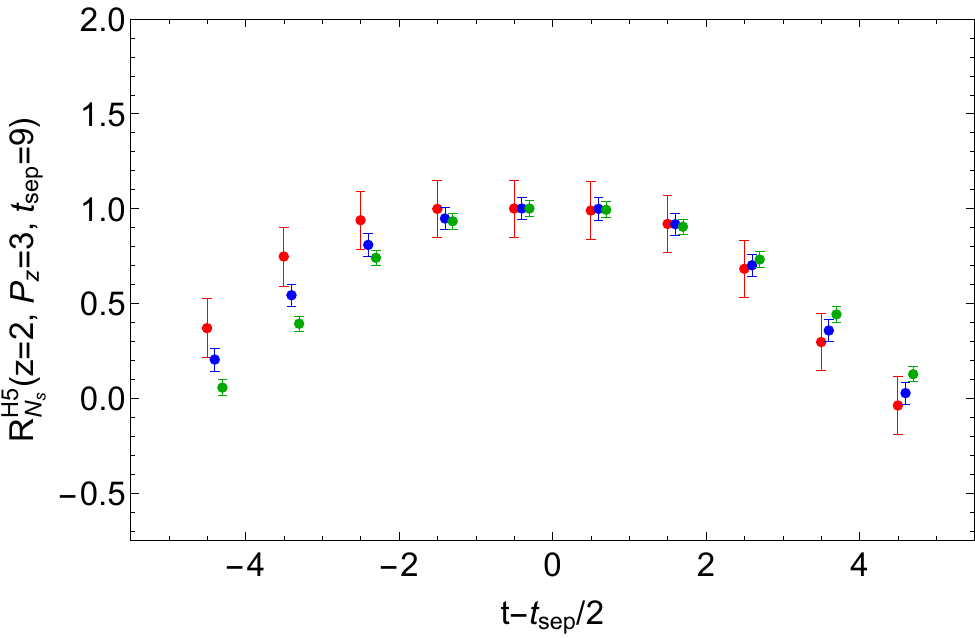}
\centering
    \includegraphics[width=0.3\textwidth]{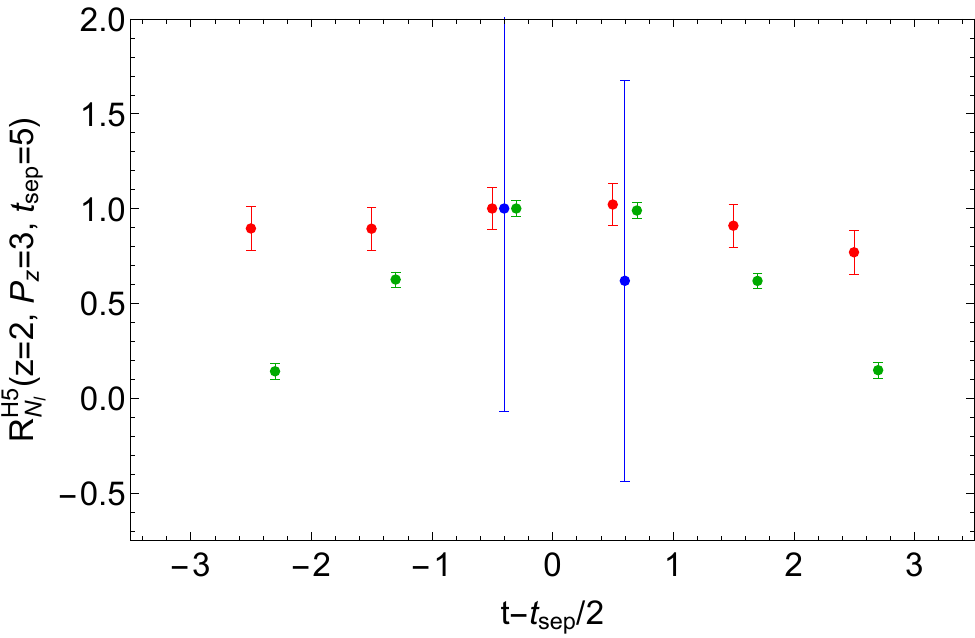}
\centering
    \includegraphics[width=0.3\textwidth]{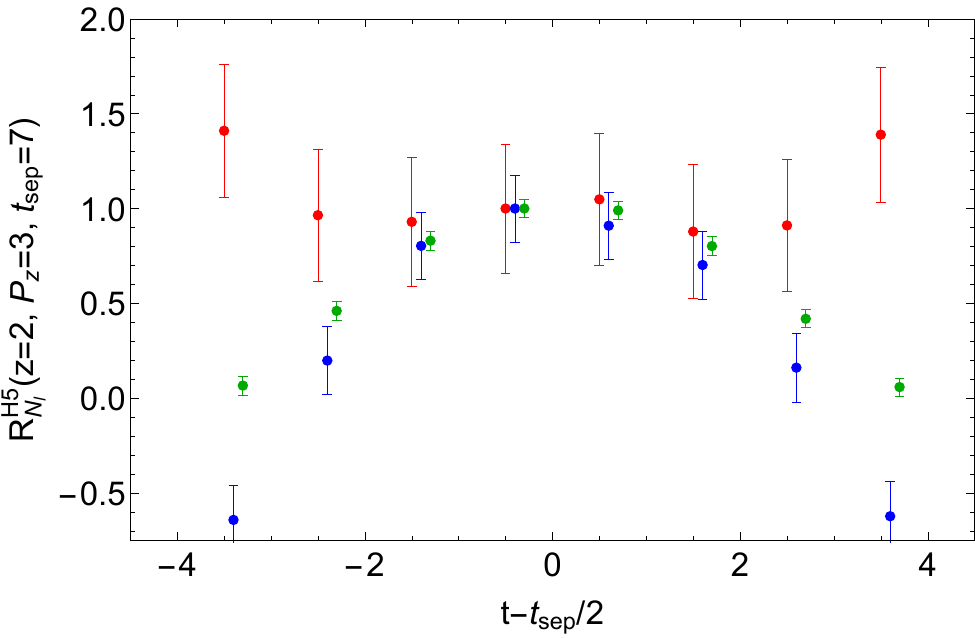}
\centering
    \includegraphics[width=0.3\textwidth]{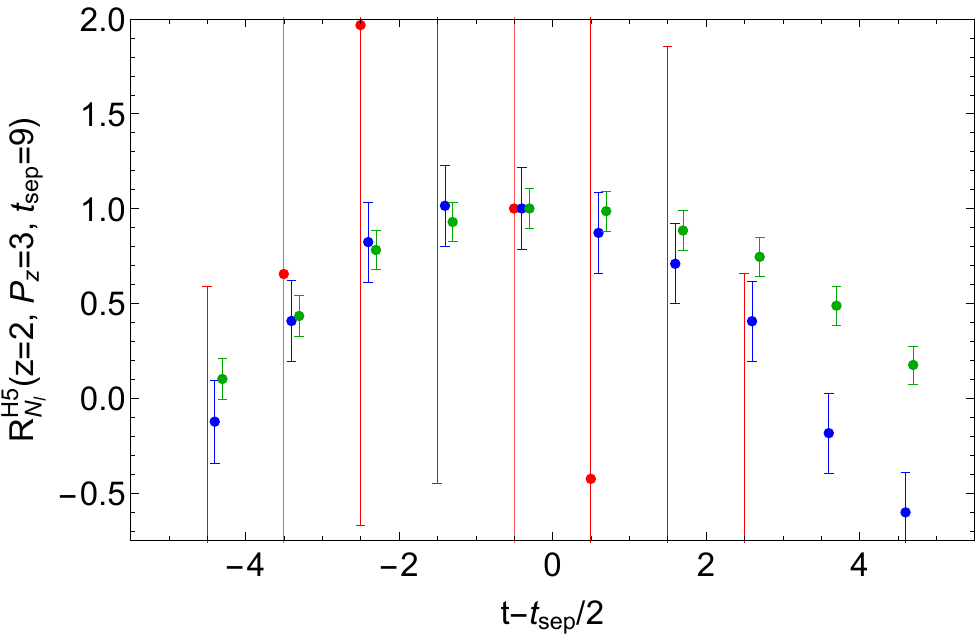}
\centering
    \includegraphics[width=0.3\textwidth]{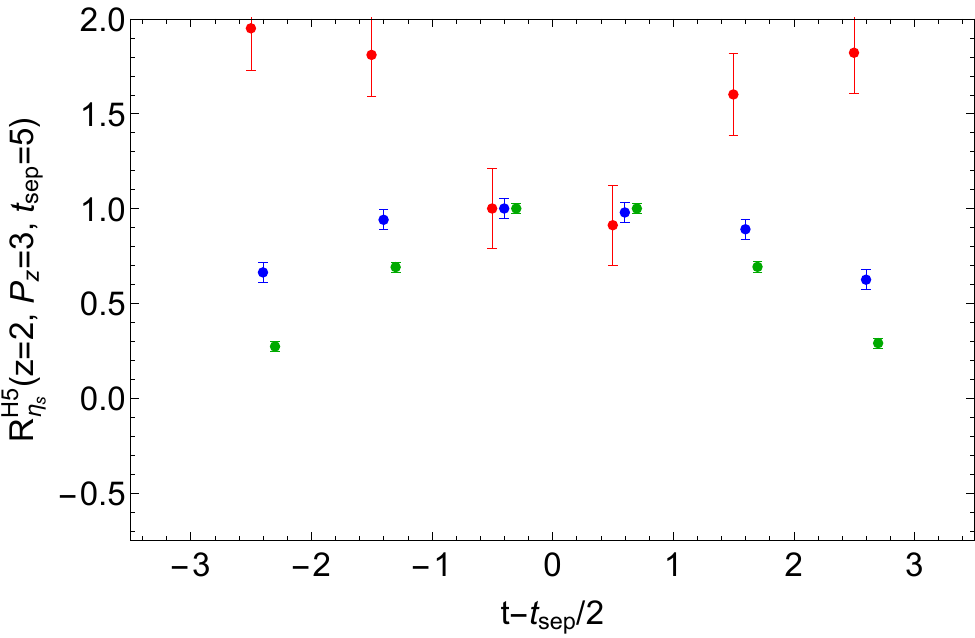}
\centering
    \includegraphics[width=0.3\textwidth]{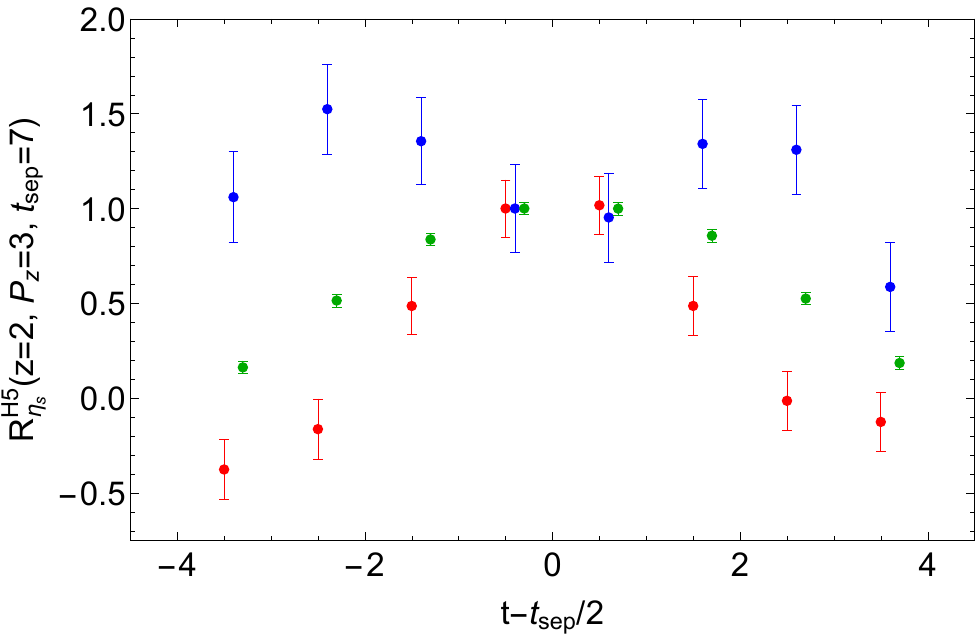}
\centering
    \includegraphics[width=0.3\textwidth]{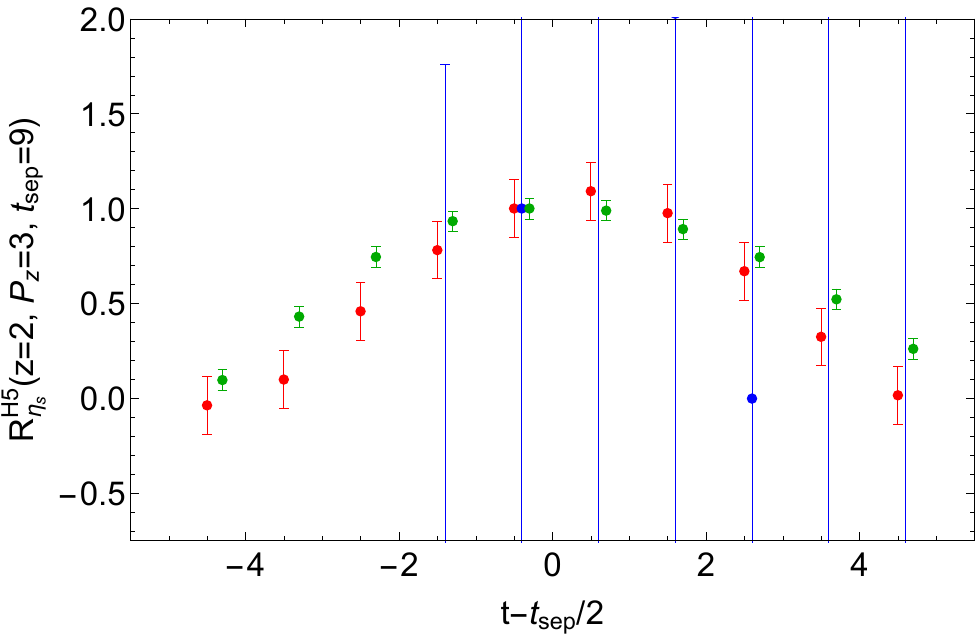}
\centering
    \includegraphics[width=0.3\textwidth]{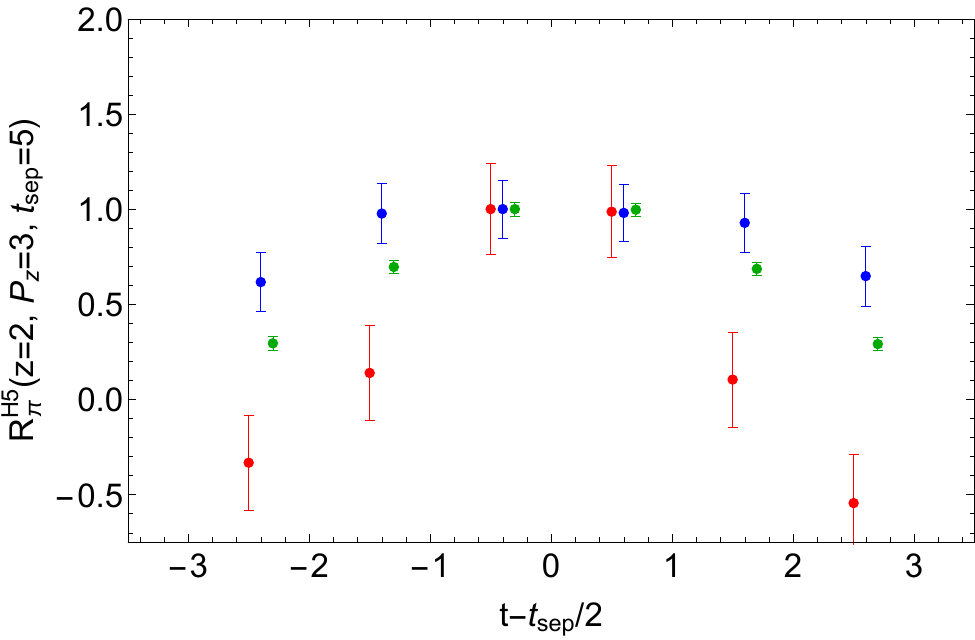}
\centering
    \includegraphics[width=0.3\textwidth]{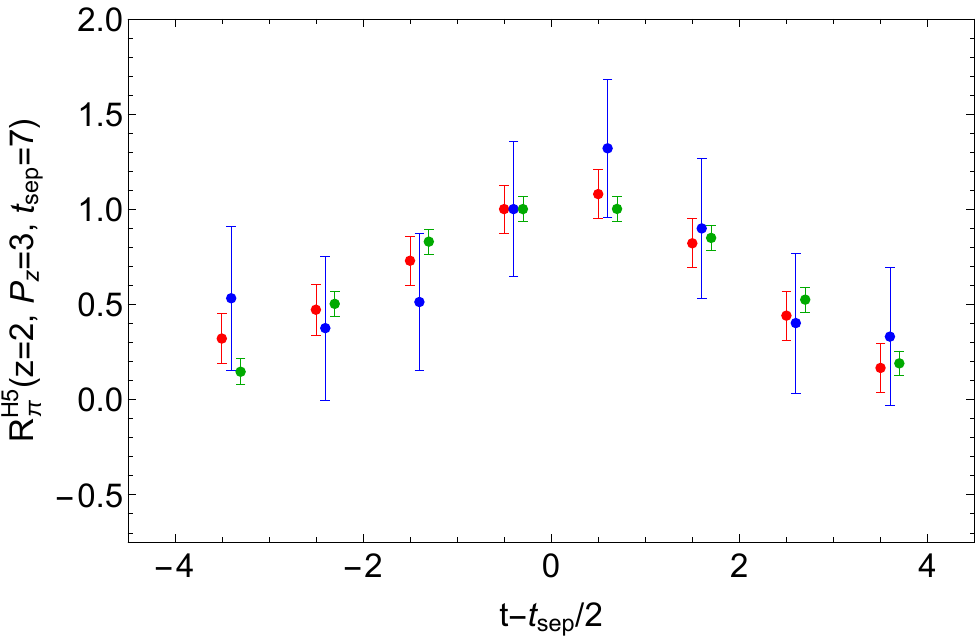}
\centering
    \includegraphics[width=0.3\textwidth]{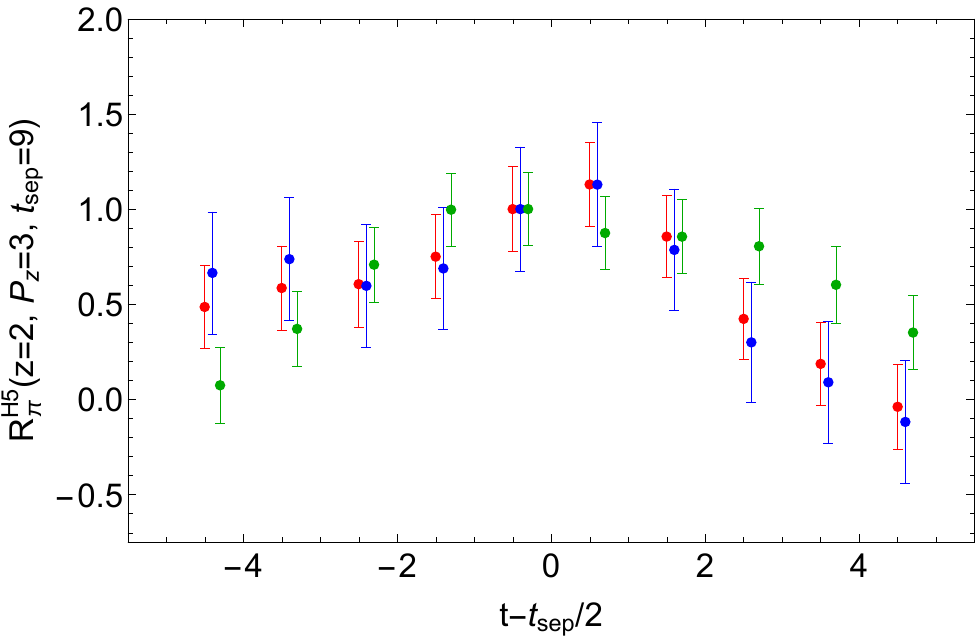}
    \caption{Ratio of the three point to the two point correlators at fixed $t_\text{sep} = 5a,7a,9a$ (left to right columns) for the strange nucleon, light nucleon, $\eta_s$, and $\pi$ with HYP5 smearing.
    In each plot, $O^{(1)}$, $O^{(2)}$, and $O^{(3)}$ are each plotted with the mean of left center-most point normalized to 1.
    The second two operators are offset to the right slightly for clarity.}
    \label{fig:H5_ratio_comp}
\end{figure*}

To judge the how well the operators perform, we may compare the signal and behaviors of the ratios of the two and three point correlators.
\begin{equation} \label{eq:ratio}
      R_H(P_z;t_\text{sep},t)=\frac{ C_H^\text{3pt}(P_z;t_\text{sep},t)}{ C_H^\text{2pt}(P_z;t_\text{sep})}
\end{equation}
We plot selected ratios for each hadron and operator for $t_\text{sep} = 5a,7a,9a$ in Figs.~\ref{fig:W3_ratio_compare} and \ref{fig:H5_ratio_comp} for the Wilson-3 and HYP5 smearings.
In these plots, we normalize such that the mean of the left center-most ratio in each plot for each operator is equal to one, otherwise, the results would not be easily comparable.
We see already at this point that in most cases, $O^{(3)}$ has the best signal compared to the other operators and often very symmetrical behavior, which is to be expected for these plots.
We see that in some cases, the smaller $t_\text{sep}$ data for $O^{(1)}$ and $O^{(2)}$ have larger error or different behavior than the other $t_\text{sep}$.
This is mostly due to these data being close to 0, so the normalization inflates some of the error and exaggerates some trends.
This is already suggestive that the best ground state matrix elements will likely come from $O^{(3)}$.

The two and three point correlators can be fit to the energy eigenstate expression as,
\begin{equation}
     C_H^\text{2pt}(P_z;t)=|A_{H,0}|^2e^{-E_{H ,0}t } + |A_{H ,1}|^2e^{-E_{H ,1}t } + .....
\end{equation}

\begin{multline}
\label{eq:3pt-fit-form}
        C_H^{\text{3pt}}(z,P_z;t_\text{sep},t) = |A_{H,0}|^2 \langle 0|O_{g}|0\rangle e^{-E_{H,0}t_\text{sep}} \\
         + |A_{H,0}| |A_{H,1}| \langle 0|O|1\rangle e^{-E_{H,0}(t_\text{sep}-t)} e^{-E_{H,0}t} \\
         + |A_{H,0}| |A_{H,1}| \langle 1|O|0\rangle e^{-E_{H,1}(t_\text{sep}-t)} e^{-E_{H,0}t} \\
         + |A_{H,1}|^2 \langle 1|O|1\rangle e^{-E_{H,1}t_\text{sep}}+ \cdots
\end{multline}

The ground (first excited) state amplitudes and energies, $A_{H ,0},E_{H ,0},  (A_{H,1},E_{H ,1})$ are obtained from the two-state fits of the two point correlators.
$\langle 0|O_{g}|0 \rangle,\langle 0|O|1\rangle= \langle 1|O|0\rangle, \langle 1|O|1\rangle$ are ground state and excited state matrix elements which are extracted from the two-state simultaneous fits to the three point correlator at multiple values of $t_\text{sep}$.

\begin{figure*}[htbp]
\centering
    \includegraphics[width=0.95\textwidth]{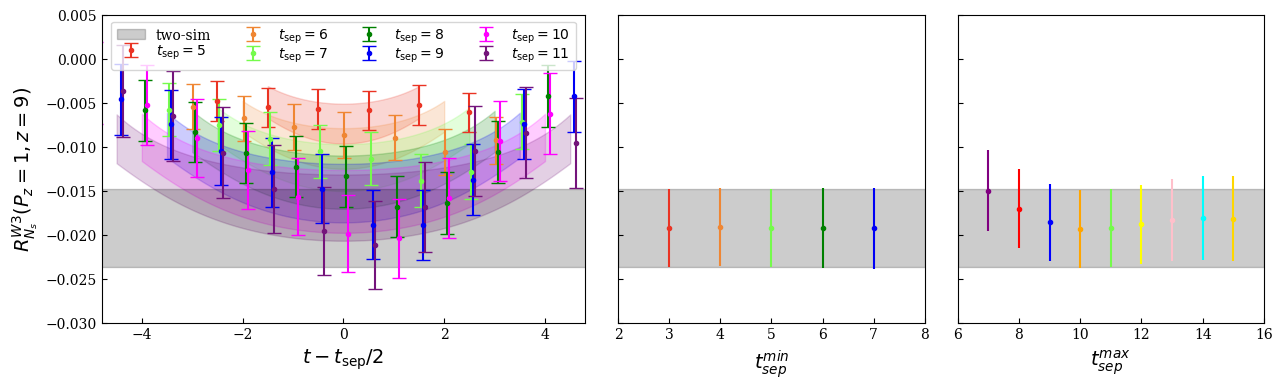}
\centering
    \includegraphics[width=0.95\textwidth]{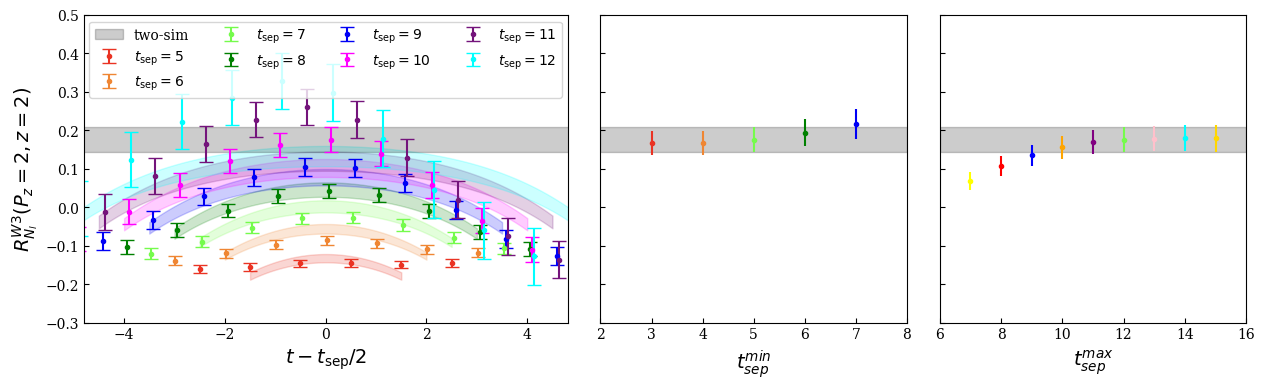}
\centering
    \includegraphics[width=0.95\textwidth]{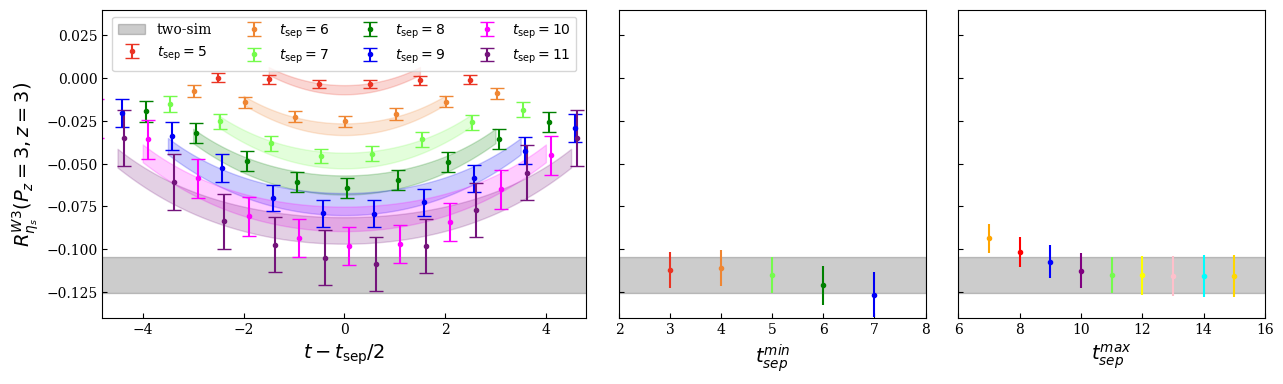}
\centering
    \includegraphics[width=0.95\textwidth]{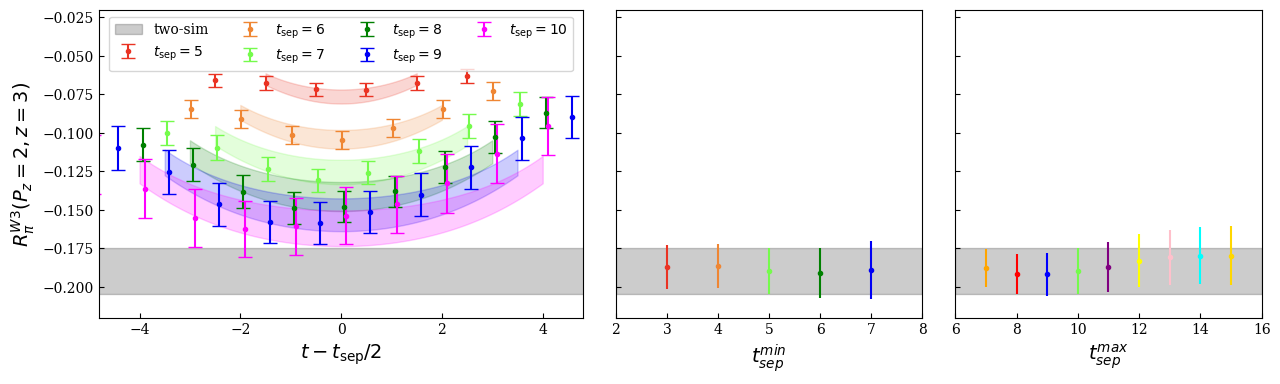}

\caption{\raggedright This figure shows example ratio plots in the left column , the extracted matrix elements as we vary the $t_\text{sep}^\text{min}$ in the middle column, and the extracted matrix elements as we vary the $t_\text{sep}^\text{max}$ in the right column, for strange nucleon, light nucleon, $\eta_s$ and $\pi$ from top to bottom. This analysis was done with operator $O^{(1)}(z)$ and Wilson3 smearing.}
\label{fig:O3JZ_W3}
\end{figure*}

\begin{figure*}[htbp]
\centering
    \includegraphics[width=0.95\textwidth]{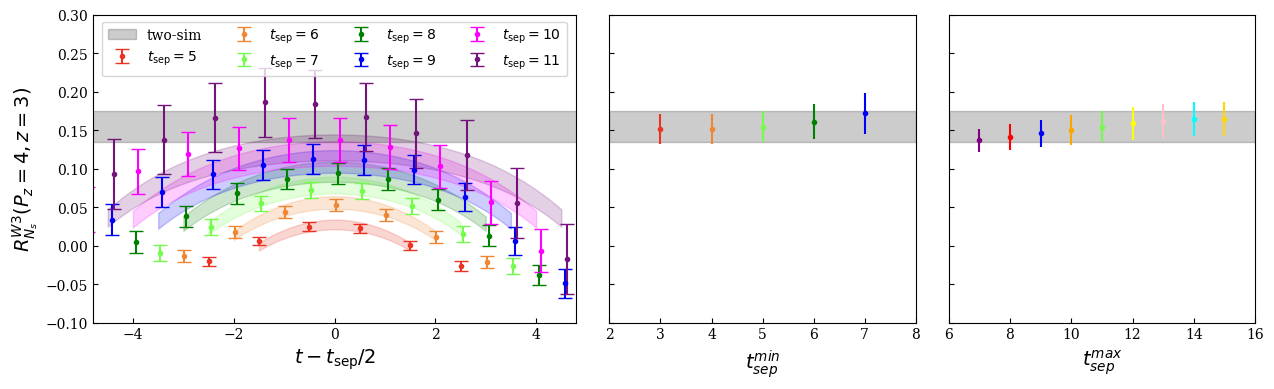}
\centering
    \includegraphics[width=0.95\textwidth]{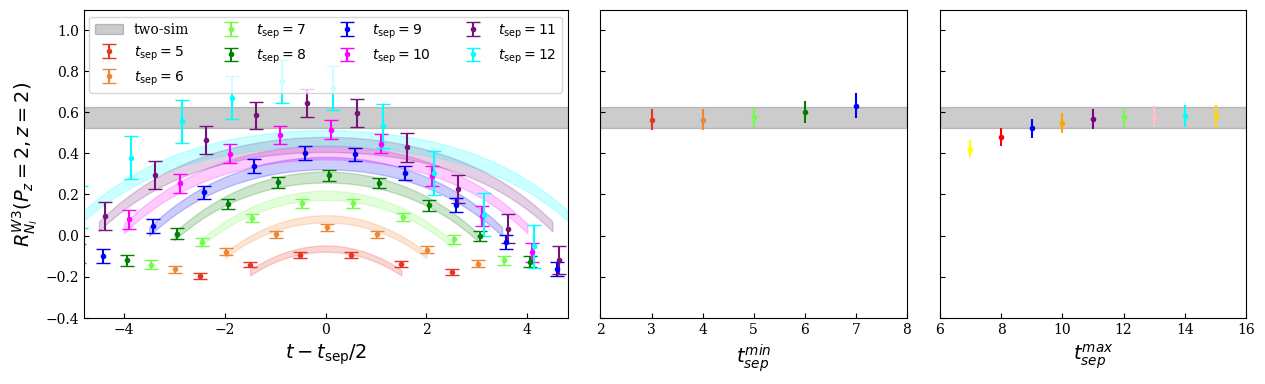}
\centering
    \includegraphics[width=0.95\textwidth]{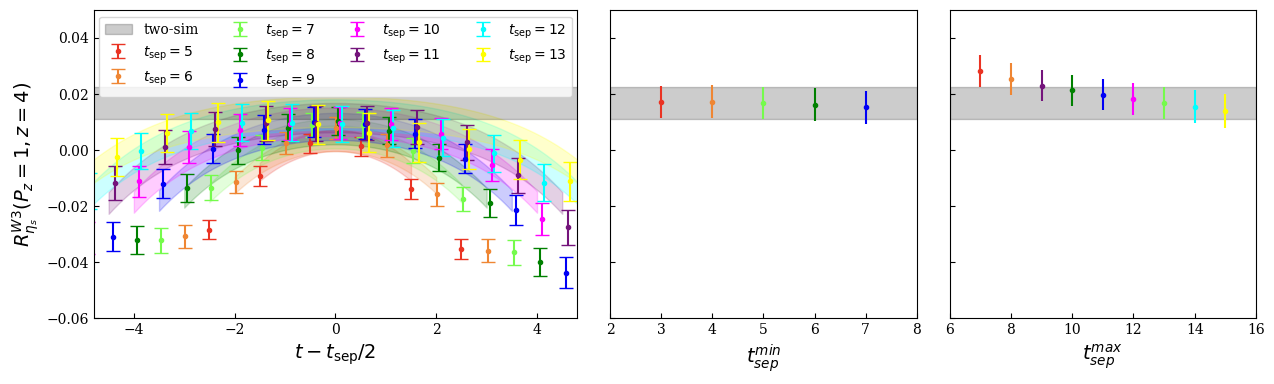}
\centering
    \includegraphics[width=0.95\textwidth]{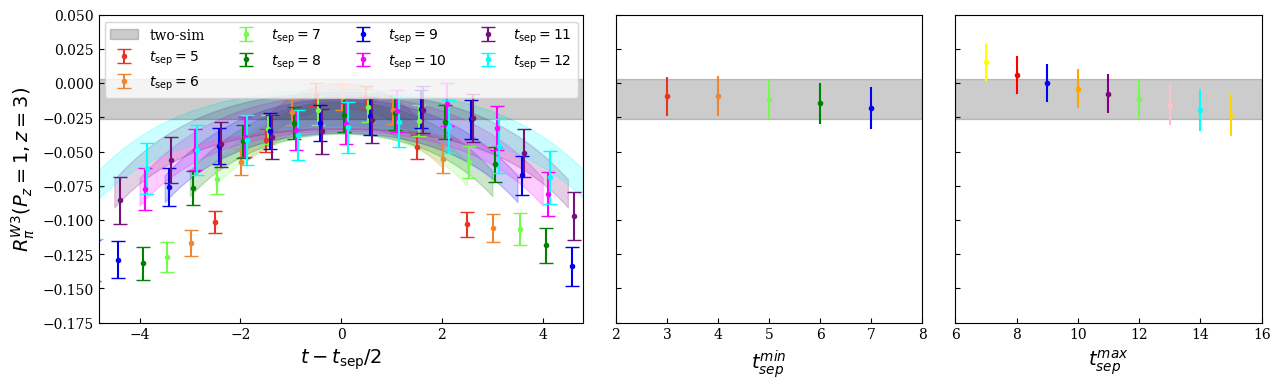}
    \caption{This figure shows example ratio plots in the left column , the extracted matrix elements as we vary the $t_\text{sep}^\text{min}$ in the middle column, and the extracted matrix elements as we vary the $t_\text{sep}^\text{max}$ in the right column, for strange nucleon, light nucleon, $\eta_s$ and $\pi$ from top to bottom. This analysis was done with operator $O^{(2)}(z)$ and Wilson3 smearing.}
    \label{fig:O3ZF_W3}
\end{figure*}

\begin{figure*}[htbp]
\centering
    \includegraphics[width=0.95\textwidth]{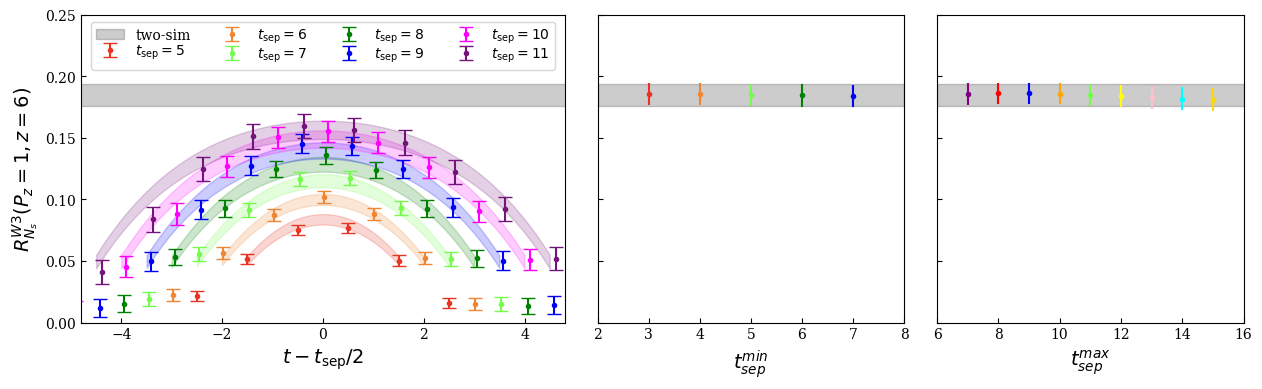}
\centering
    \includegraphics[width=0.95\textwidth]{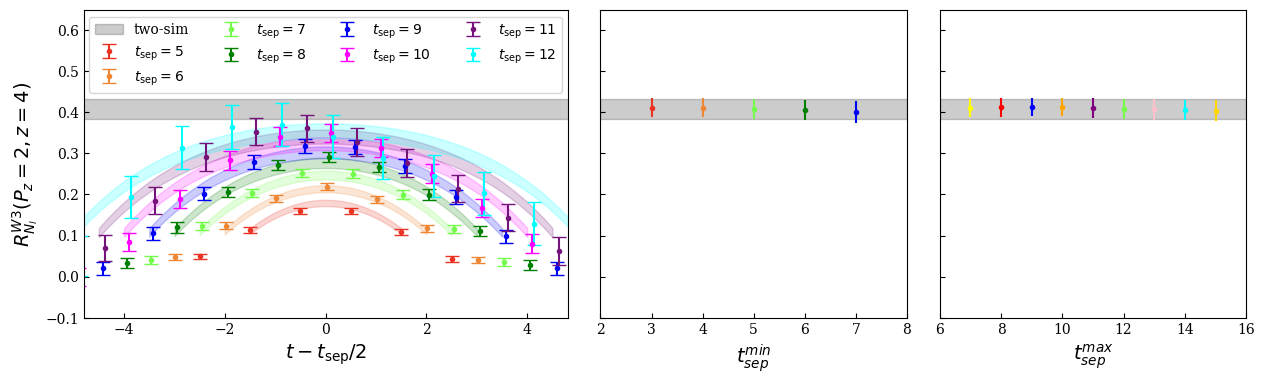}
\centering
    \includegraphics[width=0.95\textwidth]{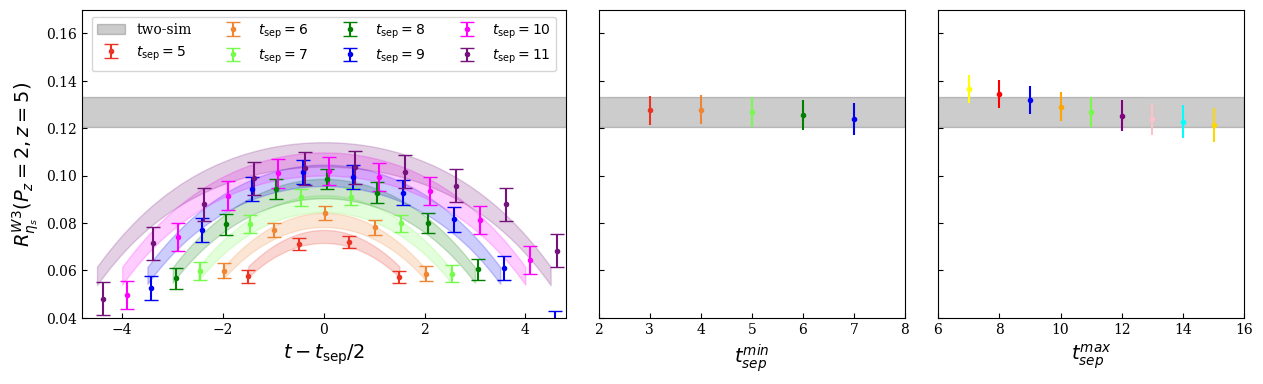}
\centering
    \includegraphics[width=0.95\textwidth]{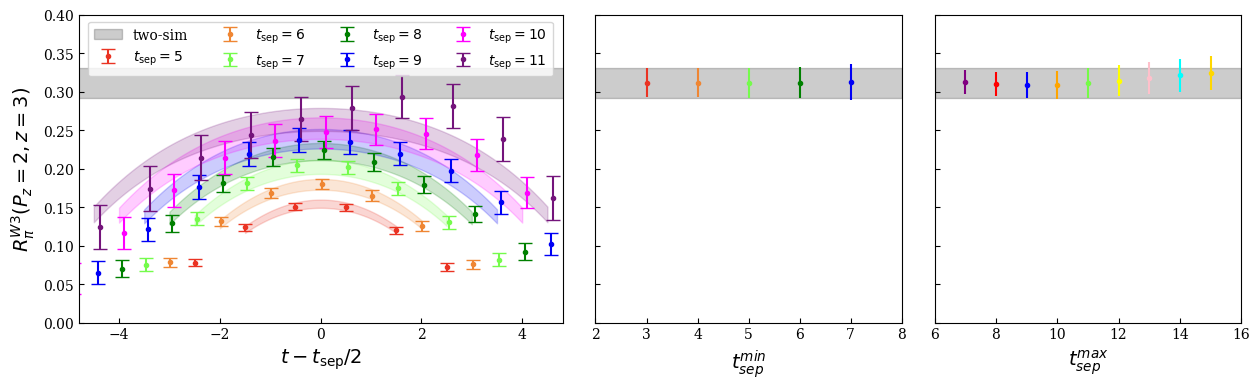}
    \caption{This figure shows example ratio plots in the left column , the extracted matrix elements as we vary the $t_\text{sep}^\text{min}$ in the middle column, and the extracted matrix elements as we vary the $t_\text{sep}^\text{max}$ in the right column, for strange nucleon, light nucleon, $\eta_s$ and $\pi$ from top to bottom. This analysis was done with operator $O^{(3)}(z)$ and Wilson3 smearing.}
    \label{fig:O0IB_W3}
\end{figure*}

\begin{figure*}[htbp]
\centering
    \includegraphics[width=0.95\textwidth]{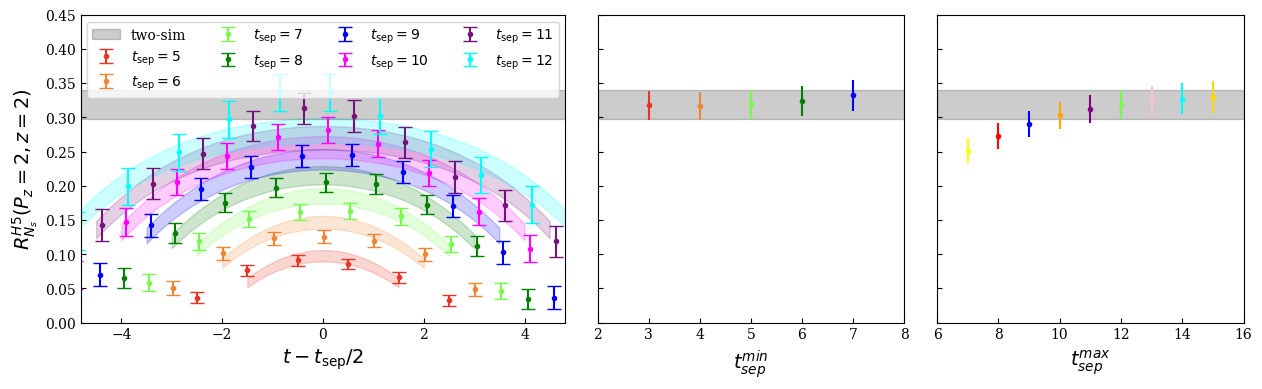}
\centering
    \includegraphics[width=0.95\textwidth]{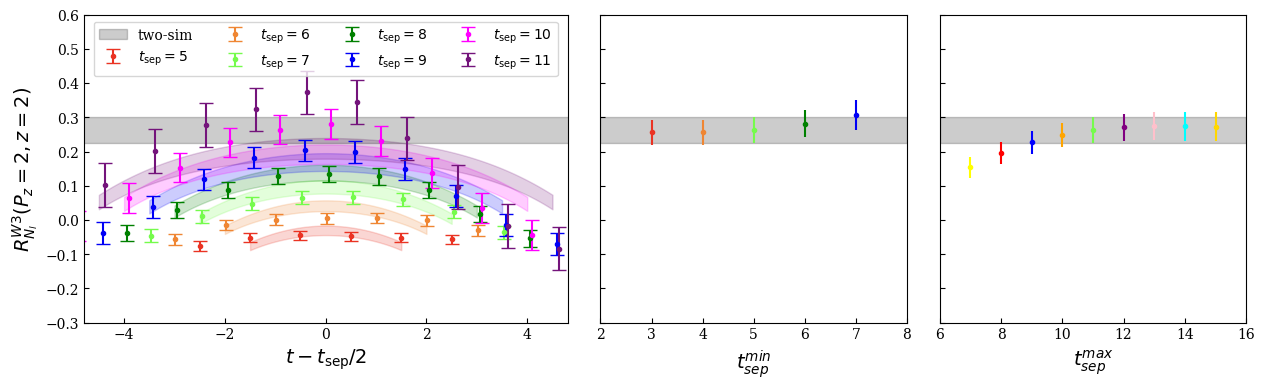}
\centering
    \includegraphics[width=0.95\textwidth]{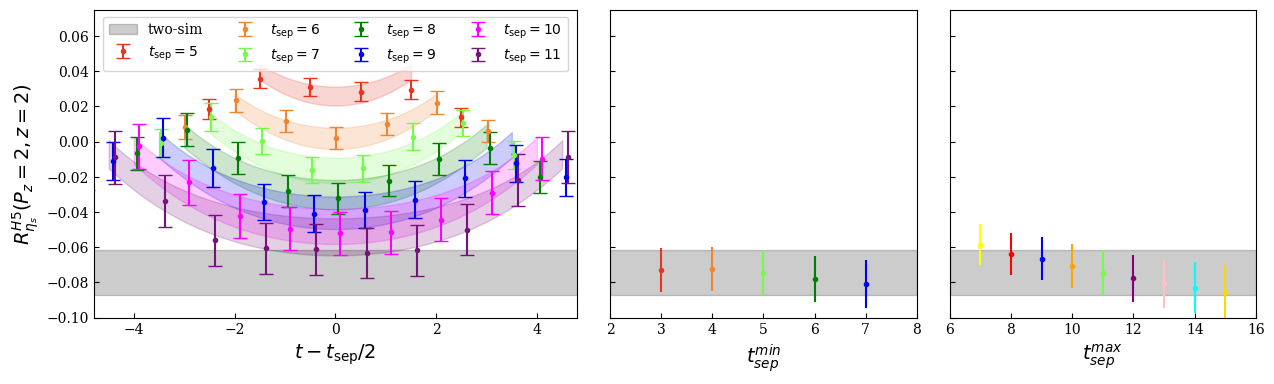}
\centering
    \includegraphics[width=0.95\textwidth]{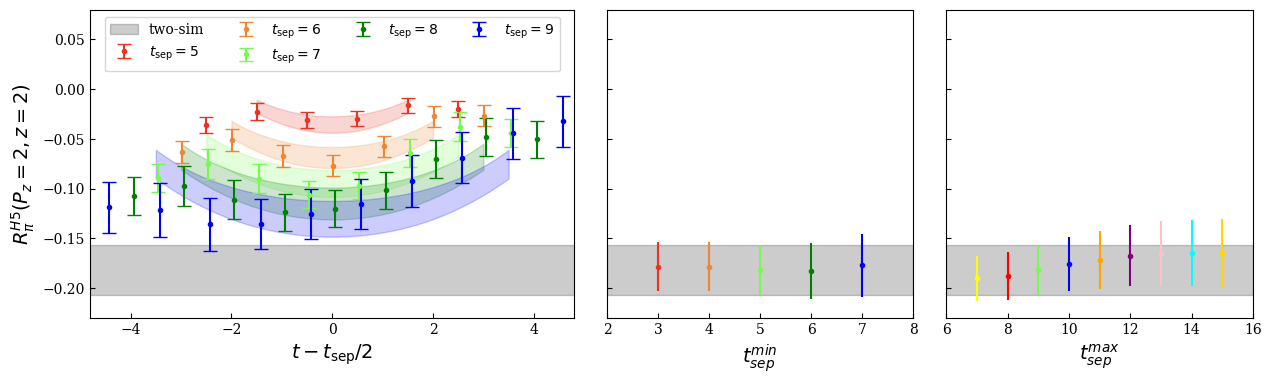}
    \caption{This figure shows example ratio plots in the left column , the extracted matrix elements as we vary the $t_\text{sep}^\text{min}$ in the middle column, and the extracted matrix elements as we vary the $t_\text{sep}^\text{max}$ in the right column, for strange nucleon, light nucleon, $\eta_s$ and $\pi$ from top to bottom. This analysis was done with operator $O^{(1)}(z)$ and HYP5 smearing.}
    \label{fig:O3JZ_H5}
\end{figure*}

\begin{figure*}[htbp]
\centering
    \includegraphics[width=0.95\textwidth]{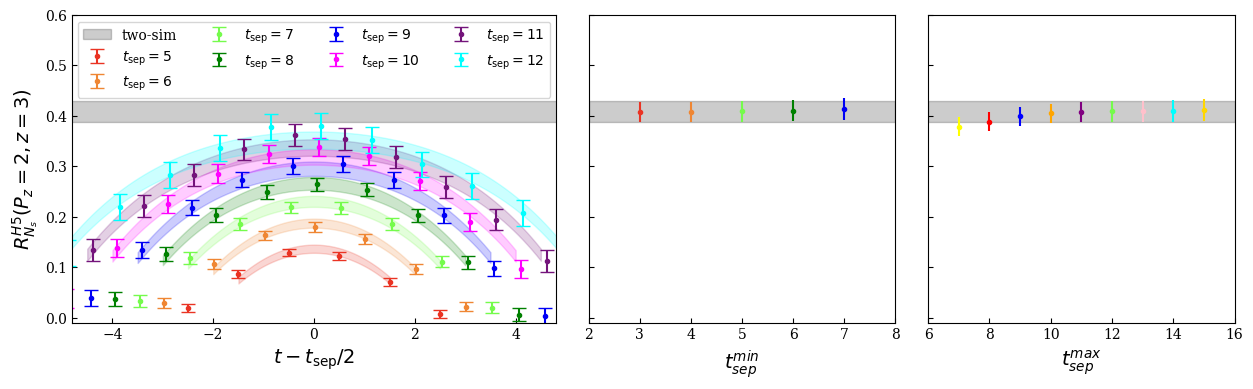}
\centering
    \includegraphics[width=0.95\textwidth]{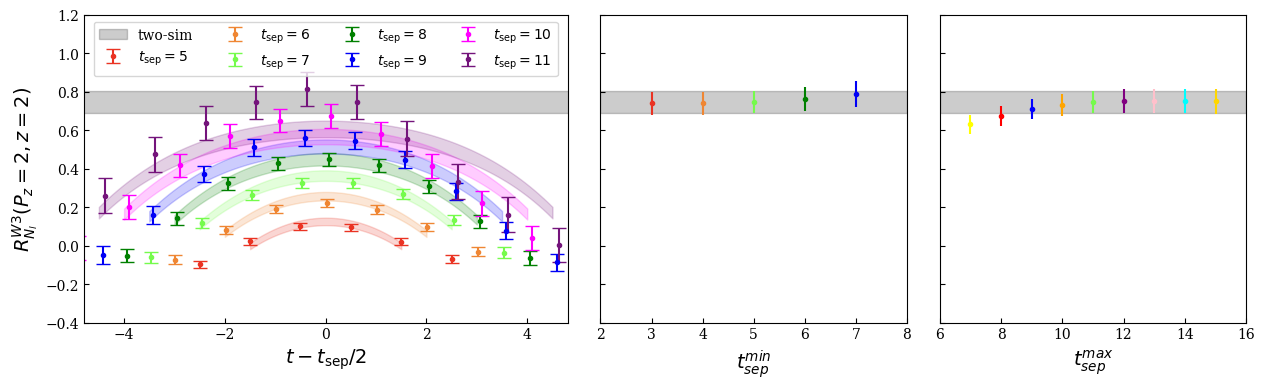}
\centering
    \includegraphics[width=0.95\textwidth]{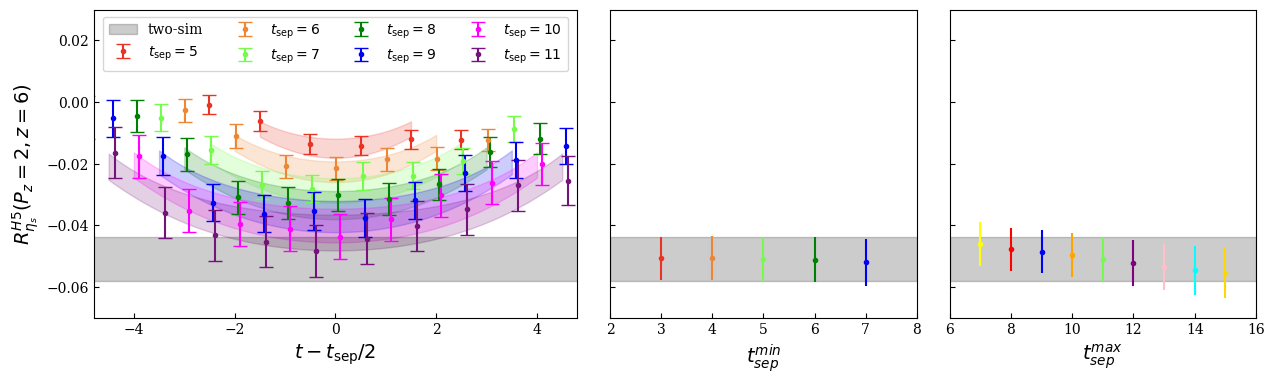}
\centering
    \includegraphics[width=0.95\textwidth]{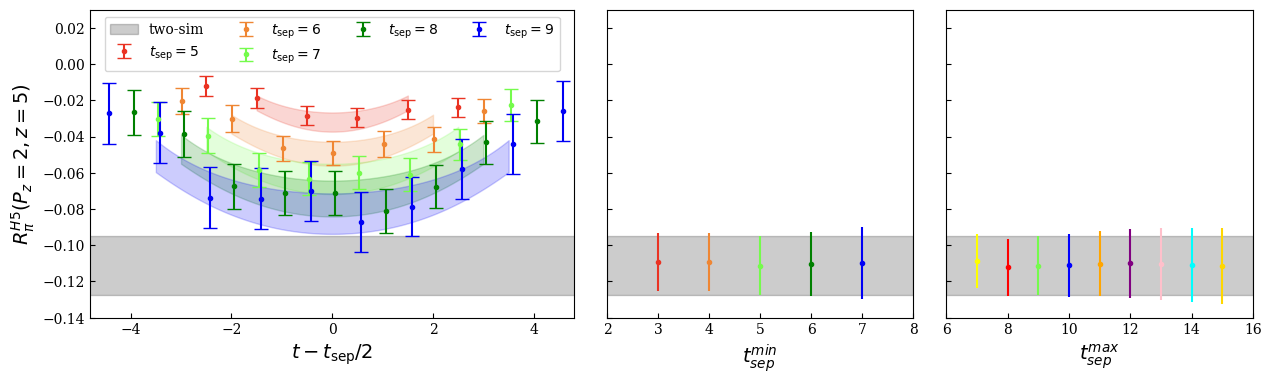}
    \caption{This figure shows example ratio plots in the left column , the extracted matrix elements as we vary the $t_\text{sep}^\text{min}$ in the middle column, and the extracted matrix elements as we vary the $t_\text{sep}^\text{max}$ in the right column, for strange nucleon, light nucleon, $\eta_s$ and $\pi$ from top to bottom. This analysis was done with operator $O^{(2)}(z)$ and HYP5 smearing.}
    \label{fig:O3ZF_H5}
\end{figure*}

\begin{figure*}[htbp]
\centering
    \includegraphics[width=0.95\textwidth]{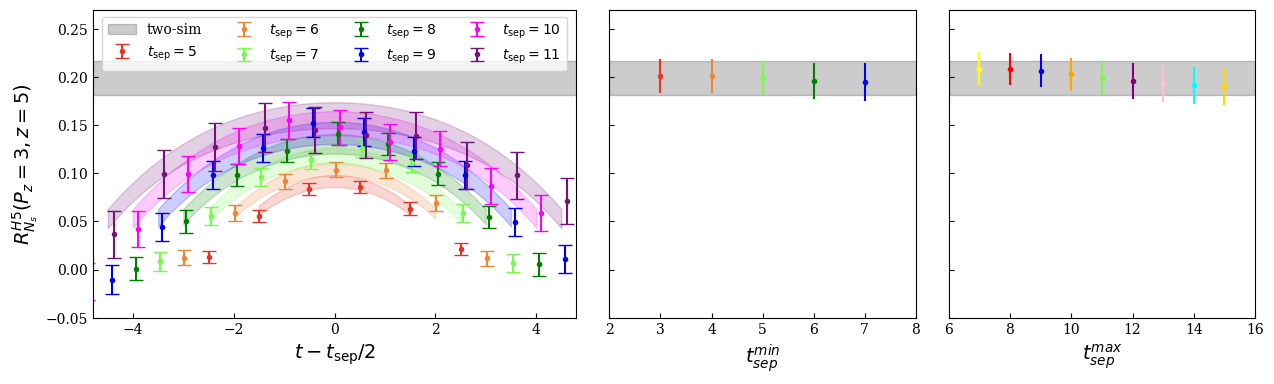}
\centering
    \includegraphics[width=0.95\textwidth]{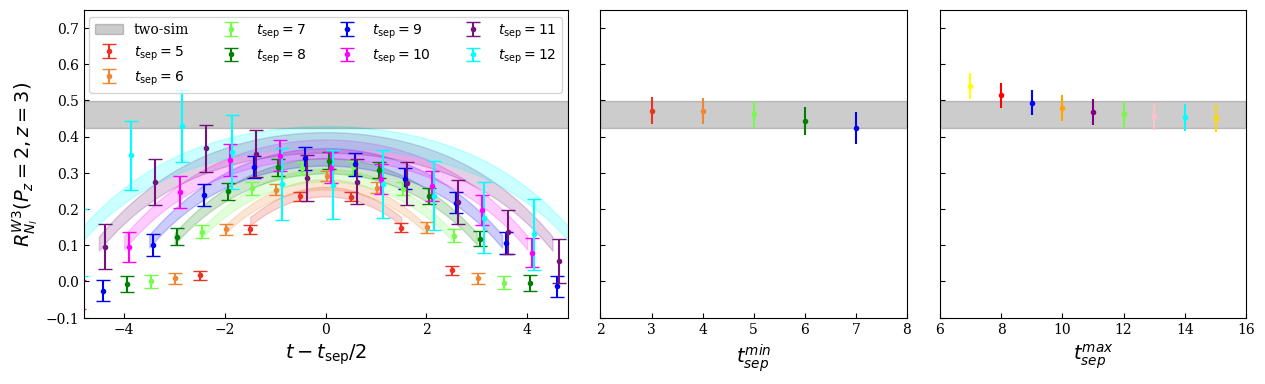}
\centering
    \includegraphics[width=0.95\textwidth]{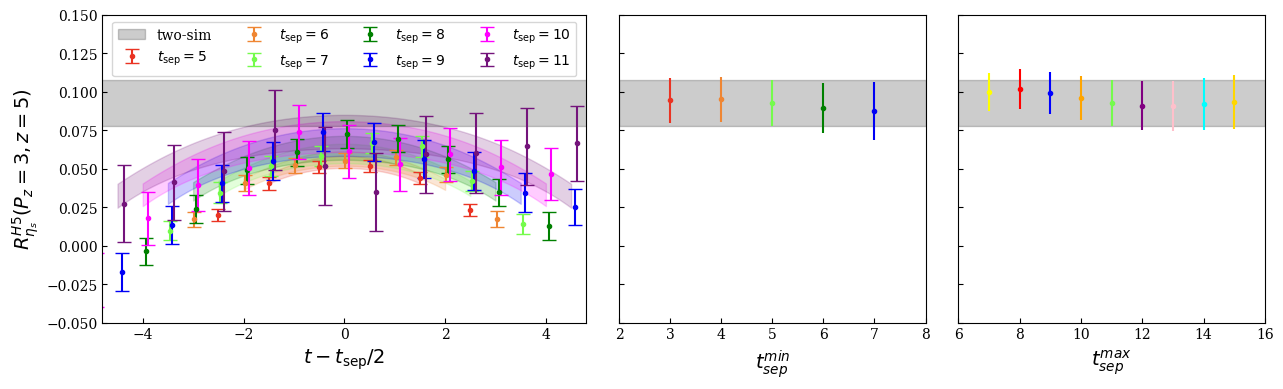}
\centering
    \includegraphics[width=0.95\textwidth]{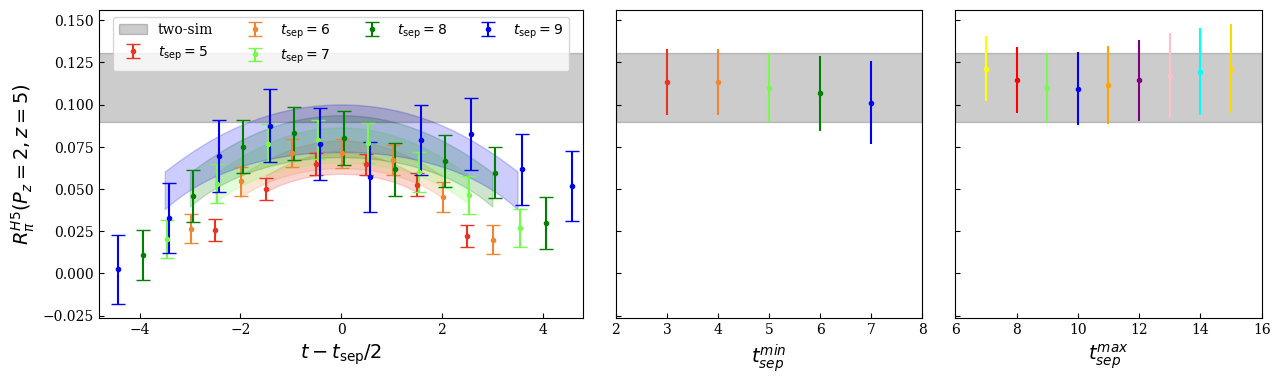}
    \caption{This figure shows example ratio plots in the left column , the extracted matrix elements as we vary the $t_\text{sep}^\text{min}$ in the middle column, and the extracted matrix elements as we vary the $t_\text{sep}^\text{max}$ in the right column, for strange nucleon, light nucleon, $\eta_s$ and $\pi$ from top to bottom. This analysis was done with operator $O^{(3)}(z)$ and HYP5 smearing.}
    \label{fig:O0IB_H5}
\end{figure*}

The reliability of our fits for extracting the matrix elements can be checked by comparing the fits to the ratios as defined in Eq.~\ref{eq:ratio}.
If the excited state contamination is small, the ratios would eventually approach the ground state matrix element.
This is shown in the example ratio plots outlined in Figs.~\ref{fig:O3JZ_W3}, \ref{fig:O3ZF_W3}, \ref{fig:O0IB_W3}, \ref{fig:O3JZ_H5}, \ref{fig:O3ZF_H5} and \ref{fig:O0IB_H5}.
Each figure represents one operator and one smearing type for strange nucleon, light nucleon, $\eta_s$ and $\pi$.
The left most column of the example ratio plot shows $R_H$ at different source-sink separations $t_\text{sep}$, along with reconstructions of the fit shown in the colored bands and the fitted ground state matrix elements represented by the grey band.
We observe that as we increase the $t_\text{sep}$ the ratios and their respective reconstructed bands move towards the grey bands, upwards if the fitted matrix element is positive and downwards if the matrix element is negative.
As per Eq.~\ref{eq:ratio} the ratios should be symmetric alongside the source and sink, we see that this is the case for most of the lower $t_\text{sep}$ however this pattern deviates as we get to higher $t_\text{sep}$ values.
This can be mainly due statistical fluctuations and lower signal to noise ratio at higher $t_\text{sep}$ that causes this deviance.
But in general the ratio plots do display symmetry and approach the ground state matrix element (gray band) attesting to the reliability of our fitting process.

Our choice of source-sink separation used in the fits plays a crucial role in the simultaneous fitting process.
We need to determine if our extracted ground state matrix element is stable for our choice of $t^\text{min}_\text{sep}$ and $t^\text{max}_\text{sep}$.
In order to do so we study the $t^\text{min}_\text{sep}$ and $t^\text{max}_\text{sep}$ dependence. The middle column of the ratio plots outlined in Figs.~\ref{fig:O3JZ_W3}, \ref{fig:O3ZF_W3}, \ref{fig:O0IB_W3}, \ref{fig:O3JZ_H5}, \ref{fig:O3ZF_H5} and \ref{fig:O0IB_H5} show the extracted matrix element as we vary the $t^\text{min}_\text{sep}$.
Our final choice for $t^\text{min}_\text{sep}$ is indicated by the light green point in the middle column.
The plots demonstrate that the extracted matrix elements converge as we decrease the $t^\text{min}_\text{sep}$ and our within error of our final choice of ground state matrix element, this shows that our choice for $t^\text{min}_\text{sep}$ is reliable.
We performed a similar analysis to determine the $t^\text{max}_\text{sep}$. The right most column of the ratio plots outlined in Figs.~\ref{fig:O3JZ_W3}, \ref{fig:O3ZF_W3}, \ref{fig:O0IB_W3}, \ref{fig:O3JZ_H5}, \ref{fig:O3ZF_H5} and \ref{fig:O0IB_H5} show the extracted matrix element as we vary the $t^\text{max}_\text{sep}$.
Our final $t^\text{max}_\text{sep}$ choice and the ground state matrix element used in the rest of the analysis is outlined by the light green point.
As the plots show, when we increase the $t^\text{max}_\text{sep}$ the extracted matrix elements converge and stay within the error range (grey band) of our final matrix elements.
This shows that our choice for $t^\text{max}_\text{sep}$ is consistent across various $t^\text{max}_\text{seps}$ and therefore reliable.
Using the same process we determined $t^\text{max}_\text{sep}$ and $t^\text{min}_\text{sep}$ ranges for all hadrons across the three operators, two smearings, and various Wilson link displacement $z$s and hadron boosted momenta $P_z$s.

\begin{figure*}[htbp]
\centering
        \includegraphics[width=.3\textwidth]{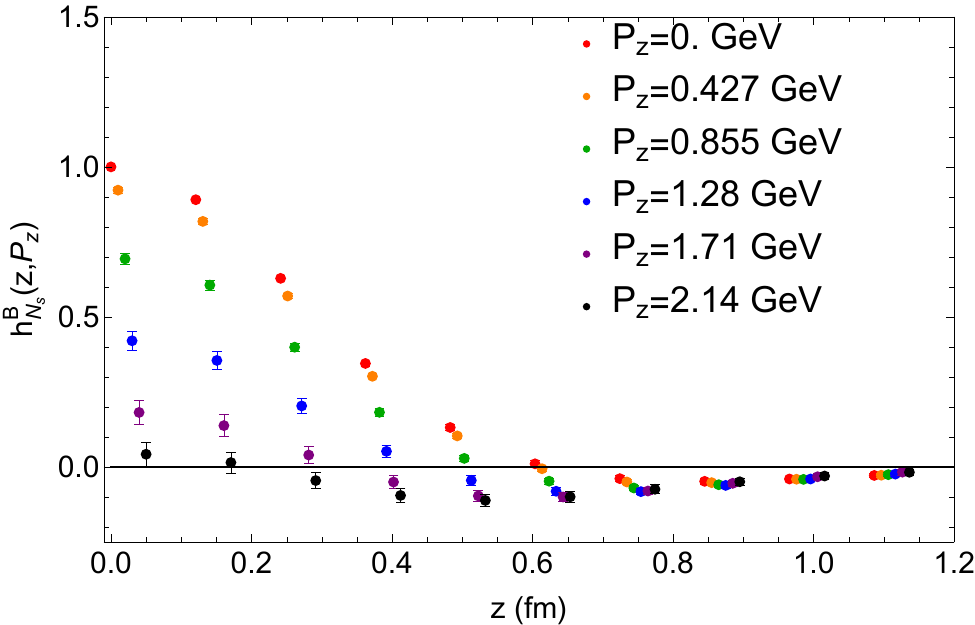}
\centering
        \includegraphics[width=.3\textwidth]{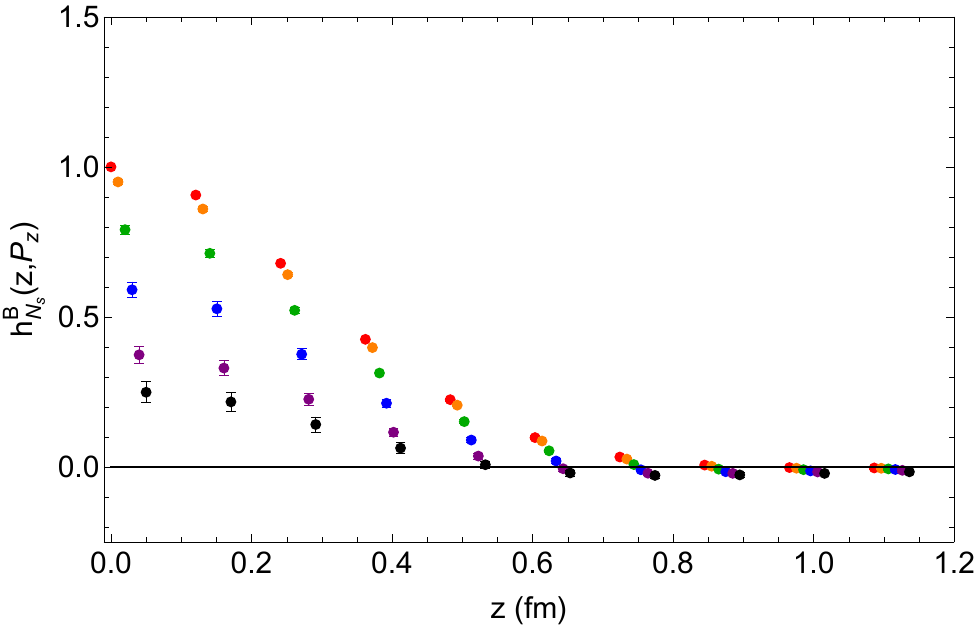}
\centering
        \includegraphics[width=.3\textwidth]{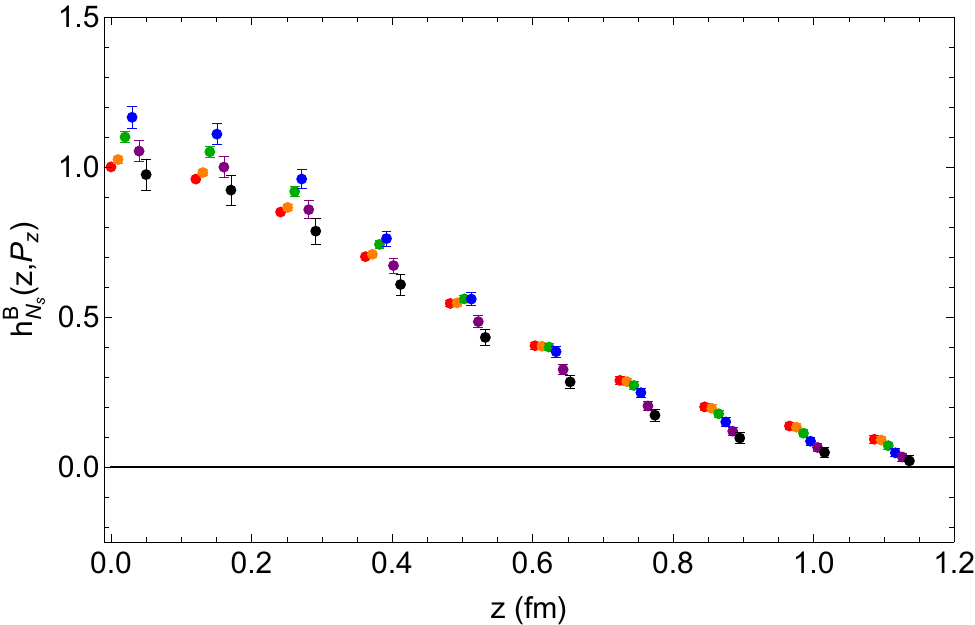}
\centering
        \includegraphics[width=.3\textwidth]{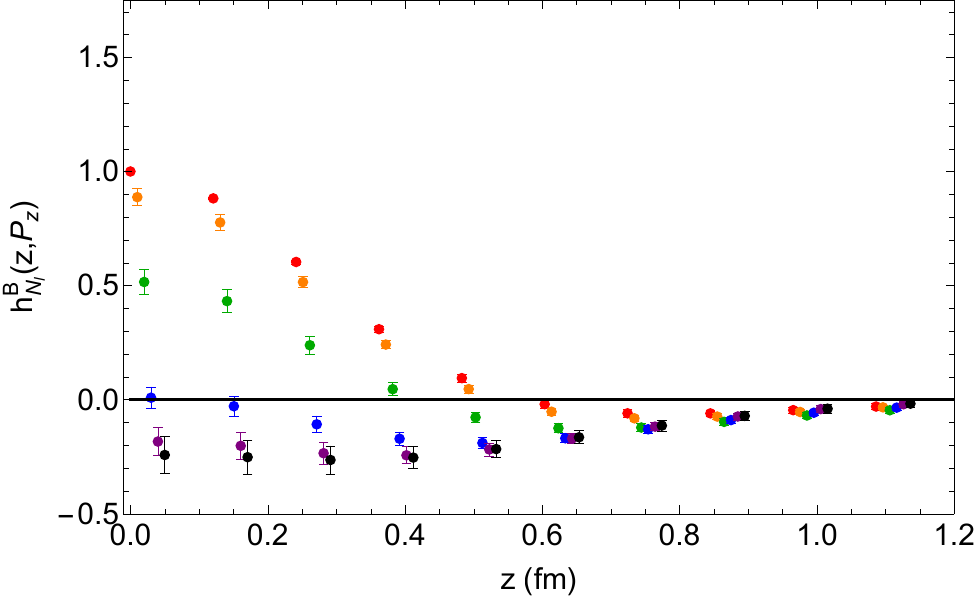}
\centering
        \includegraphics[width=.3\textwidth]{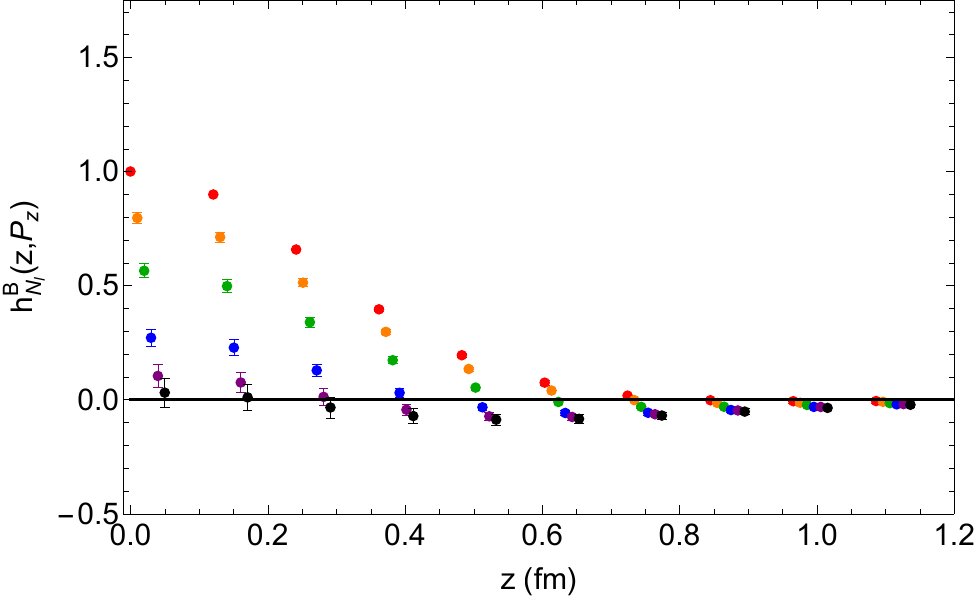}
\centering
        \includegraphics[width=.3\textwidth]{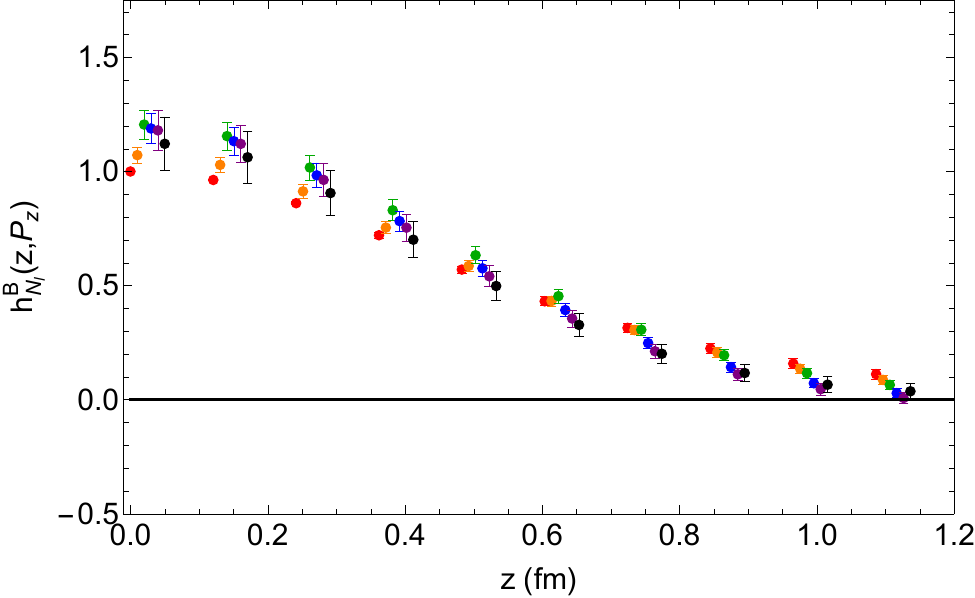}
\centering
        \includegraphics[width=.3\textwidth]{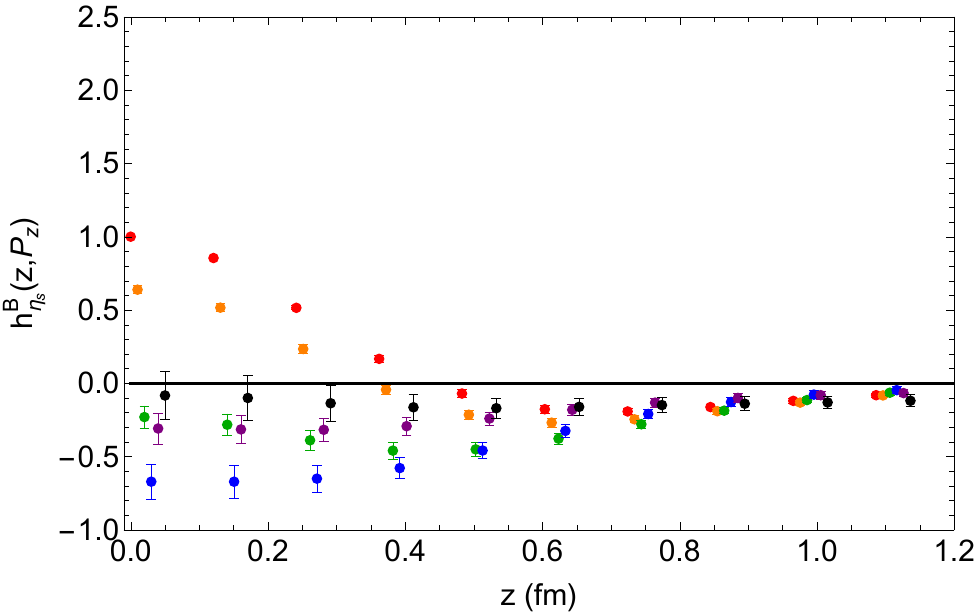}
\centering
        \includegraphics[width=.3\textwidth]{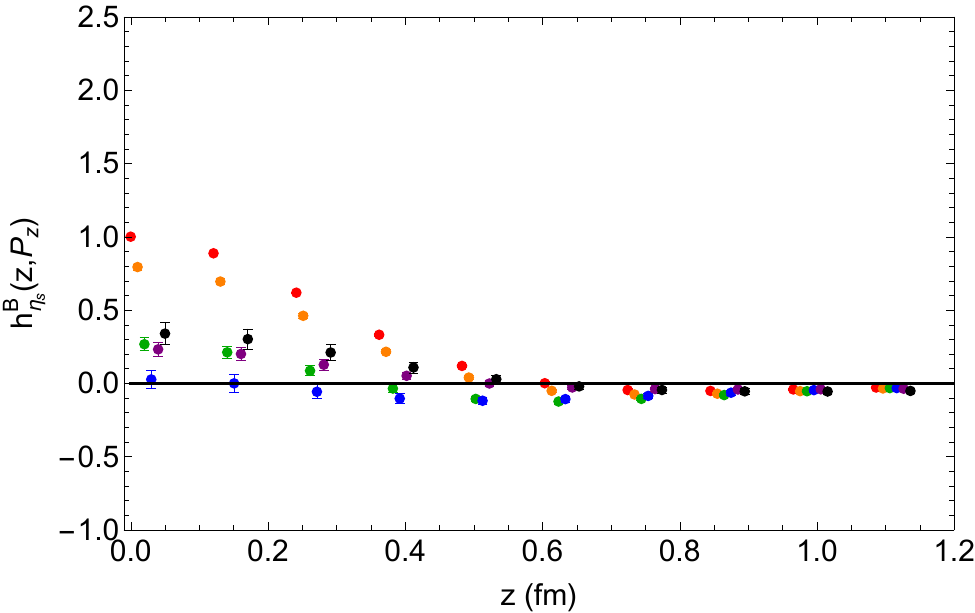}
\centering
        \includegraphics[width=.3\textwidth]{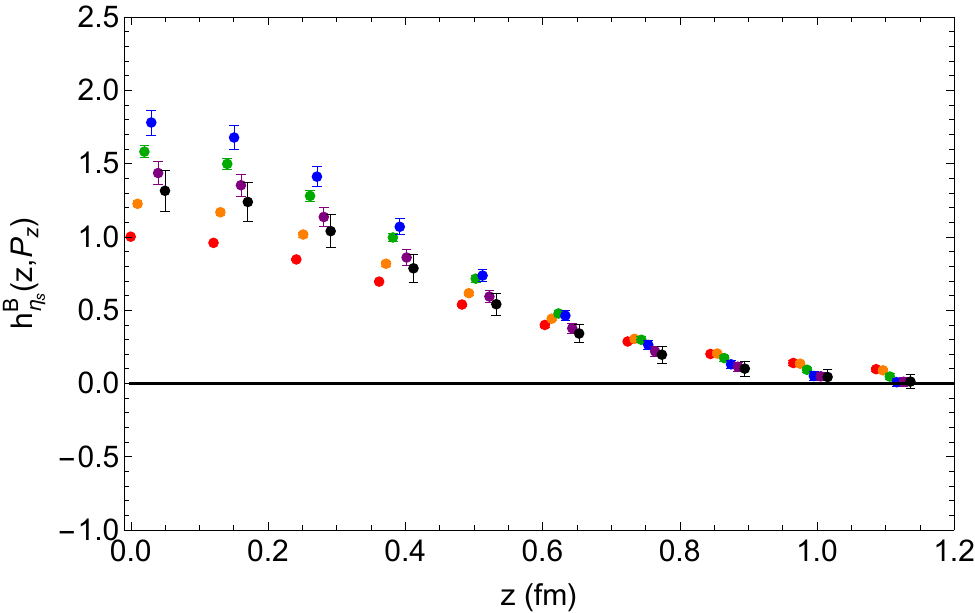}
\centering
        \includegraphics[width=.3\textwidth]{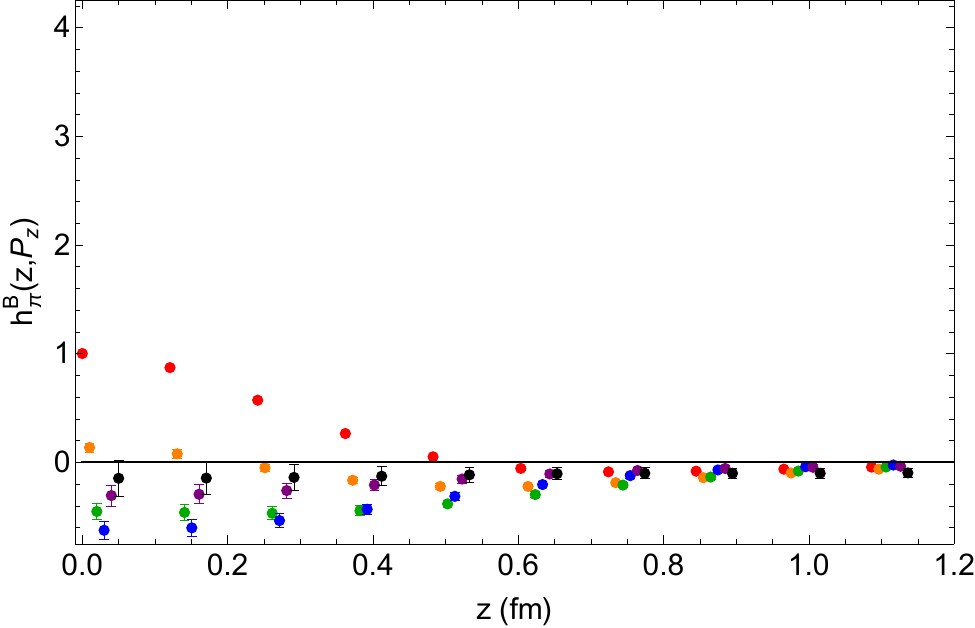}
\centering
        \includegraphics[width=.3\textwidth]{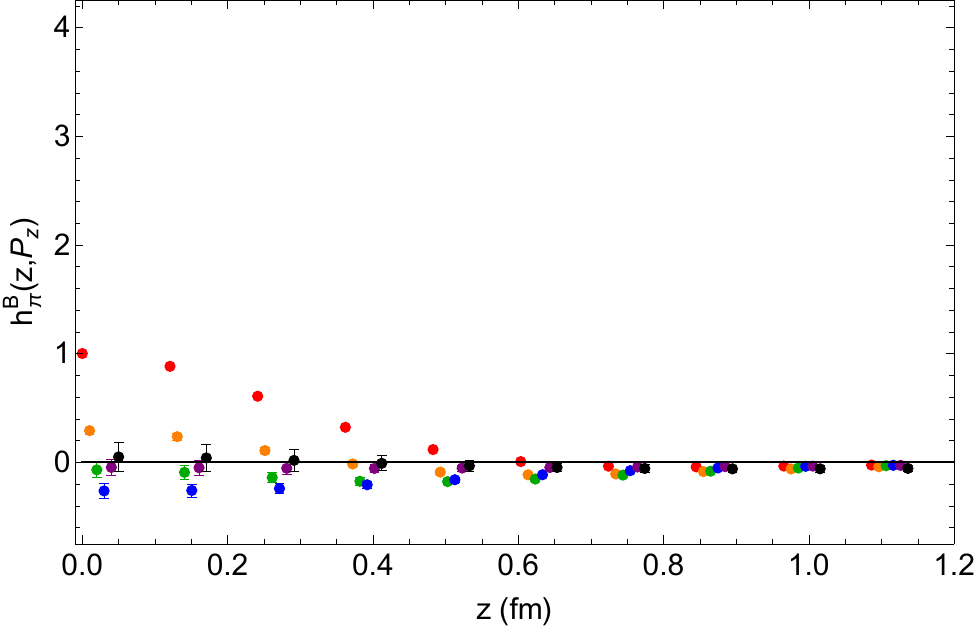}
\centering
        \includegraphics[width=.3\textwidth]{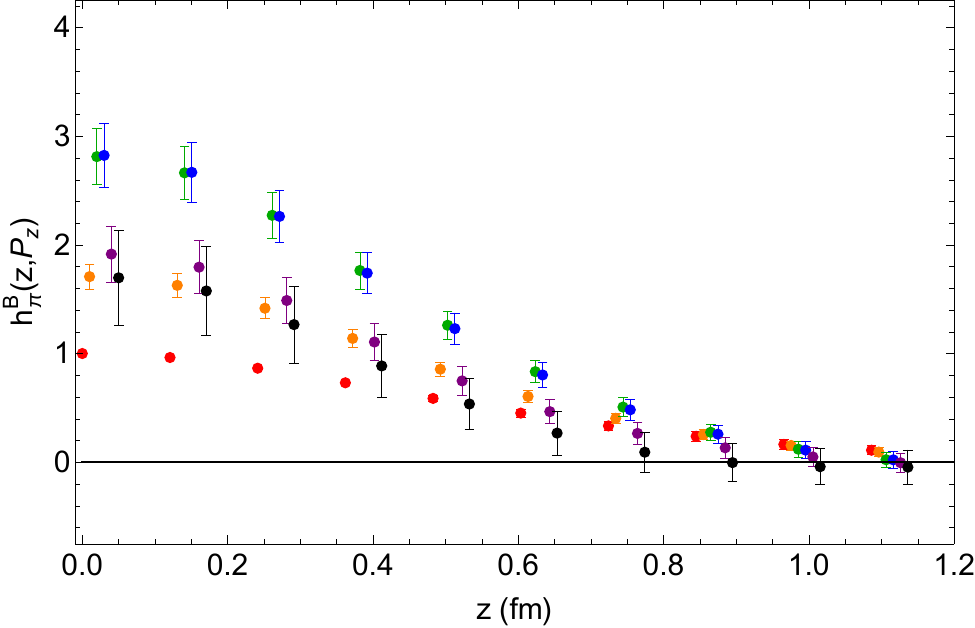}

    \caption{Bare matrix elements for the Wilson3 smearing data, normalized such that $h^\text{B}(0,0)=1$ for the strange nucleon, light nucleon, $\eta_s$, and $\pi$ (rows top to bottom) for each operator $O^{(1,2,3)}(z)$ (columns left to right).}
    \label{fig:Bare_MEs_W3}
\end{figure*}

\begin{figure*}[htbp]
\centering
        \includegraphics[width=.3\textwidth]{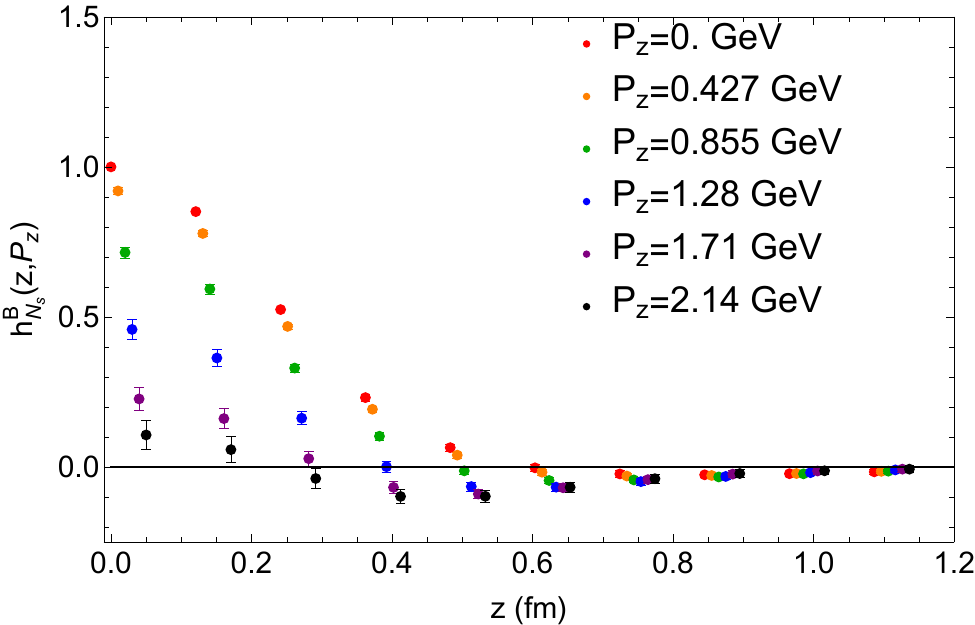}
\centering
        \includegraphics[width=.3\textwidth]{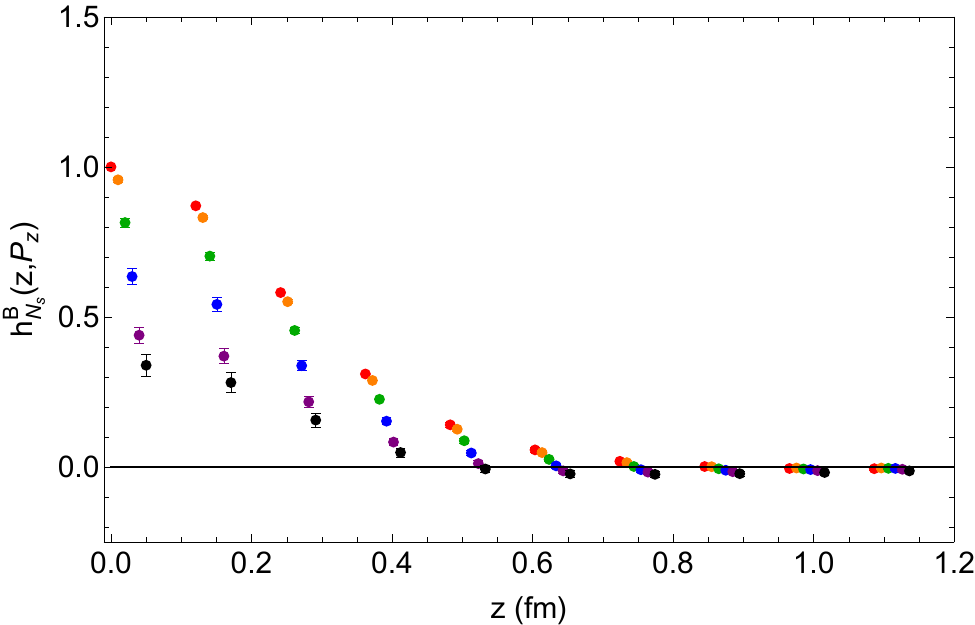}
\centering
        \includegraphics[width=.3\textwidth]{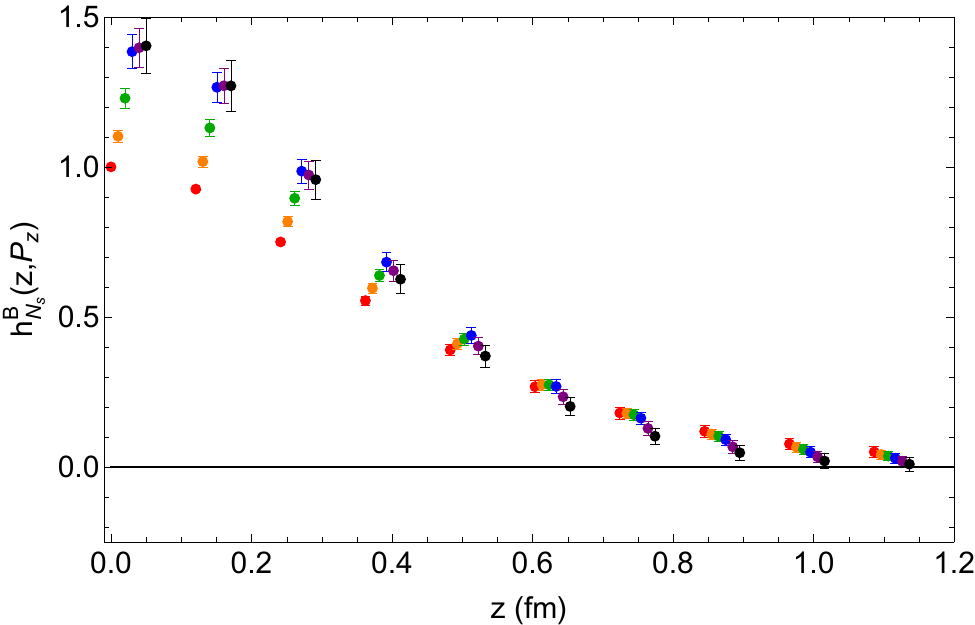}
\centering
        \includegraphics[width=.3\textwidth]{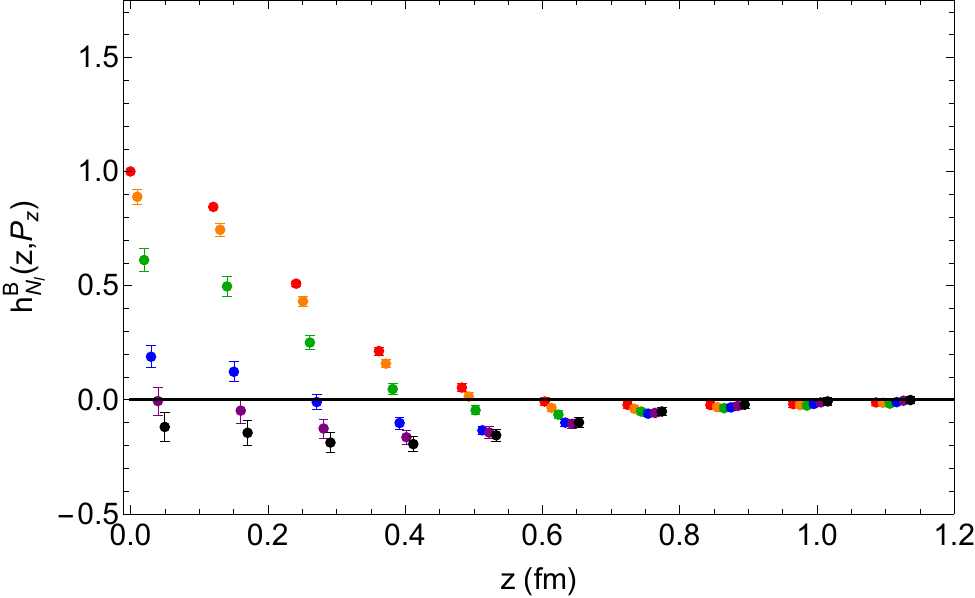}
\centering
        \includegraphics[width=.3\textwidth]{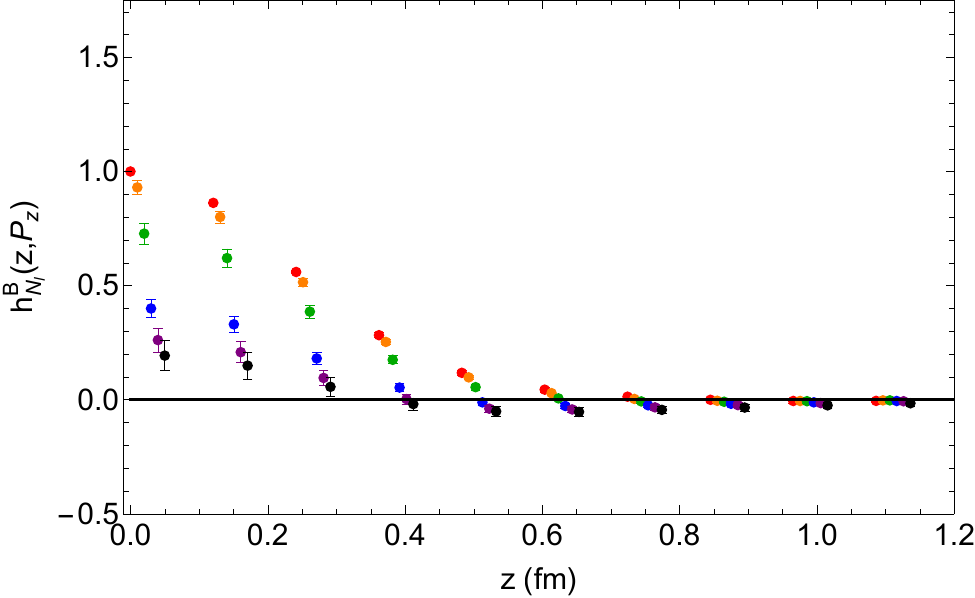}
\centering
        \includegraphics[width=.3\textwidth]{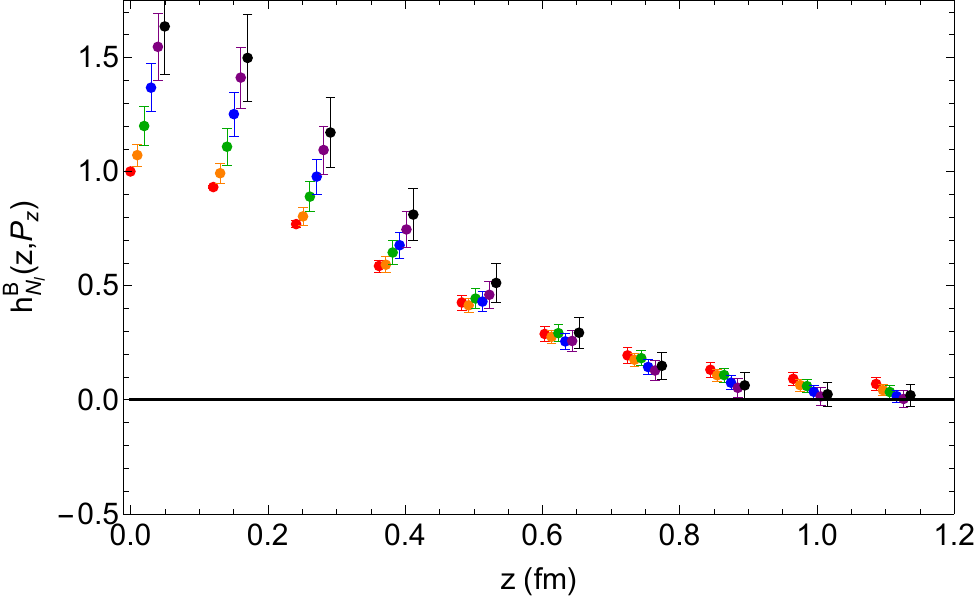}
\centering
        \includegraphics[width=.3\textwidth]{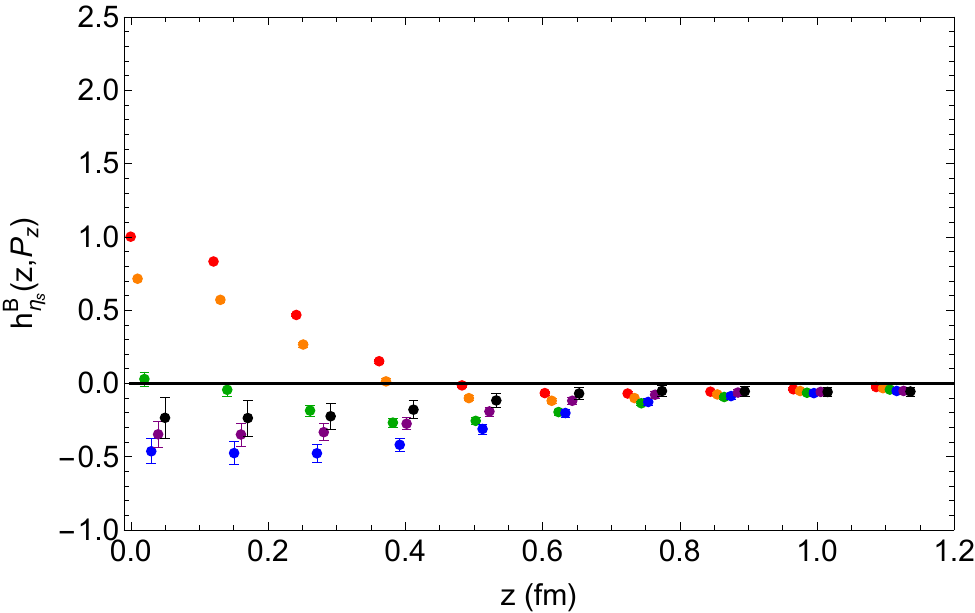}
\centering
        \includegraphics[width=.3\textwidth]{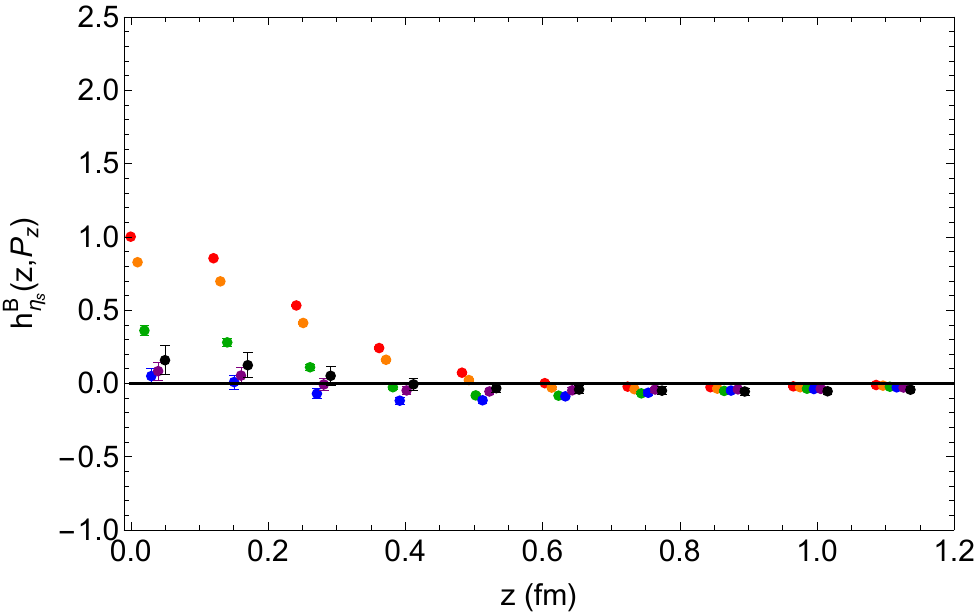}
\centering
        \includegraphics[width=.3\textwidth]{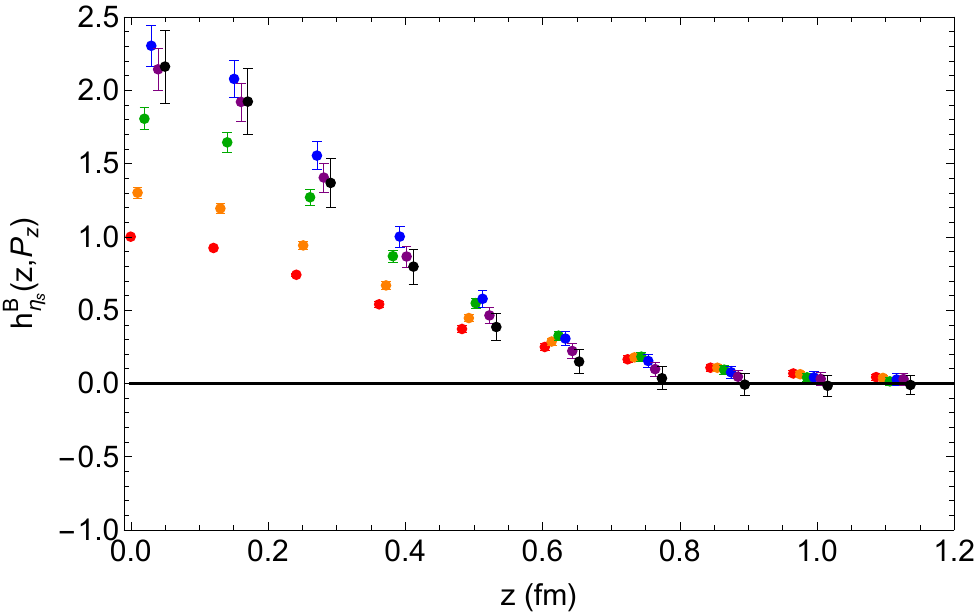}
\centering
        \includegraphics[width=.3\textwidth]{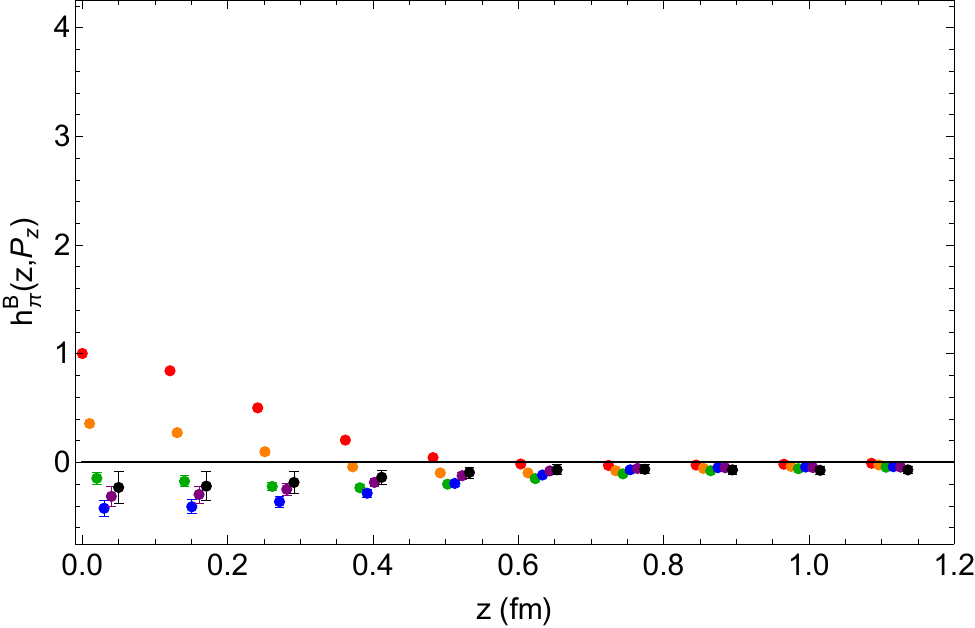}
\centering
        \includegraphics[width=.3\textwidth]{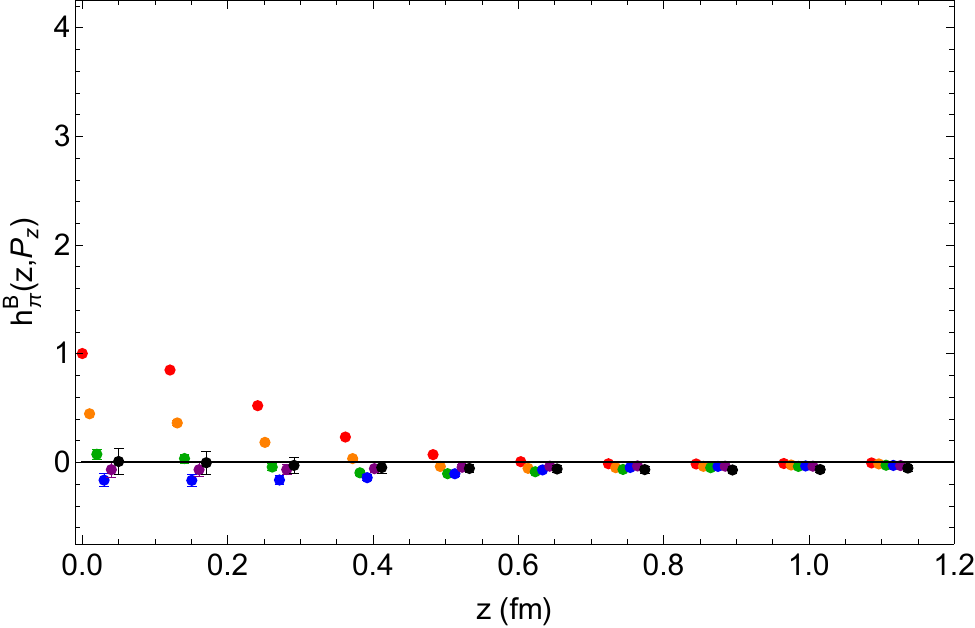}
\centering
        \includegraphics[width=.3\textwidth]{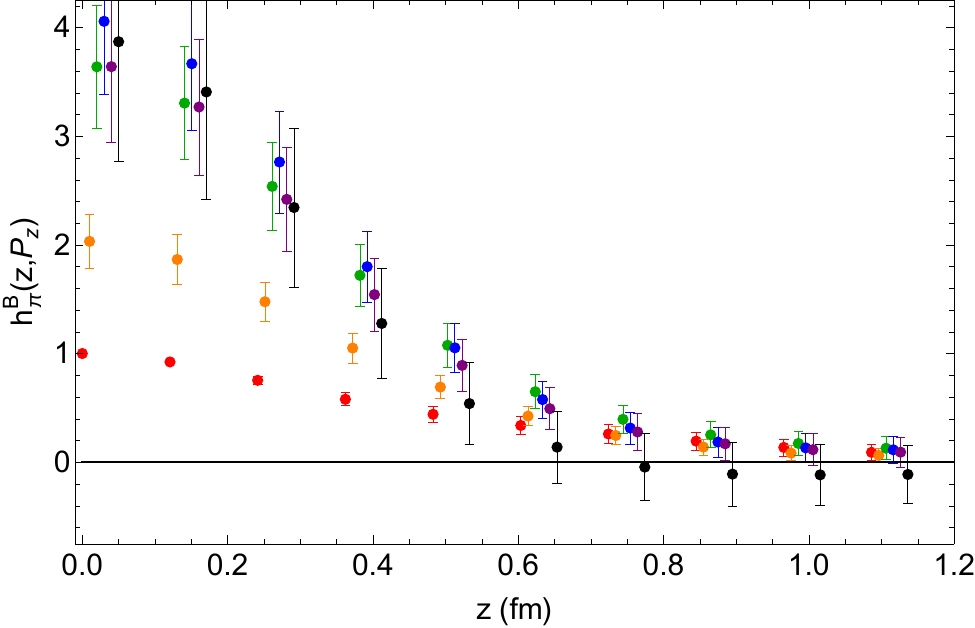}

    \caption{Bare matrix elements for the HYP5 smearing data, normalized such that $h^\text{B}(0,0)=1$ for the strange nucleon, light nucleon, $\eta_s$, and $\pi$ (rows top to bottom) for each operator $O^{(1,2,3)}(z)$ (columns left to right).}
    \label{fig:Bare_MEs_H5}
\end{figure*}

With the bare matrix elements fit for each hadron, operator, and smearing, we can compare their behavior.
In Figs.~\ref{fig:Bare_MEs_W3} and~\ref{fig:Bare_MEs_H5}, we show the bare matrix elements $h^\text{B}(z,P_z)$ for the Wilson-3 and HYP5 smearings.
The matrix elements are normalized such that $h^\text{B}(0,0)=1$ and not divided out by any kinematic factors.
The behavior of the matrix elements at fixed-$z$ for different $P_z$ is not necessarily monotonic in every case.
These effects are particularly apparent for $O^{(3)}$ in the Wilson-3 case for all hadrons and for the mesons in both smearing cases (bottom two rows of each figure) for $O^{(1)}$ and $O^{(2)}$ operators.
Nonetheless, we expect renormalization to remove any factors (kinematic or otherwise) that could be producing this behavior.
More concerningly, we see in both smearing cases that $O^{(1)}$ and $O^{(2)}$ (first and second columns) both cross zero at different momenta and distances, while $O^{(3)}$ (third column) stays above zero aside from some ambiguity due to statistical noise at large distances.
From Eq.~\ref{eq:hyb_def}, we can see that some of these noisy, near-zero, matrix elements may bring large errors into the renormalized matrix elements, especially those at $z=0$ or $P_z = 0$.
This is highly suggestive that $O^{(3)}$ will likely produce more consistent renormalized matrix elements while $O^{(1)}$ and $O^{(2)}$ may not work as well.

\section{Results and Discussion}\label{sec:result}

\subsection{Ratio Renormalized Matrix Elements}

\begin{figure*}
\centering
        \includegraphics[width=.3\textwidth]{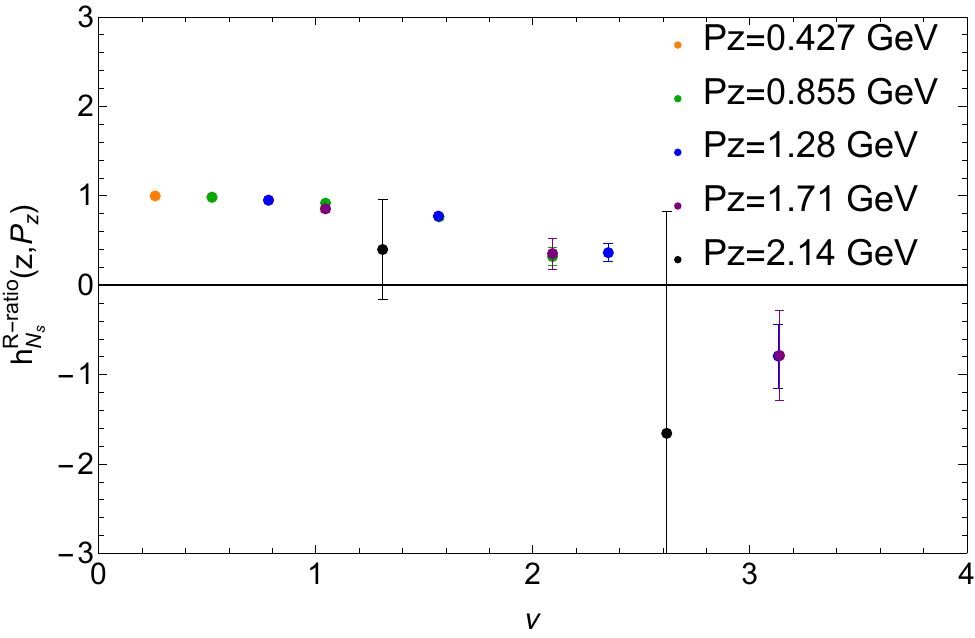}
\centering
        \includegraphics[width=.3\textwidth]{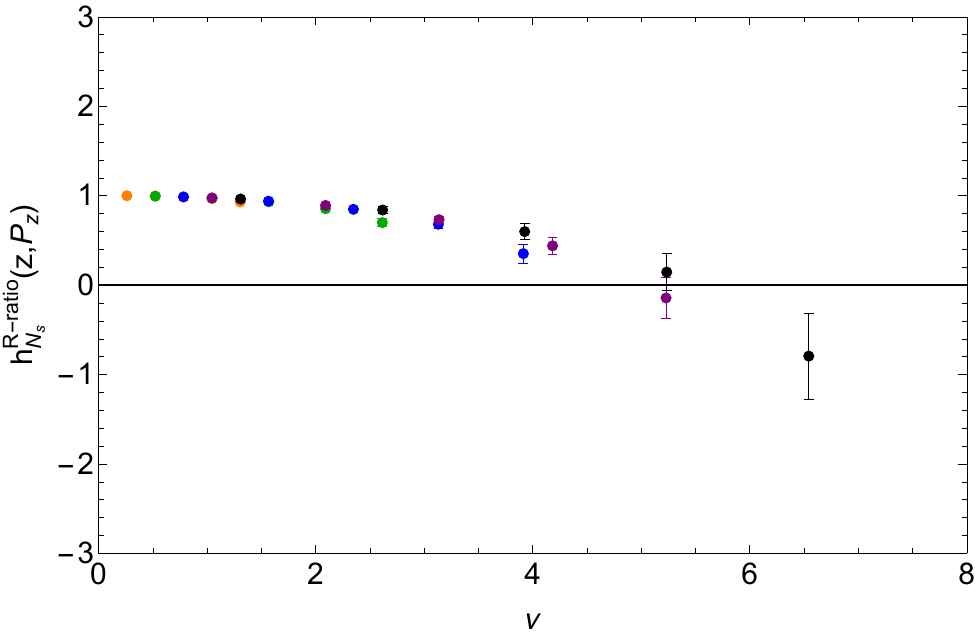}
\centering
        \includegraphics[width=.3\textwidth]{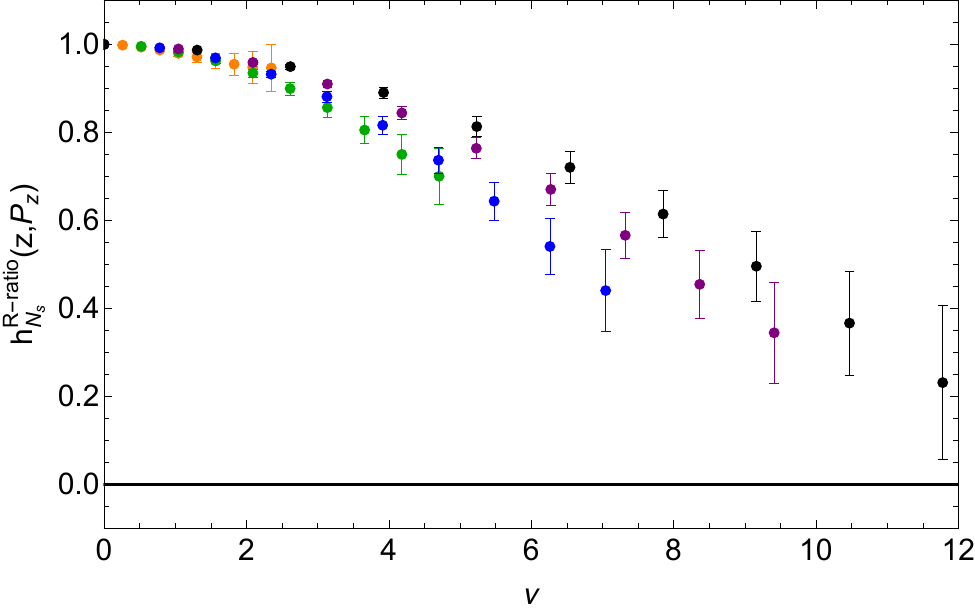}
\centering
        \includegraphics[width=.3\textwidth]{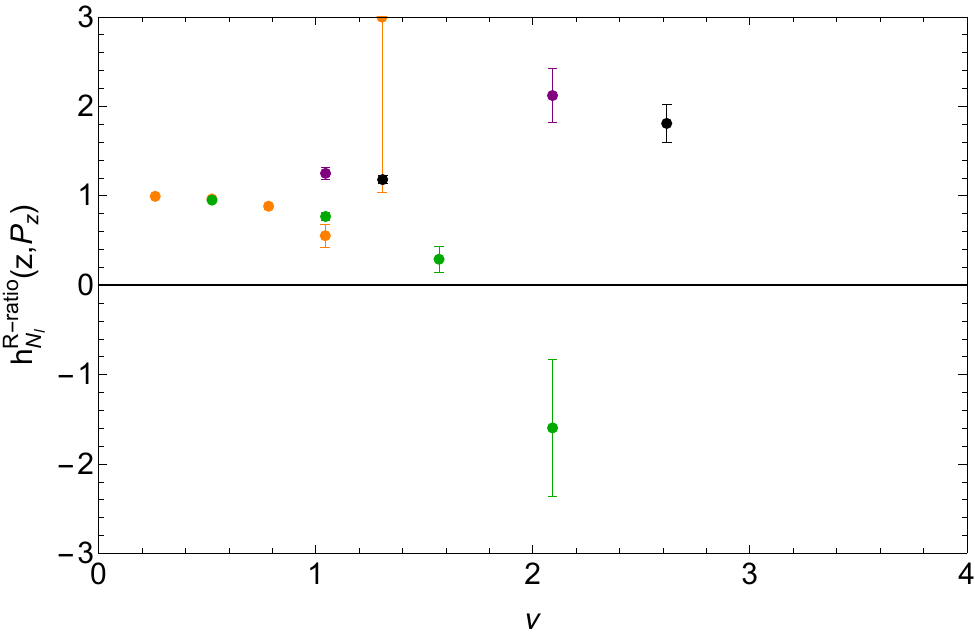}
\centering
        \includegraphics[width=.3\textwidth]{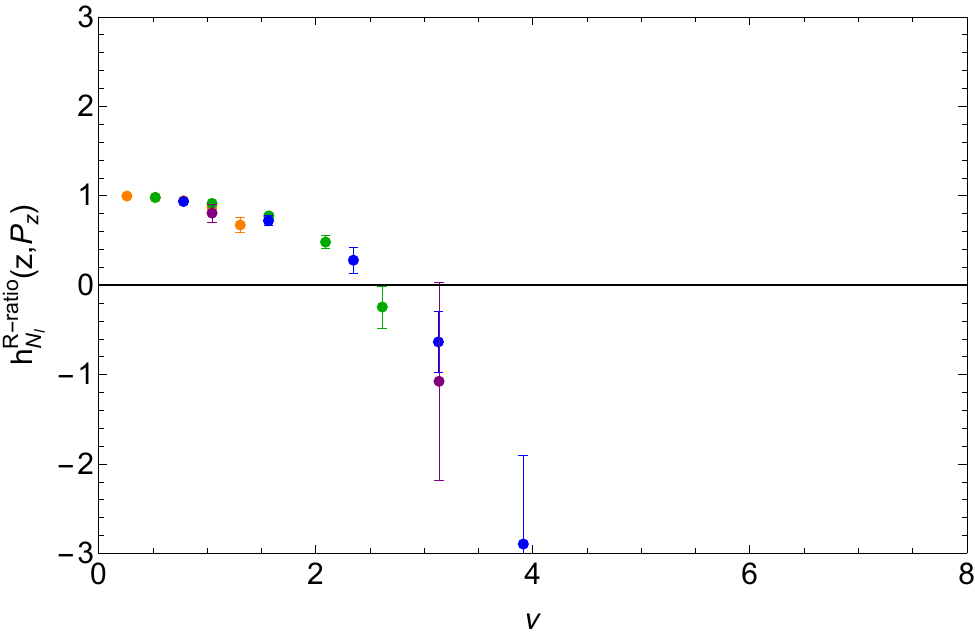}
\centering
        \includegraphics[width=.3\textwidth]{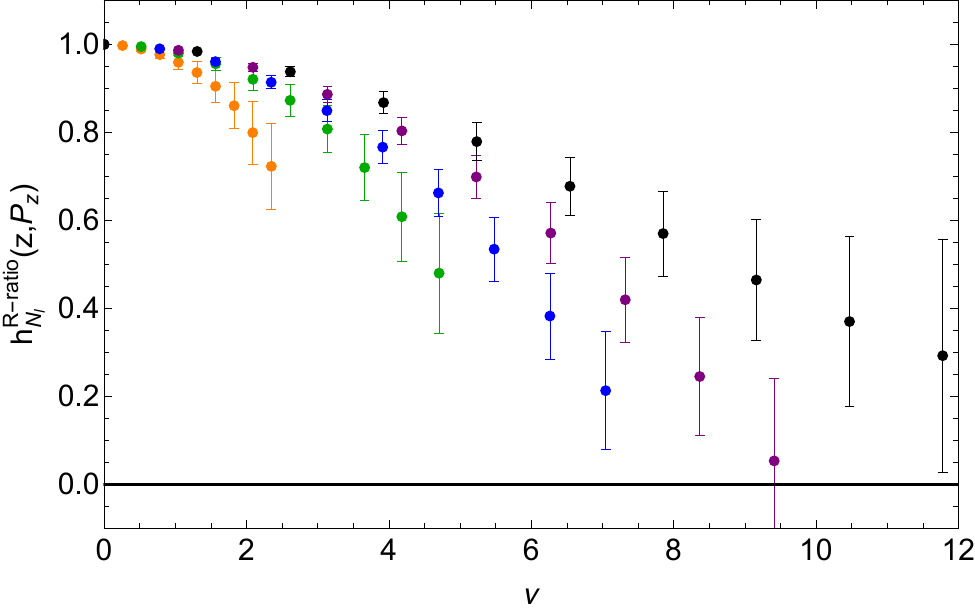}
\centering
        \includegraphics[width=.3\textwidth]{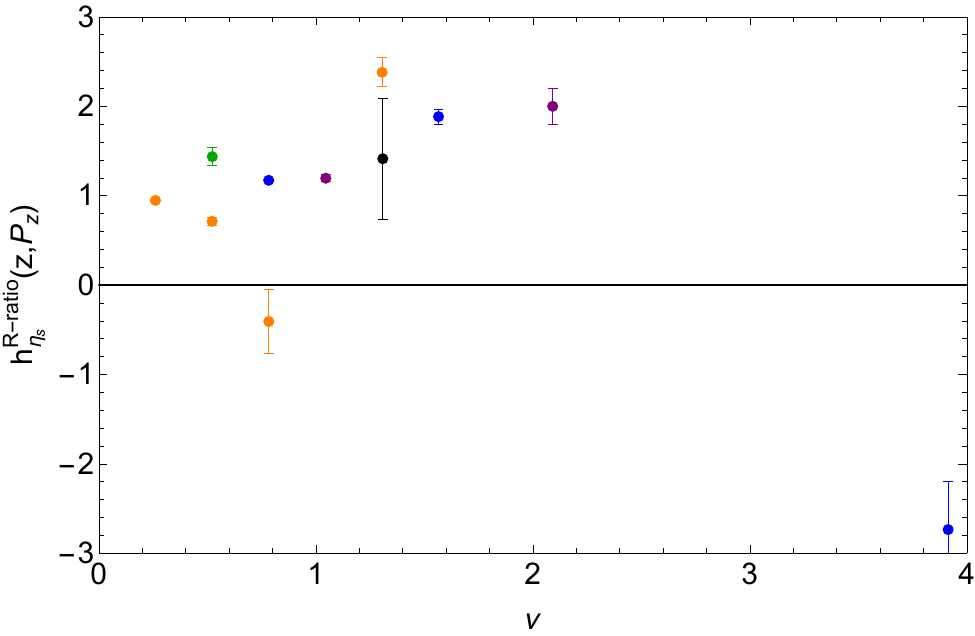}
\centering
        \includegraphics[width=.3\textwidth]{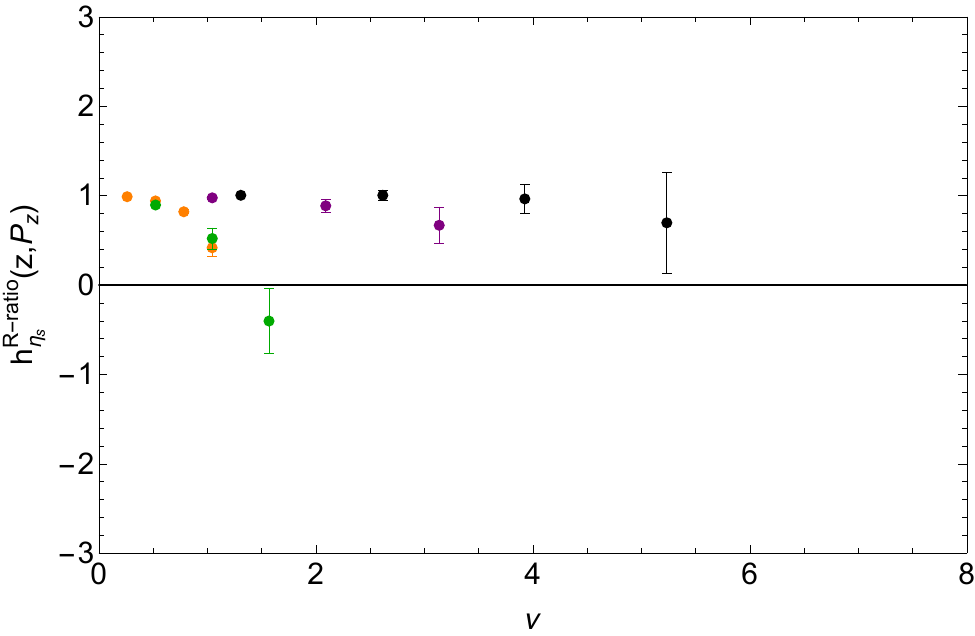}
\centering
        \includegraphics[width=.3\textwidth]{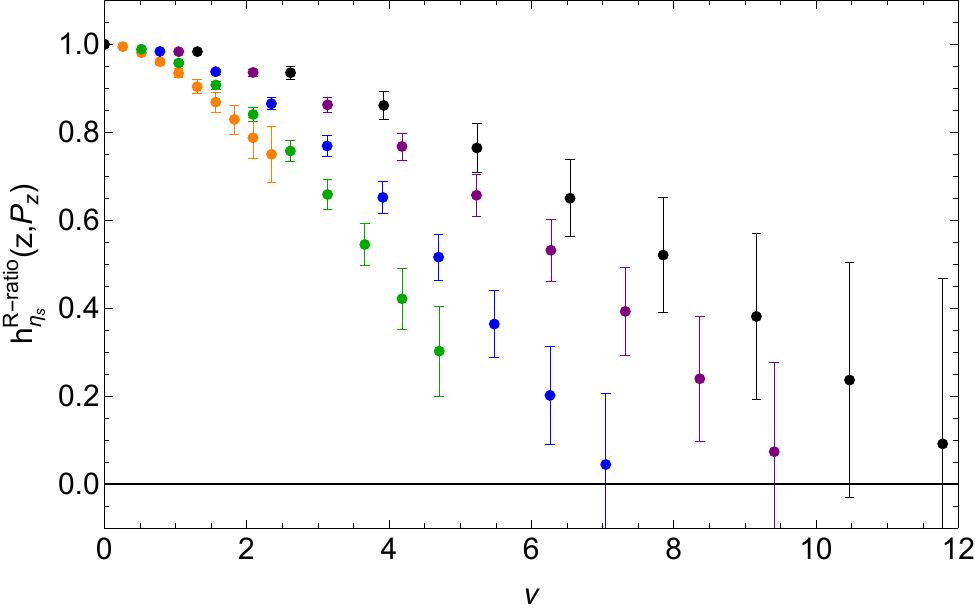}
\centering
        \includegraphics[width=.3\textwidth]{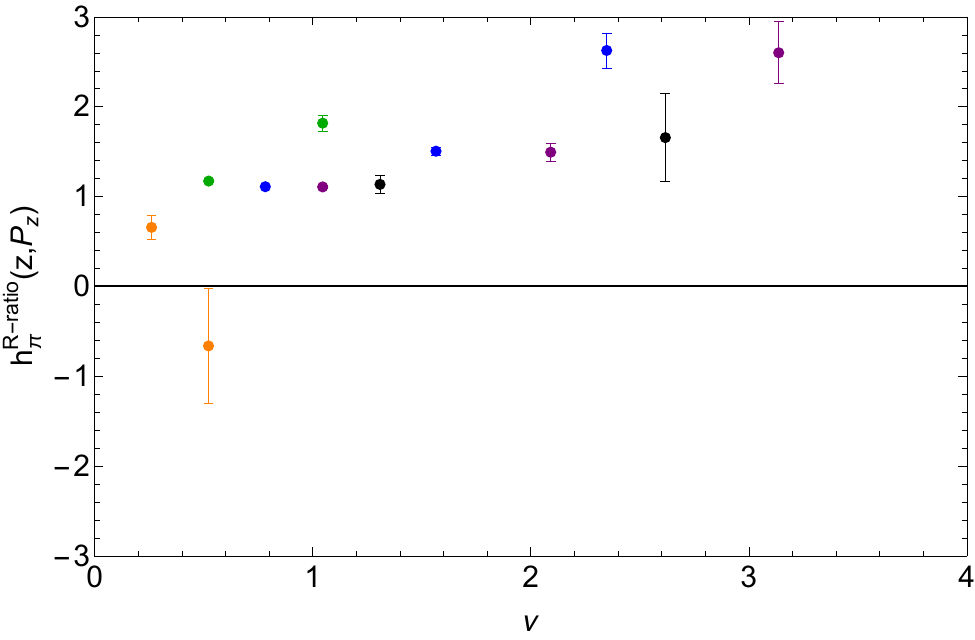}
\centering
        \includegraphics[width=.3\textwidth]{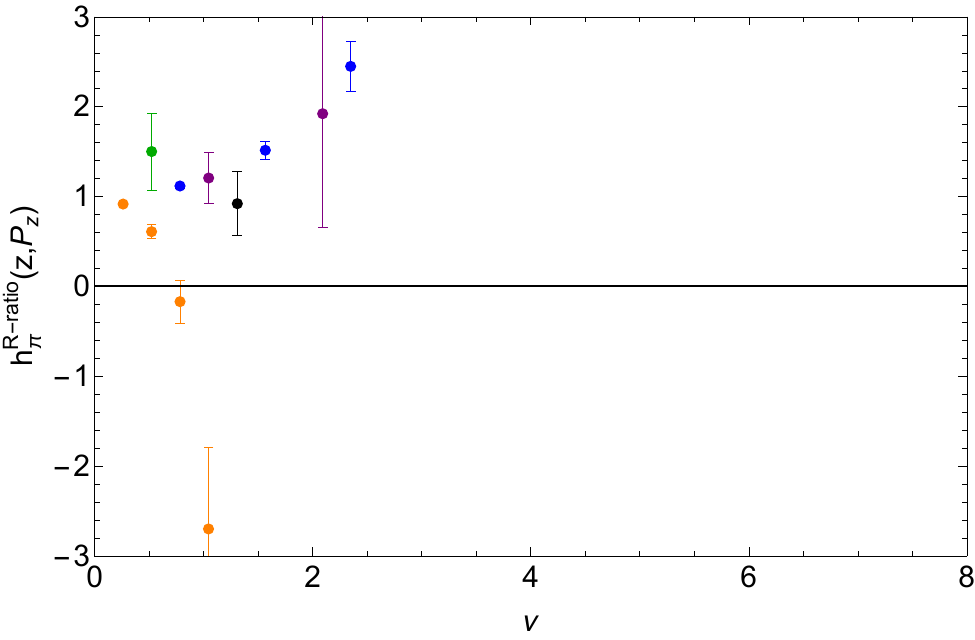}
\centering
        \includegraphics[width=.3\textwidth]{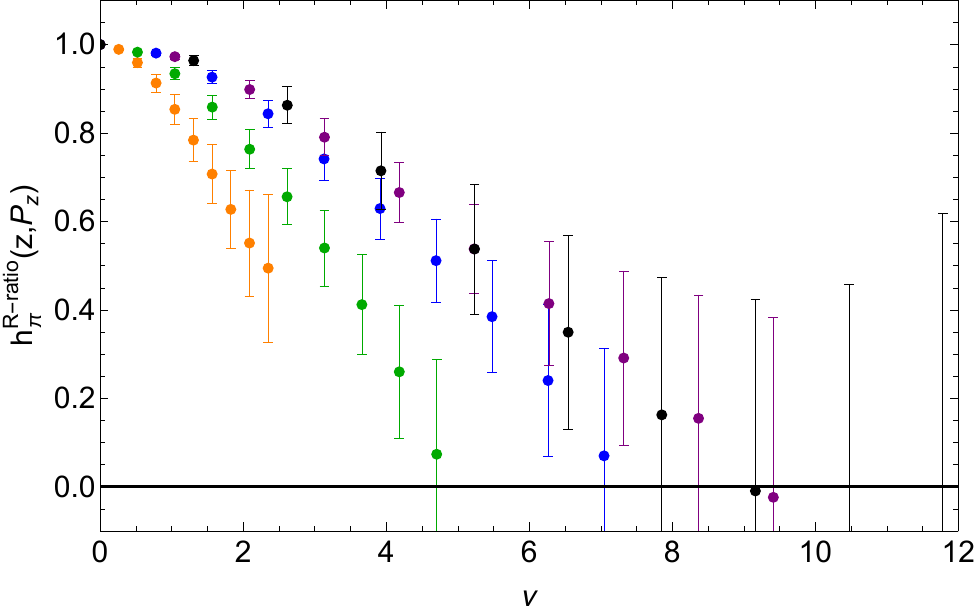}
    \caption{Ratio renormalized matrix elements for the Wilson3 smearing data for the strange nucleon, light nucleon, $\eta_s$, and $\pi$ (rows top to bottom) for each operator $O^{(1,2,3)}(z)$ (columns left to right).}
    \label{fig:ratio-MEs-W3}
\end{figure*}

\begin{figure*}
\centering
        \includegraphics[width=.3\textwidth]{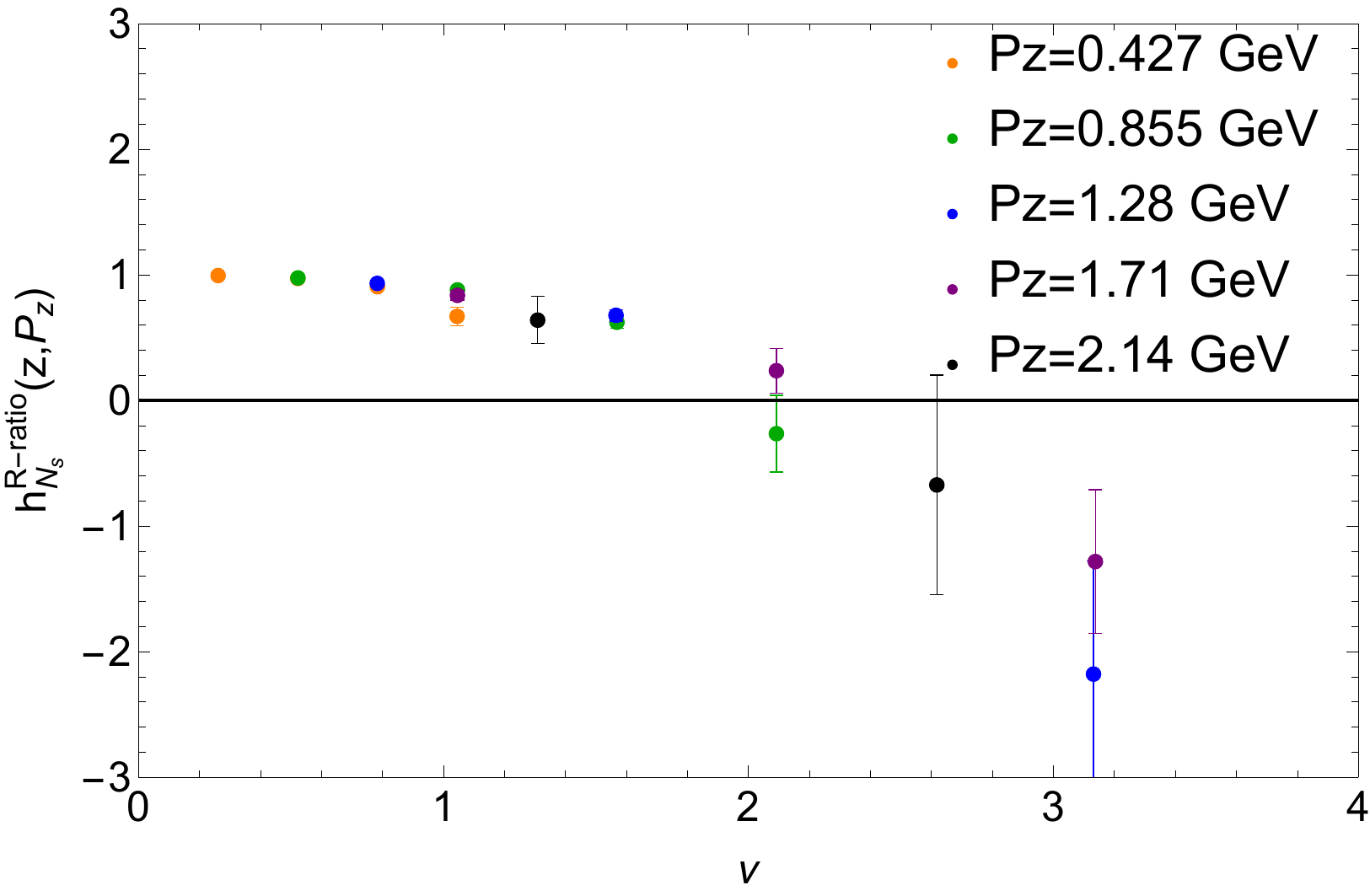}
\centering
        \includegraphics[width=.3\textwidth]{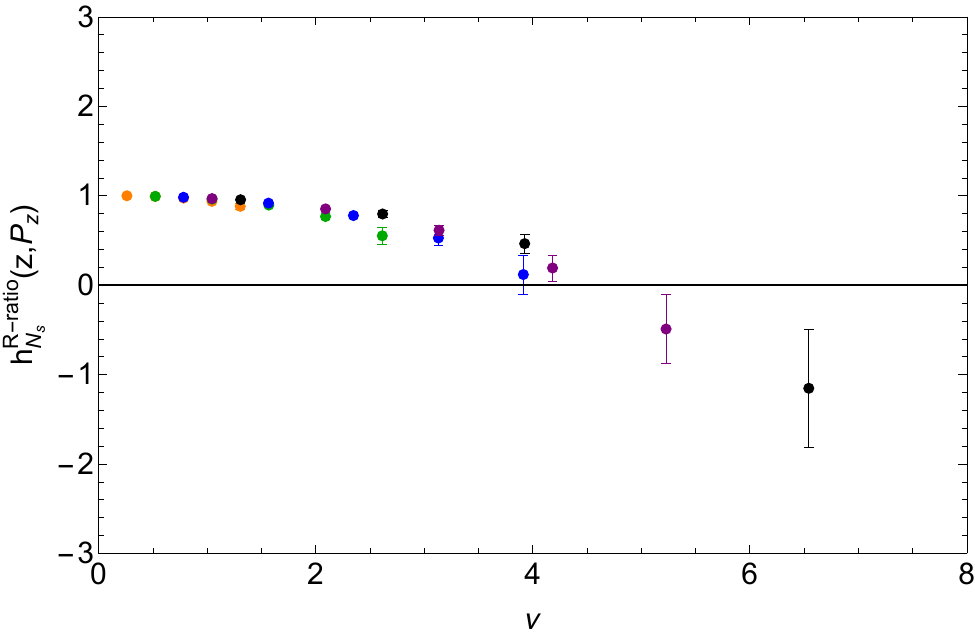}
\centering
        \includegraphics[width=.3\textwidth]{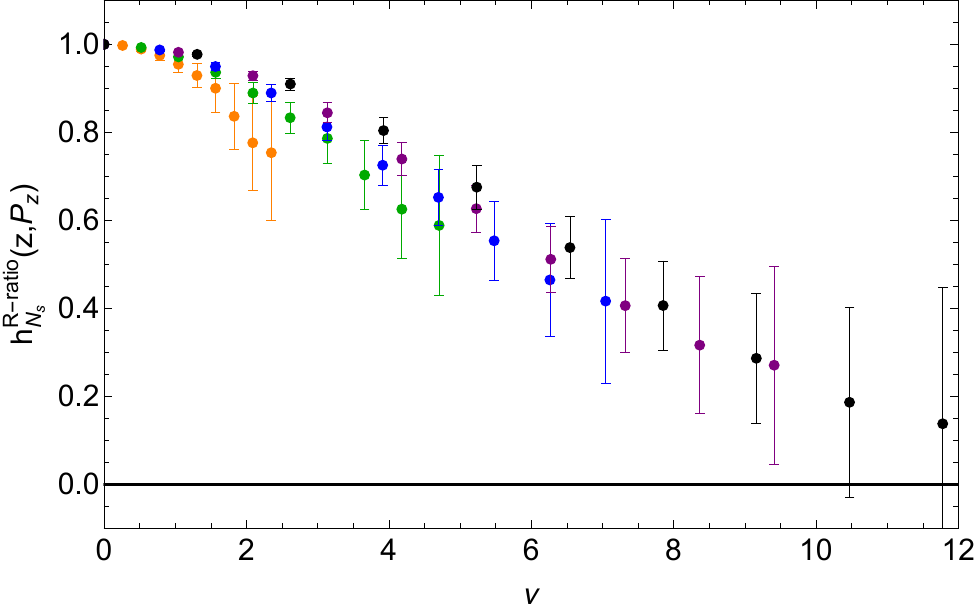}
\centering
        \includegraphics[width=.3\textwidth]{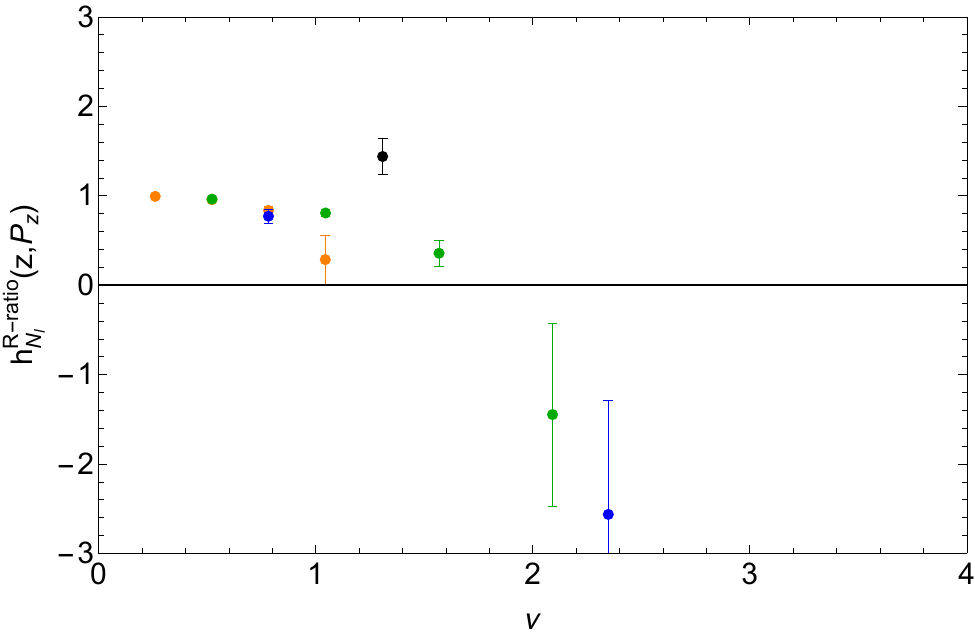}
\centering
        \includegraphics[width=.3\textwidth]{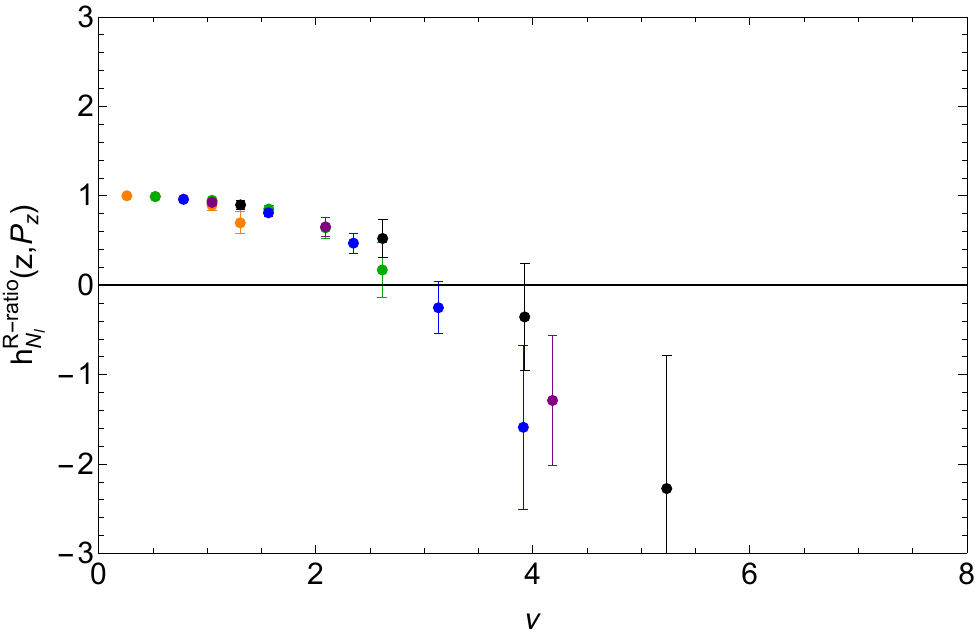}
\centering
        \includegraphics[width=.3\textwidth]{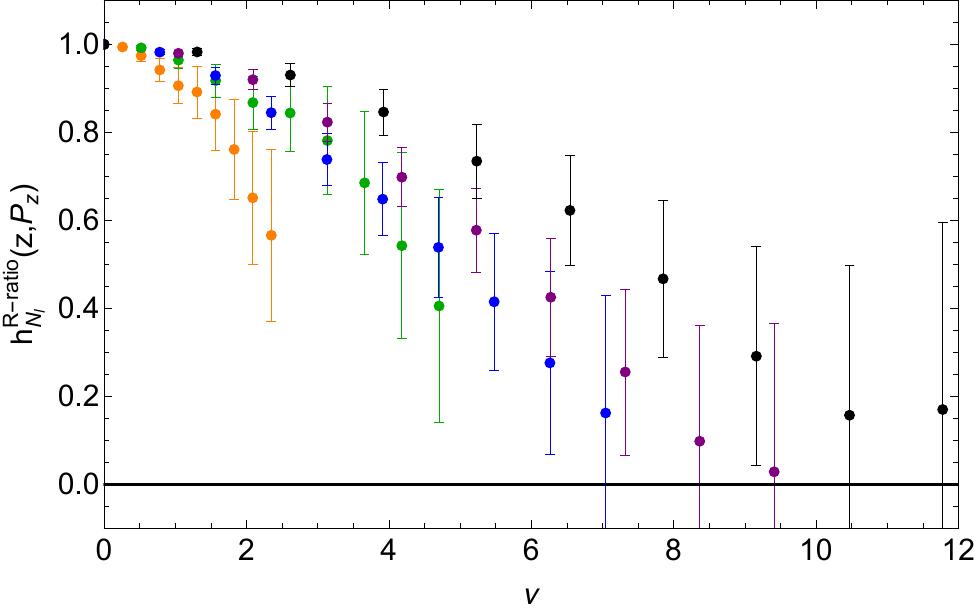}
\centering
        \includegraphics[width=.3\textwidth]{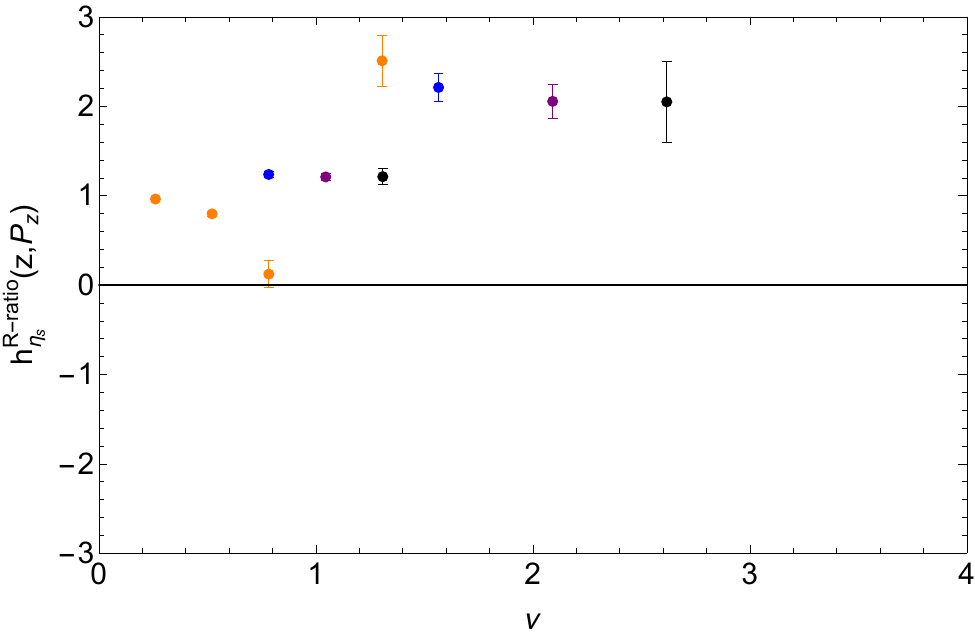}
\centering
        \includegraphics[width=.3\textwidth]{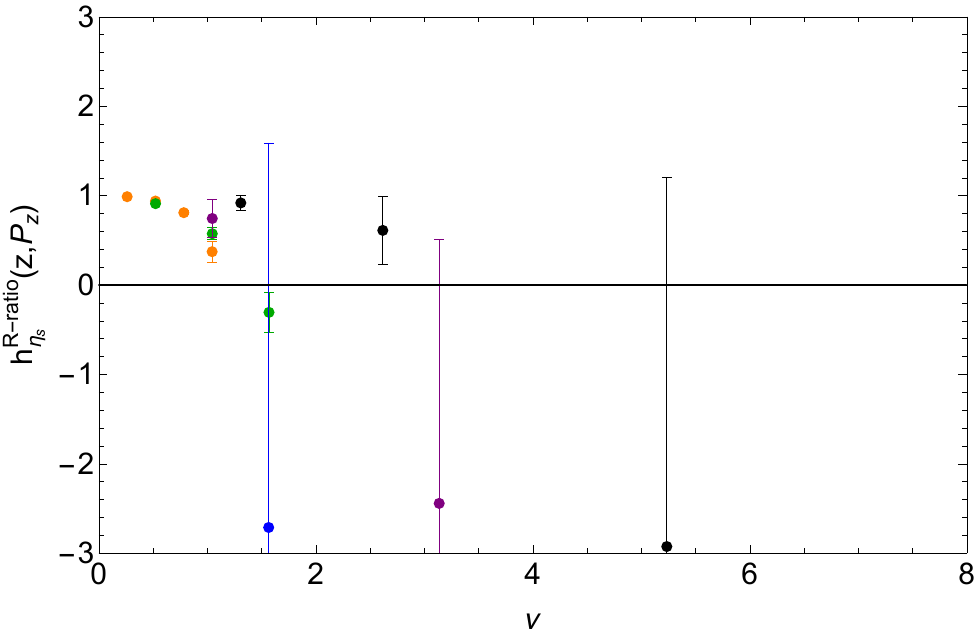}
\centering
        \includegraphics[width=.3\textwidth]{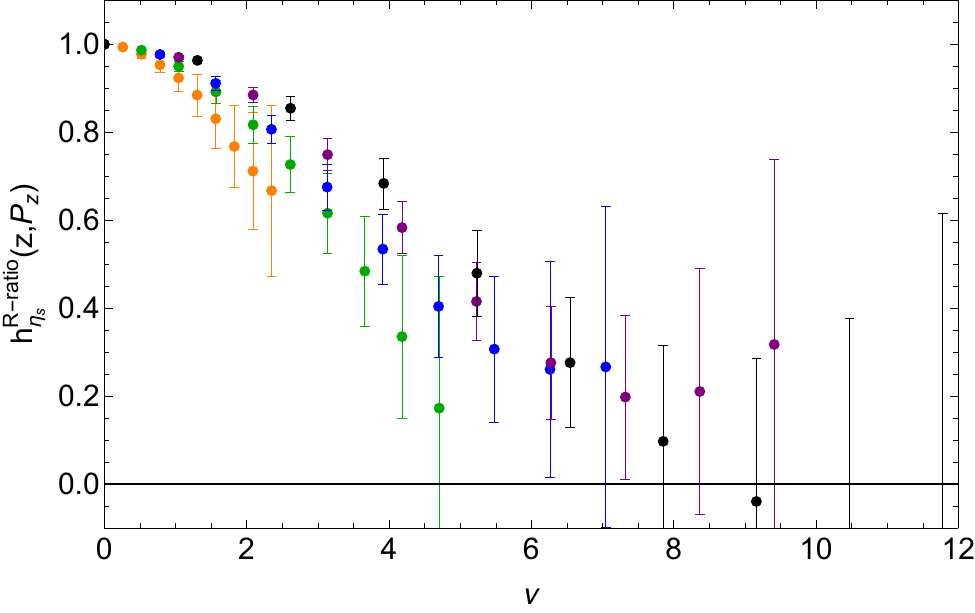}
\centering
        \includegraphics[width=.3\textwidth]{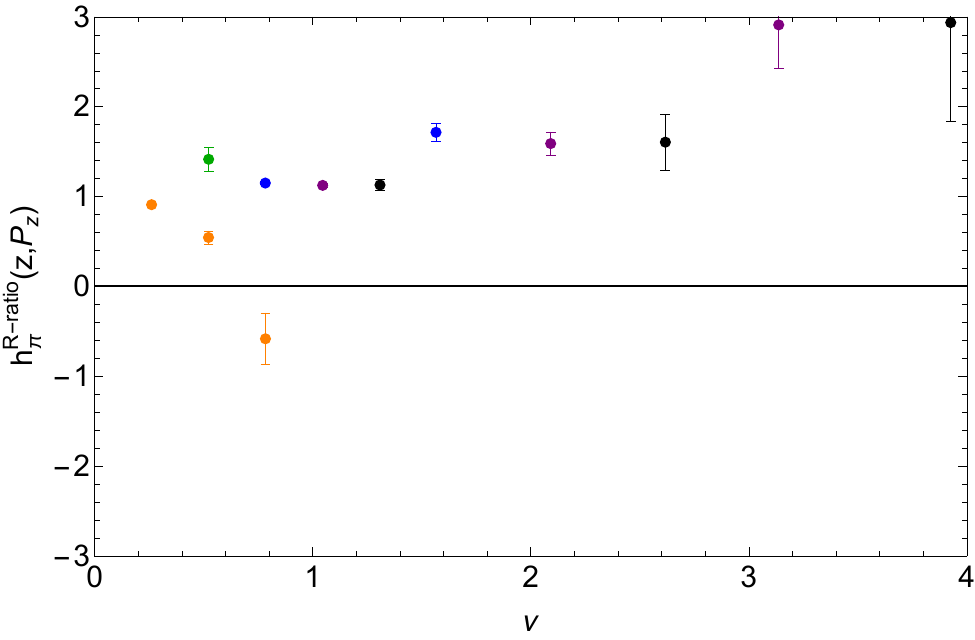}
\centering
        \includegraphics[width=.3\textwidth]{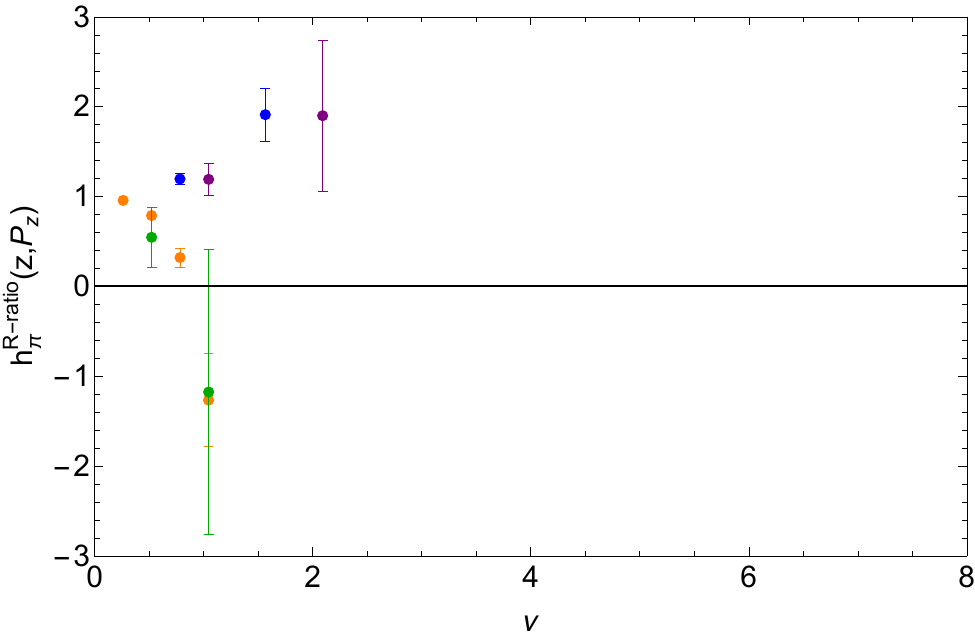}
\centering
        \includegraphics[width=.3\textwidth]{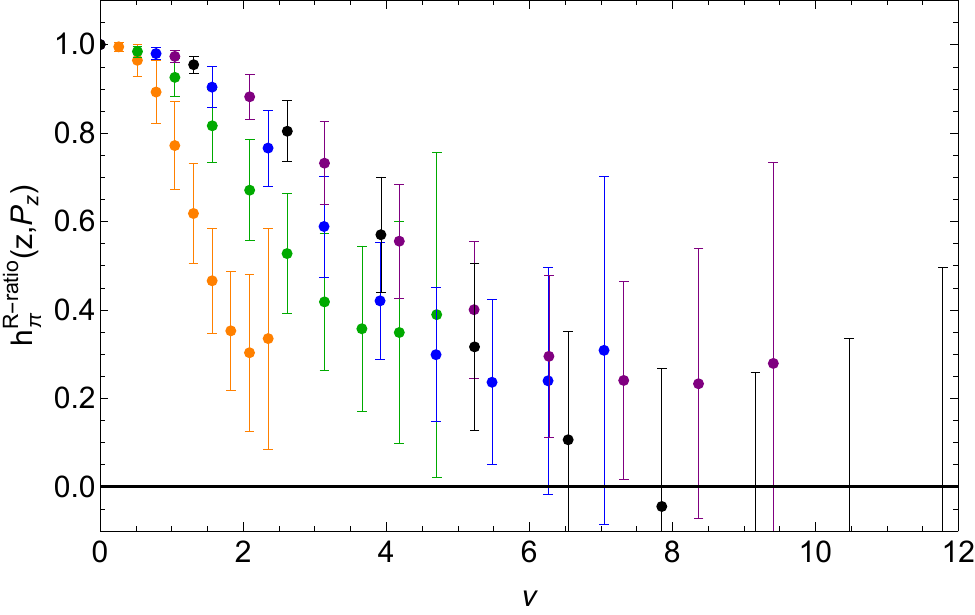}

    \caption{Ratio renormalized matrix elements for the HYP5 smearing data for the strange nucleon, light nucleon, $\eta_s$, and $\pi$ (rows top to bottom) for each operator $O^{(1,2,3)}(z)$ (columns left to right).}
    \label{fig:ratio-MEs_H5}
\end{figure*}

With the bare matrix elements, we may follow Eq.~\ref{eq:hyb_def} with $z_s \rightarrow \infty$ to get ratio renormalized matrix elements.
For each hadron and operator, we plot our results for Wilson-3 and HYP5 smearing in Figs.~\ref{fig:ratio-MEs-W3} and \ref{fig:ratio-MEs_H5}, respectively.
We plot the data against the unitless and invariant Ioffe time $\nu = zP_z$ so as to be able to compare results from different $P_z$.
To improve the clarity of the graph, we remove the many points with error over 200\% or with means of magnitude greater than three.
Note that the horizontal range increases in the plots from left to right and that the vertical range of the $O^{(3)}$ plots (rightmost column) is smaller than the first two.
In these two figures, we can immediately see that $O^{(1)}$ and $O^{(2)}$ (left two columns) both seem to have poor signal, the matrix elements diverge to infinity, and primarily in the meson cases, the matrix elements are very inconsistent between different momenta.
These effects likely come from zero-crossings in the bare matrix elements.
We see that $O^{(2)}$ (middle column) has reasonably smooth behavior in the nucleon cases (top two rows).
However, $O^{(3)}$ has by far the smoothest behavior and does not cross zero at a magnitude of more than $1\sigma$.

\begin{figure*}
\centering
    \includegraphics[width=.45\textwidth]{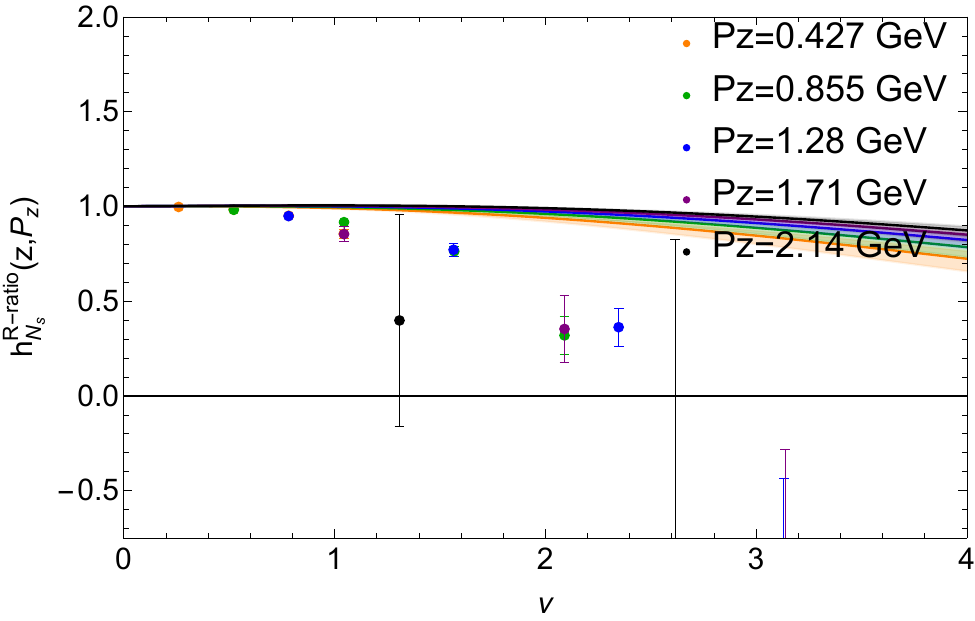}
\centering
    \includegraphics[width=.45\textwidth]{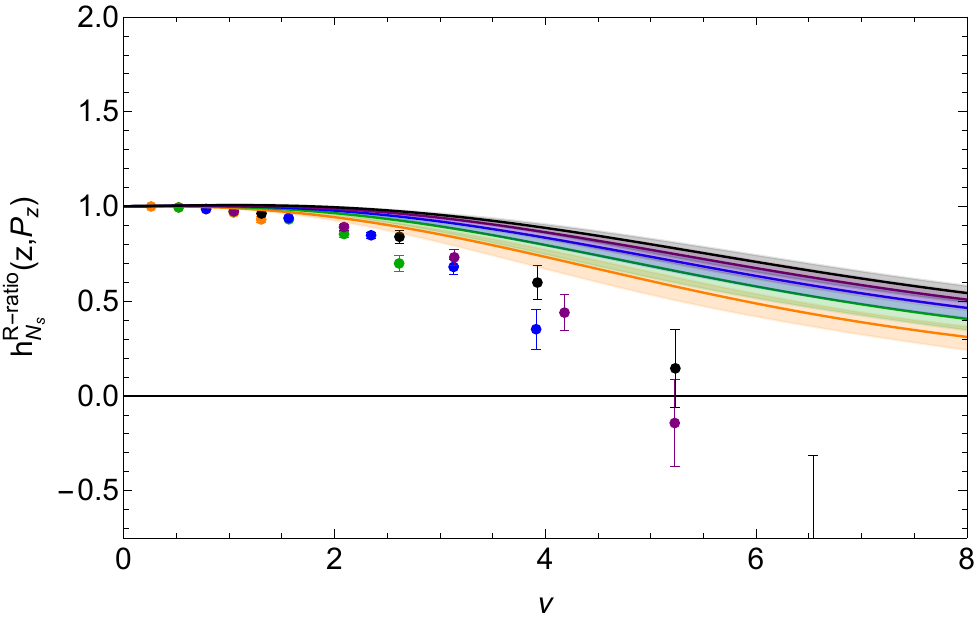}

\centering
    \includegraphics[width=.45\textwidth]{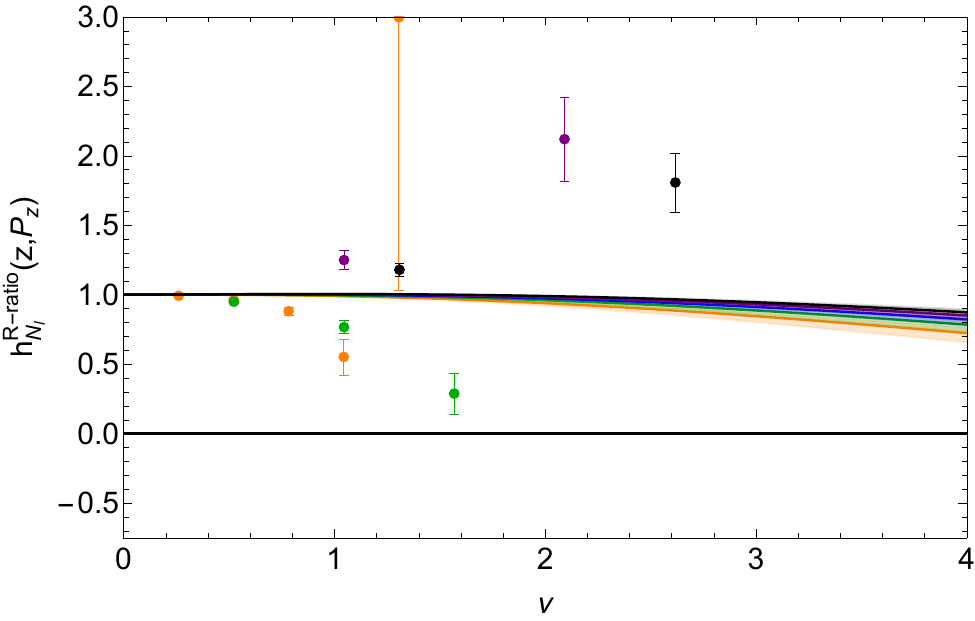}
\centering
    \includegraphics[width=.45\textwidth]{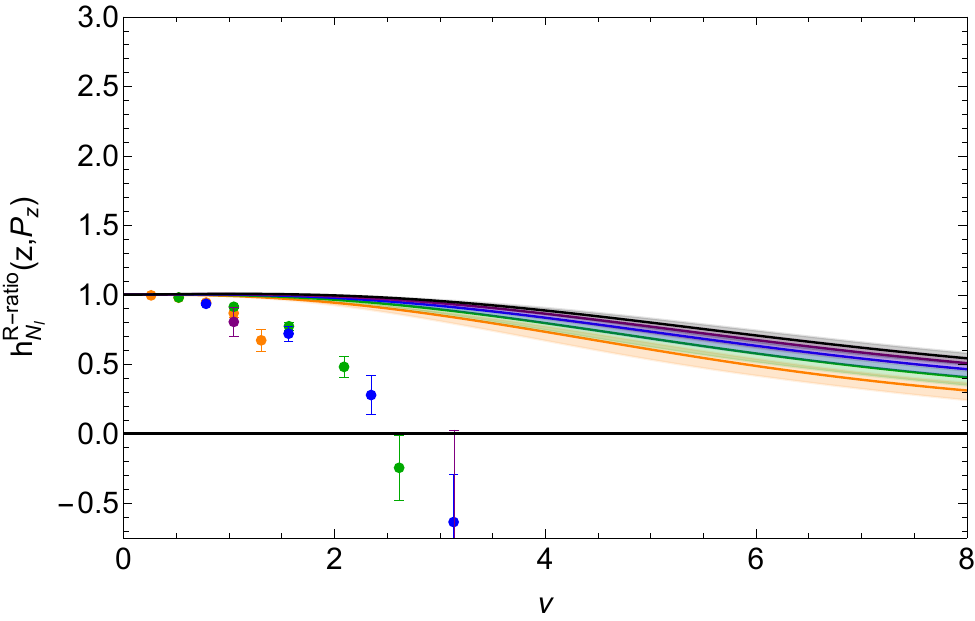}
    \caption{The lattice ratio renormalized matrix elements for the strange (top) and light (bottom) nucleon for operator $O^{(1)}$ (left) and $O^{(2)}$ (right) with Wilson-3 smearing compared to the respective, reconstructed phenomenological matrix elements from the CT18 nucleon gluon PDF~\cite{Hou:2019efy}.}\label{fig:W3_ratio_pheno_lat_comp}
\end{figure*}

\begin{figure*}
\centering
    \includegraphics[width=.45\textwidth]{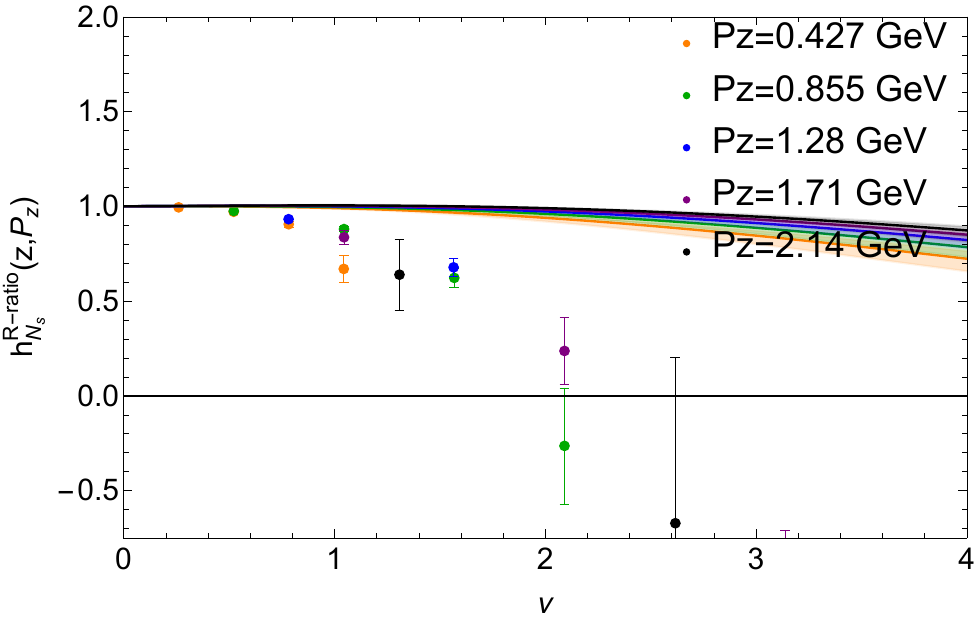}
\centering
    \includegraphics[width=.45\textwidth]{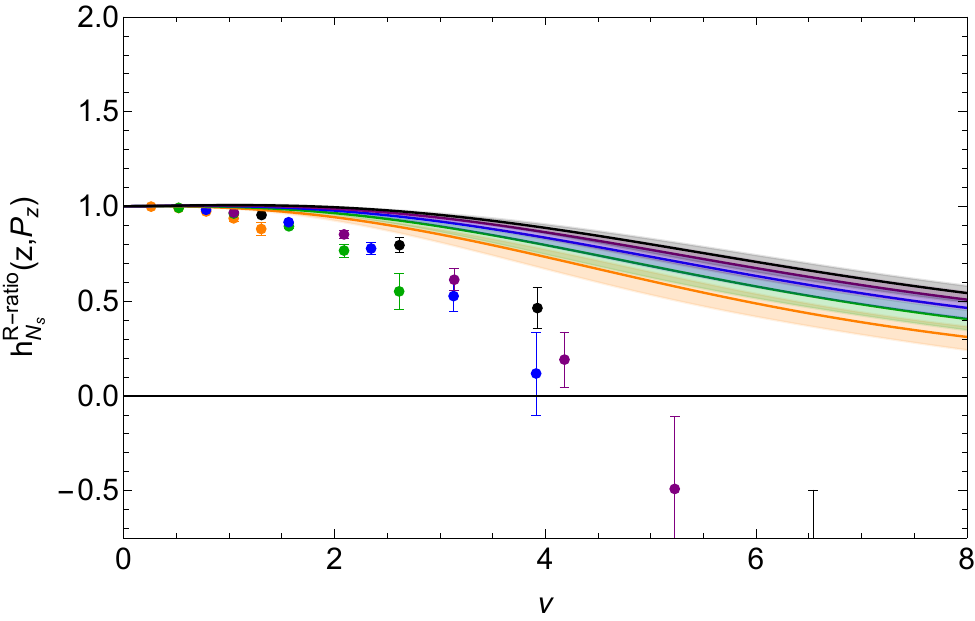}

\centering
    \includegraphics[width=.45\textwidth]{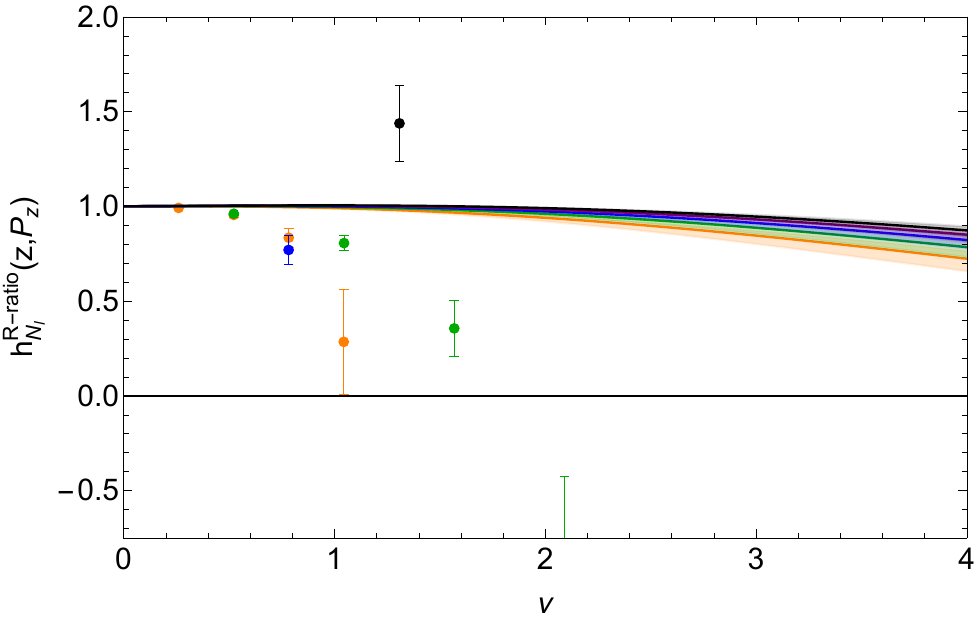}
\centering
    \includegraphics[width=.45\textwidth]{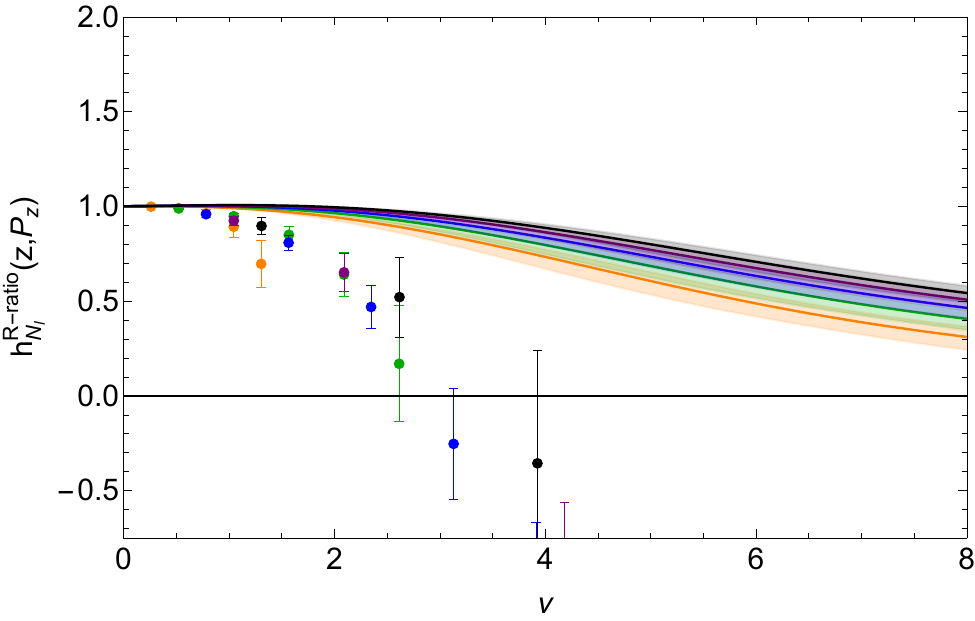}
    \caption{The lattice ratio renormalized matrix elements for the strange (top) and light (bottom) nucleon for operator $O^{(1)}$ (left) and $O^{(2)}$ (right) with HYP5 smearing compared to the respective, reconstructed phenomenological matrix elements from the CT18 nucleon gluon PDF~\cite{Hou:2019efy}.}\label{fig:H5_ratio_pheno_lat_comp}
\end{figure*}

At this level, it is clear that the signal and behavior of the ratio renormalized $O^{(3)}$ matrix elements are much better than the other operators.
We also know from the many previous studies of the gluon PDF through pseudo-PDF matching that $O^{(3)}$ produces matrix elements and PDFs comparable to phenomenological result~\cite{Fan:2020cpa,Fan:2021bcr,HadStruc:2021wmh,Fan:2022kcb,Delmar:2023agv,Good:2023ecp}.
It is worth exploring whether the behavior of the first two operators captures phenomenological behavior in any way.
We narrow down to the more commonly phenomenologically studied nucleon gluon PDF, taking the CT18~\cite{Hou:2019efy} gluon PDF at $\overline{\text{MS}}$ scheme scale $\mu=2.0$~GeV, and use Eq.~\ref{eq:matching} with the ratio matching kernels from Ref.~\cite{Yao:2022vtp} to obtain a quasi-PDF.
We ignore the glue-quark mixing term in this case, as it has been shown to be small in the pseudo-PDF studies.
We Fourier transform the quasi-PDF back to position space so that we have ``phenomenological matrix elements'' with which to compare the $O^{(1)}$ and $O^{(2)}$ matrix elements.
We plot the $O^{(1)}$ and $O^{(2)}$ operator results for the strange and light nucleons compared the the phenomenological matrix elements in Figs.~\ref{fig:W3_ratio_pheno_lat_comp} and \ref{fig:H5_ratio_pheno_lat_comp} for the Wilson-3 and HYP5 smearing respectively.
We use the asymmetrical error formula to get the error bars for the phenomenological results.
We can see that the phenomenological matrix elements are reasonably consistent in this range across different different $P_z$ and that they decay much more slowly than the lattice matrix elements.
In the top right plot in each figure, we see that the strange nucleon $O^{(2)}$ results agree best with the phenomenological results at smaller $\nu$; however the phenomenological results capture no sign change.
These results together suggest that on top of the poorer signal in the raw data, there may also be systematic contaminations in these two operators as suggested by Ref.~\cite{Balitsky:2019krf} in the context of short-distance behavior.
It will be interesting to see if hybrid renormalization can make up at all for the divergent behavior of the matrix elements.

\subsection{Hybrid-Ratio Renormalized Matrix Elements }

\subsubsection{Operators $O^{(1)}$ and $O^{(2)}$}

We only have the necessary information to apply hybrid renormalization to $O^{(1)}$ and $O^{(2)}$, so we wish to explore this to see if the results change and to possibly make an ansatz for the hybrid renormalization of $O^{(3)}$.

\begin{figure*}
\centering
        \includegraphics[width=.45\textwidth]{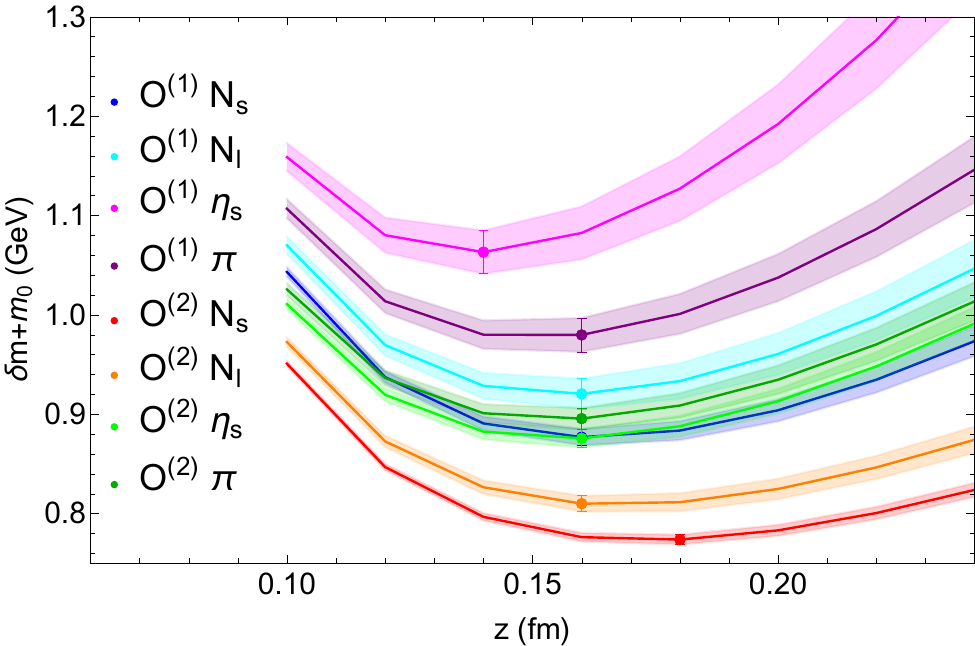}
\centering
        \includegraphics[width=.45\textwidth]{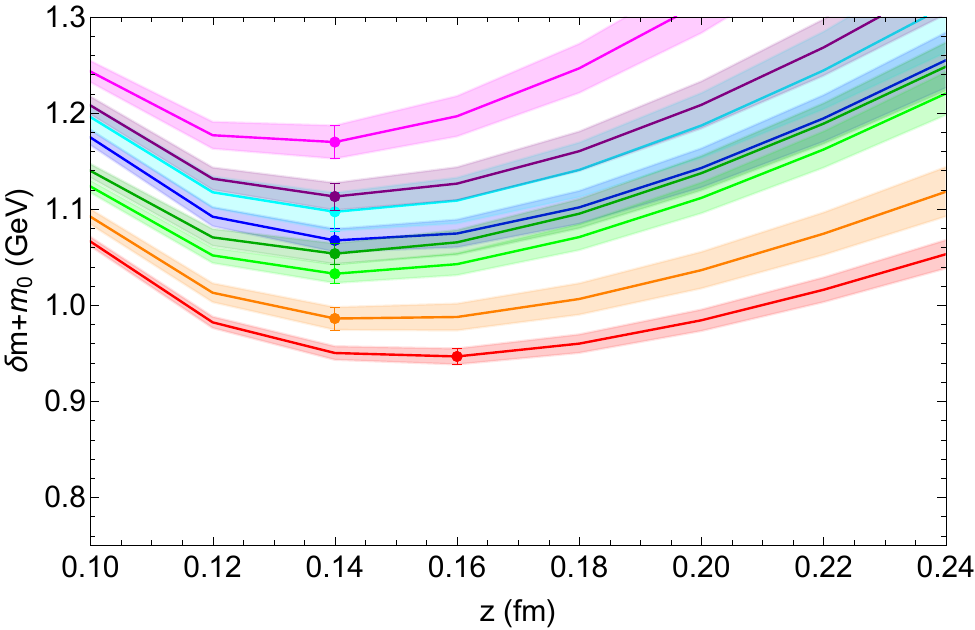}

    \caption{Fit results for the $\delta m + m_0$ for varying central point $z$ for the two operators, for each hadron for the two smearings, Wilson3 (left) and HYP5 (right)}
    \label{fig:m0_vs_z}
\end{figure*}

For operators $O^{(1)}$ and $O^{(2)}$, we have the Wilson coefficients as defined in Eq.~\ref{eq:wil-coef}, so we may fit $\delta m + m_0$ from Eq~\ref{eq:m0-fit} and apply hybrid renormalization with $\mu=2.0$~GeV.
As stated before, the lattice spacing $a\approx 0.12$~fm is too coarse to capture the range of linear behavior in the small-$z$ region, so we interpolate $h^\text{B}(z,0)$ to get finer data to apply the fit.
We fit the interpolated data to Eq.~\ref{eq:m0-fit} with points $\{z-0.2, z, z+0.2\}$ in units of fm, varying $z$.
We show these results for the $\delta m + m_0$ versus $z$ for the two operators, for each hadron for the two smearings, Wilson3 (left) and HYP5 (right), in Fig.~\ref{fig:m0_vs_z}.

We see that the fits are not consistent between the different values of smearing and for different operators, which are both expected results.
The hadron and pion mass also have noticeable effects on the fits.
The behavior is consistent in that a larger pion mass, results in a smaller $\delta m + m_0$ and that the fitted values for the nucleon are smaller than those of the mesons.
Ref.~\cite{Ji:2020brr} summarizes a few reasons why the $m_0$ fit will depend on the specific matrix element fit, and we confirm that this is non-negligible at this level.
We choose the $z$ that results in the minimum $\delta m + m_0$ for the final value for the hybrid renormalization of each respective operator, hadron, and smearing.
We make this choice because the minimum seems to correspond to the region around which the fit is most stable.
The $z$ here seems to be large enough that the logarithm in the Wilson coefficient is not diverging and small enough that perturbation theory still holds.
We leave it to future studies to consider how scale variation, leading renormalon resummation, and renormalization group resummation affect these fits~\cite{Holligan:2023rex,Su:2022fiu,Zhang:2023bxs}.

\begin{figure*}
\centering
        \includegraphics[width=.45\textwidth]{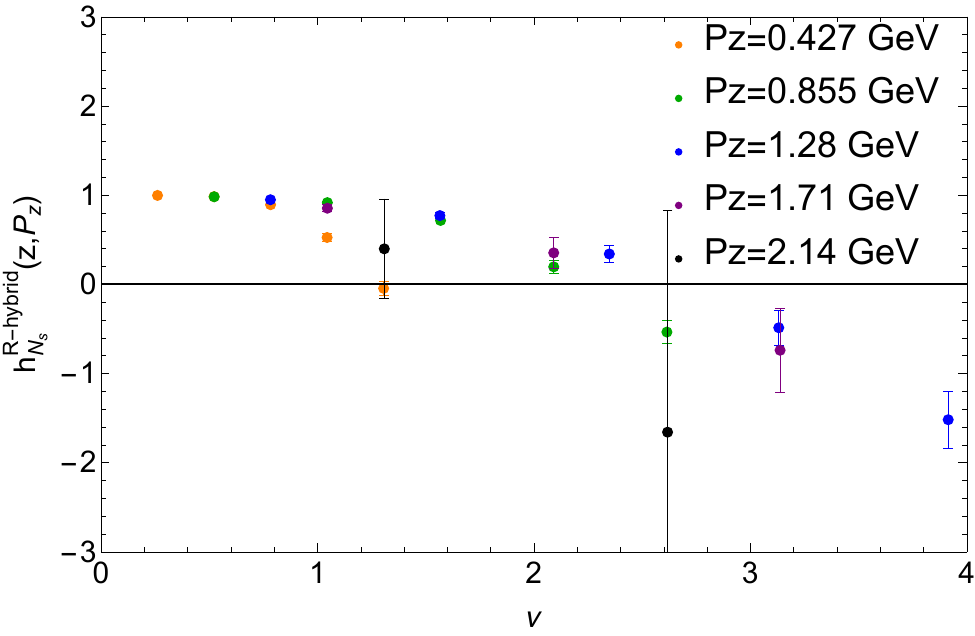}
\centering
        \includegraphics[width=.45\textwidth]{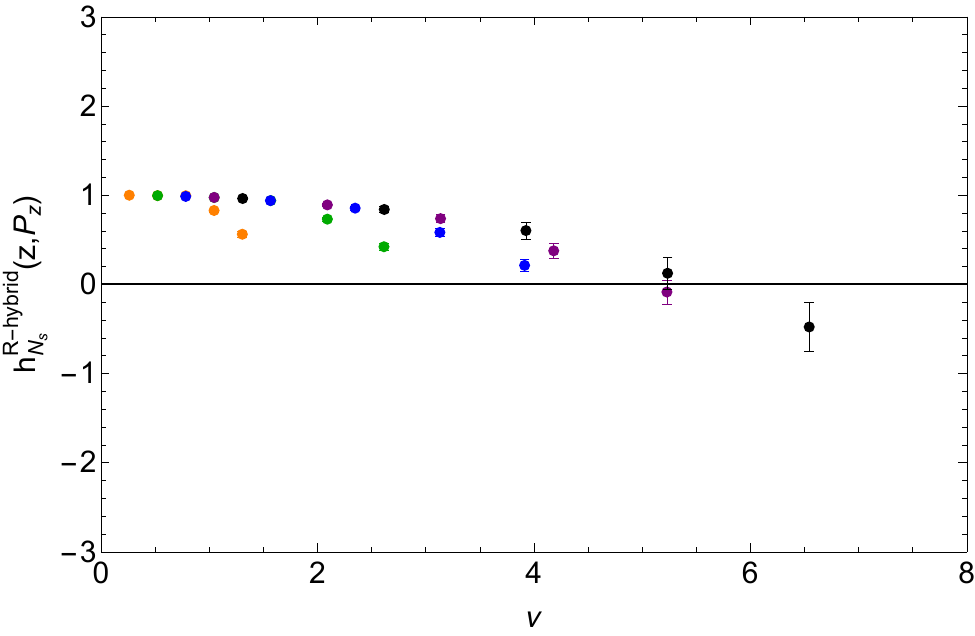}
\centering
        \includegraphics[width=.45\textwidth]{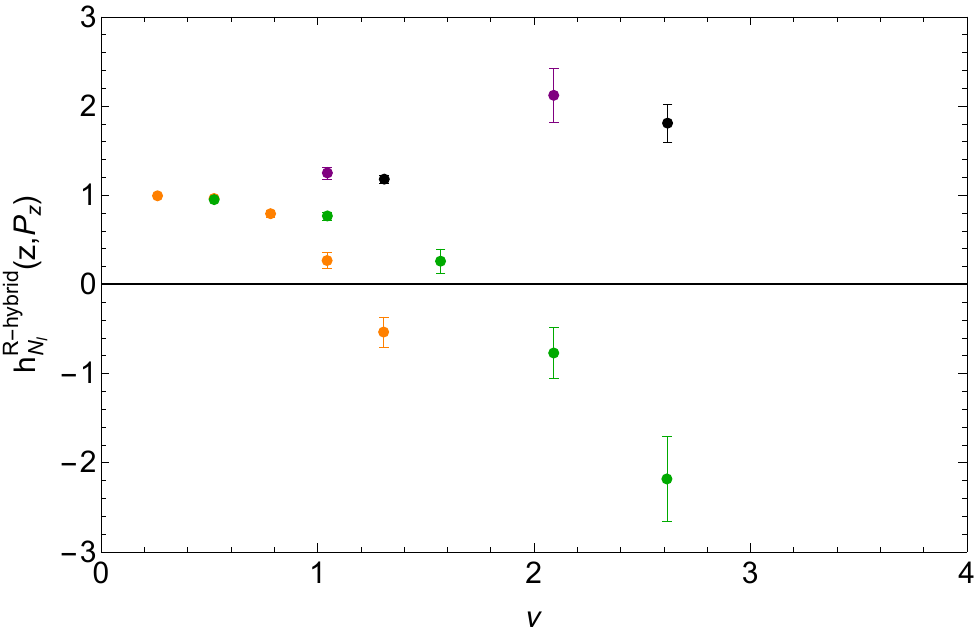}
\centering
        \includegraphics[width=.45\textwidth]{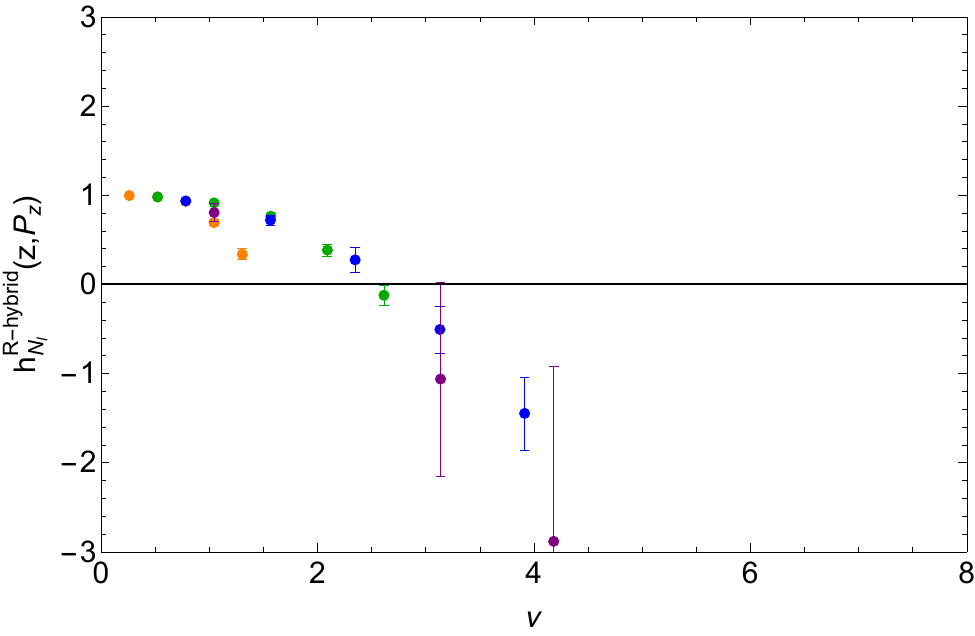}
\centering
        \includegraphics[width=.45\textwidth]{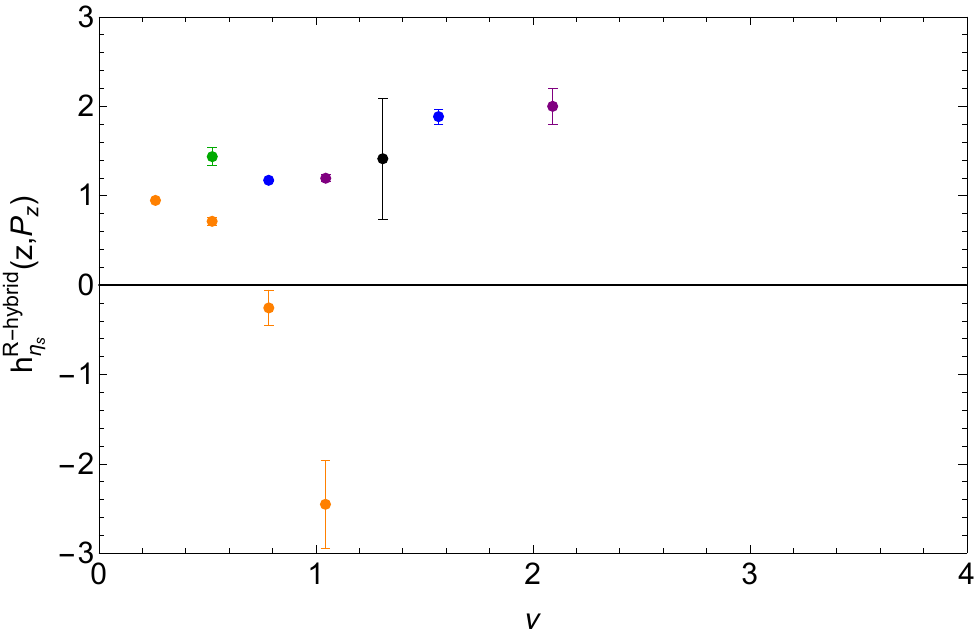}
\centering
        \includegraphics[width=.45\textwidth]{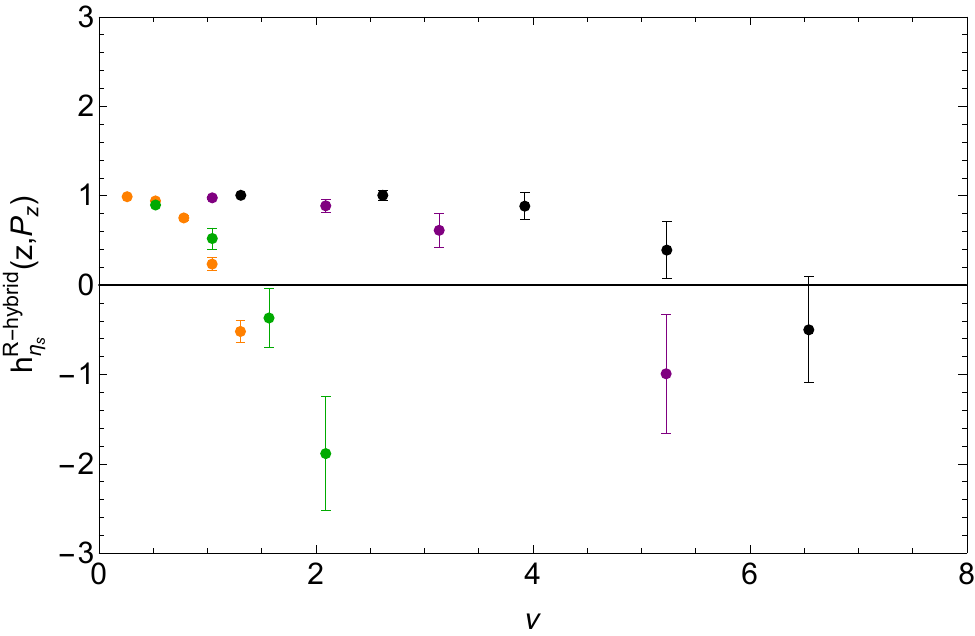}
\centering
        \includegraphics[width=.45\textwidth]{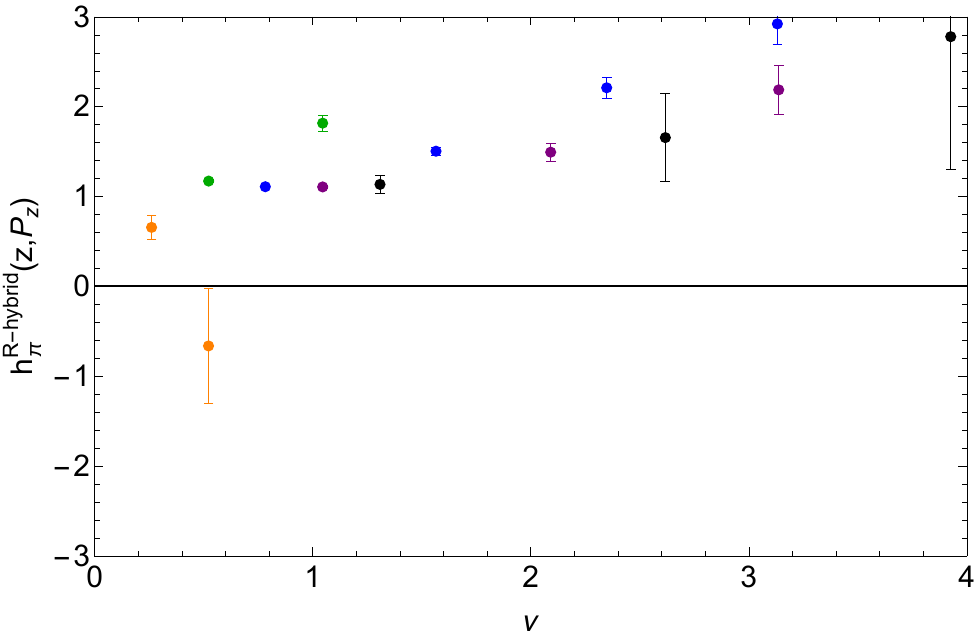}
\centering
        \includegraphics[width=.45\textwidth]{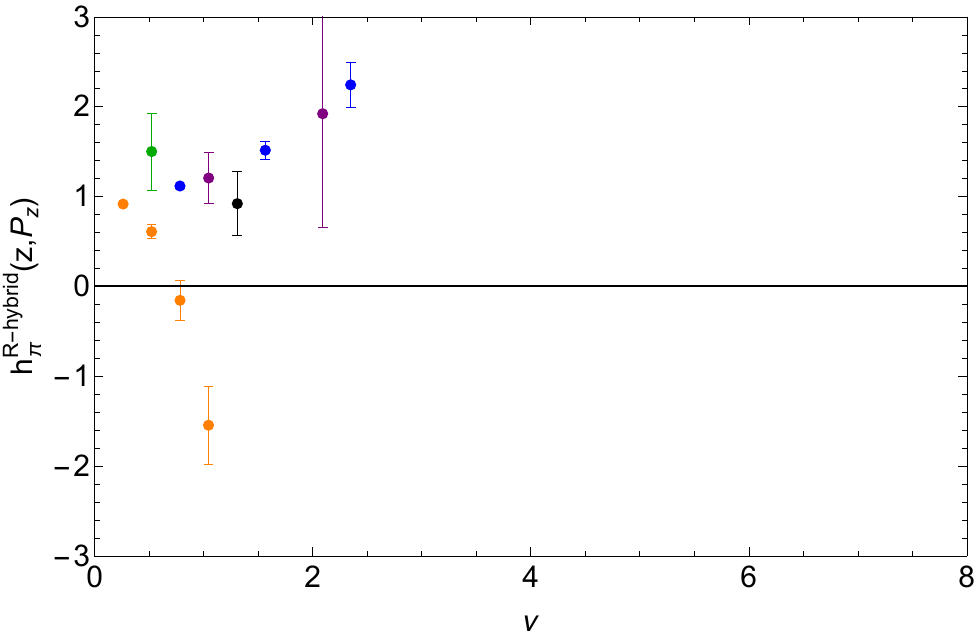}
    \caption{Hybrid-ratio renormalized matrix elements for the Wilson3 smearing data for the strange nucleon, light nucleon, $\eta_s$, and $\pi$ (rows top to bottom) for each operator $O^{(1,2,3)}(z)$ (columns left to right). The renormalization is done with $z_s = 0.24$~fm.}
    \label{fig:hybrid-MEs_W3}
\end{figure*}

\begin{figure*}
\centering
        \includegraphics[width=.45\textwidth]{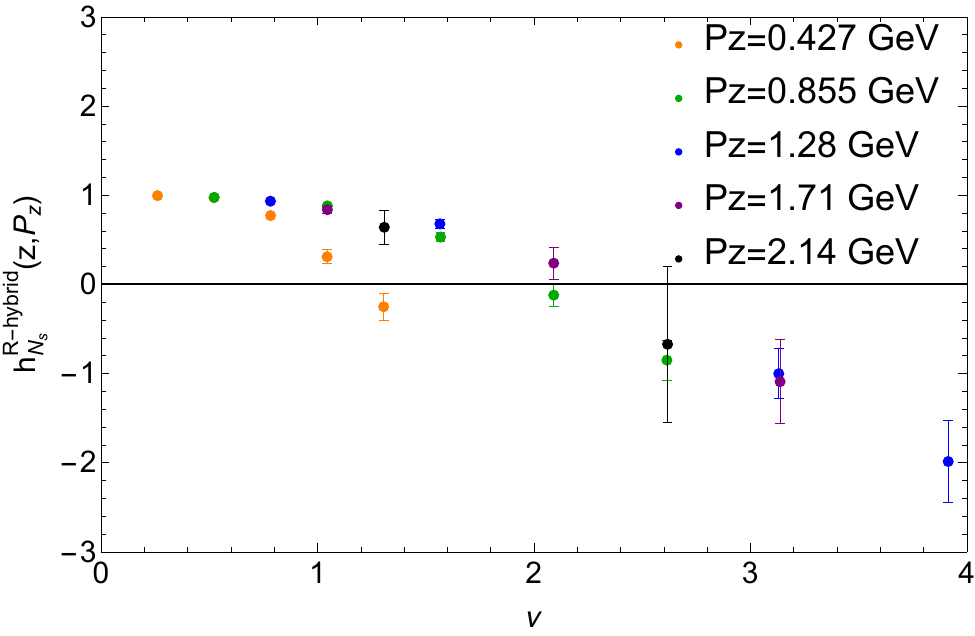}
\centering
        \includegraphics[width=.45\textwidth]{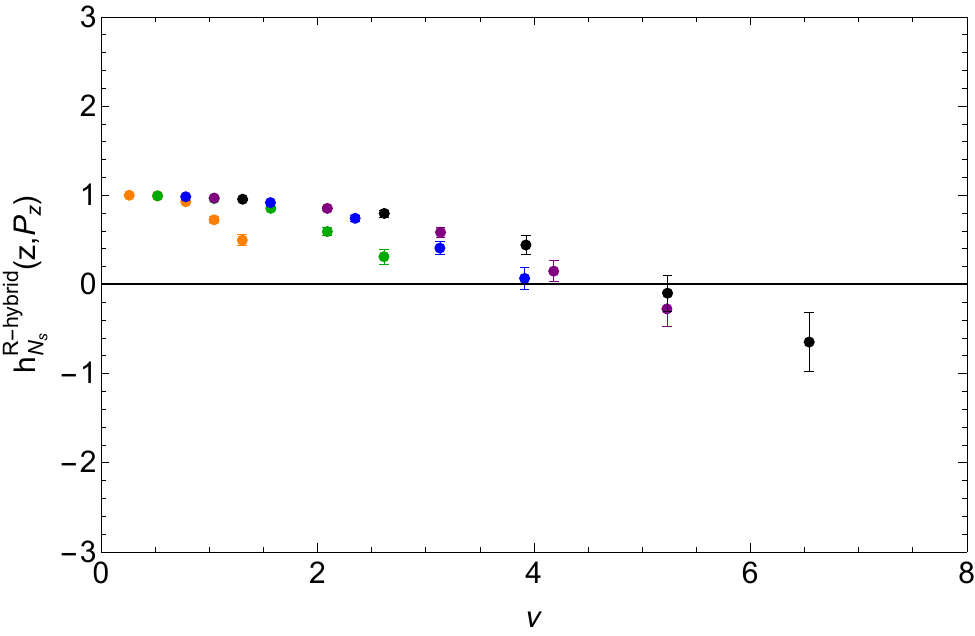}
\centering
        \includegraphics[width=.45\textwidth]{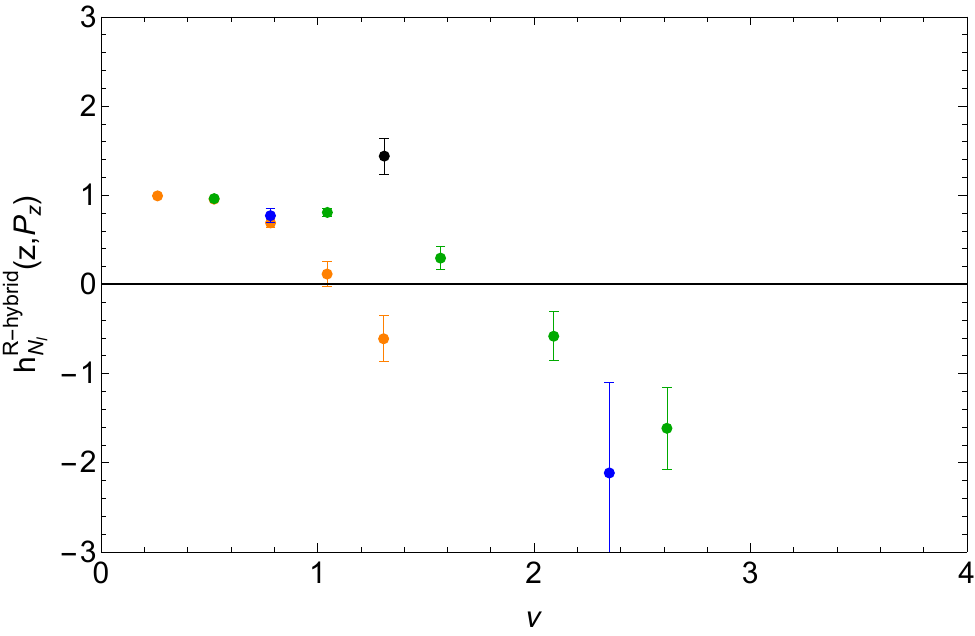}
\centering
        \includegraphics[width=.45\textwidth]{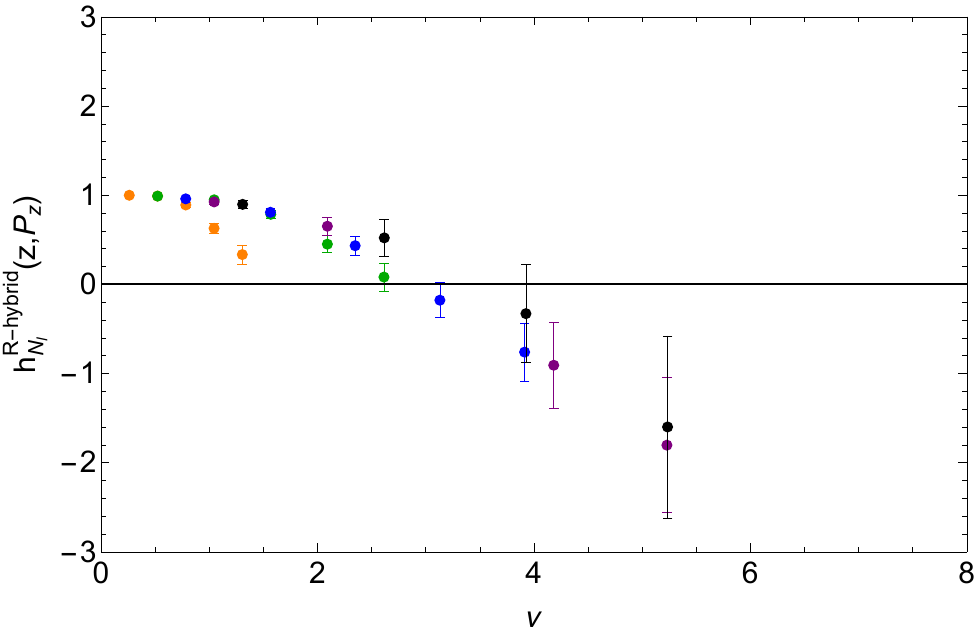}
\centering
        \includegraphics[width=.45\textwidth]{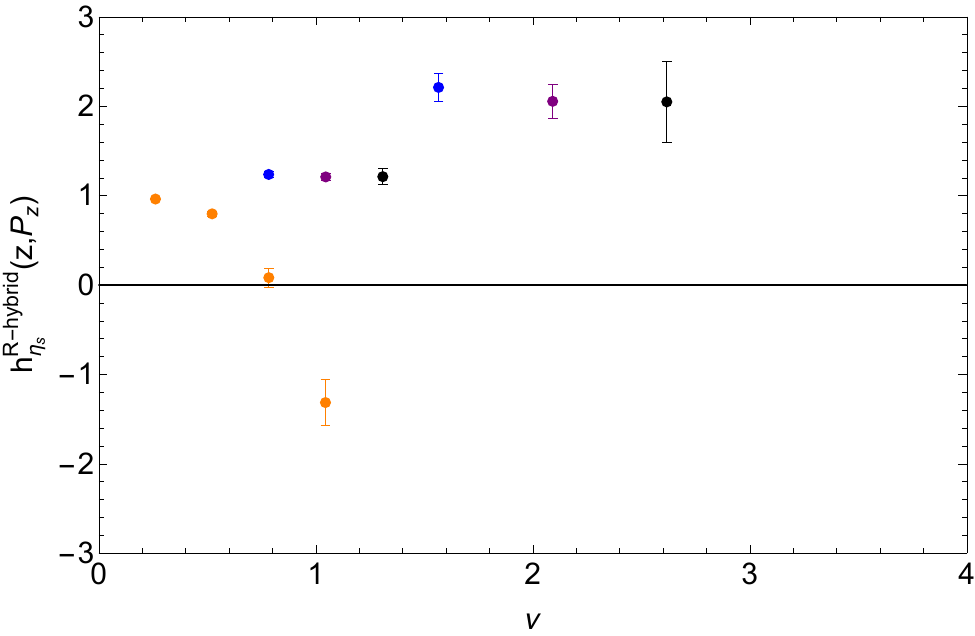}
\centering
        \includegraphics[width=.45\textwidth]{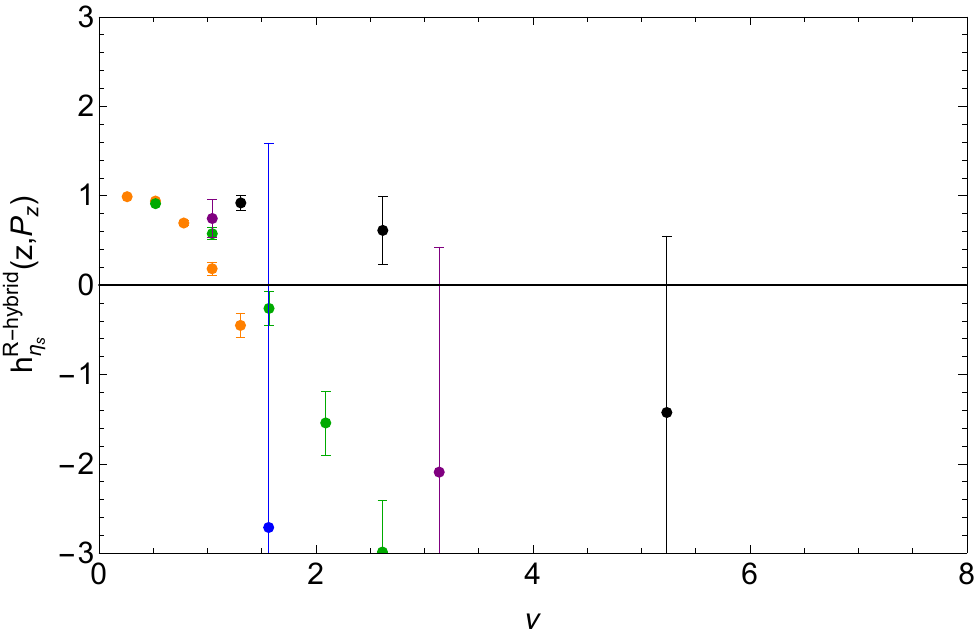}
\centering
        \includegraphics[width=.45\textwidth]{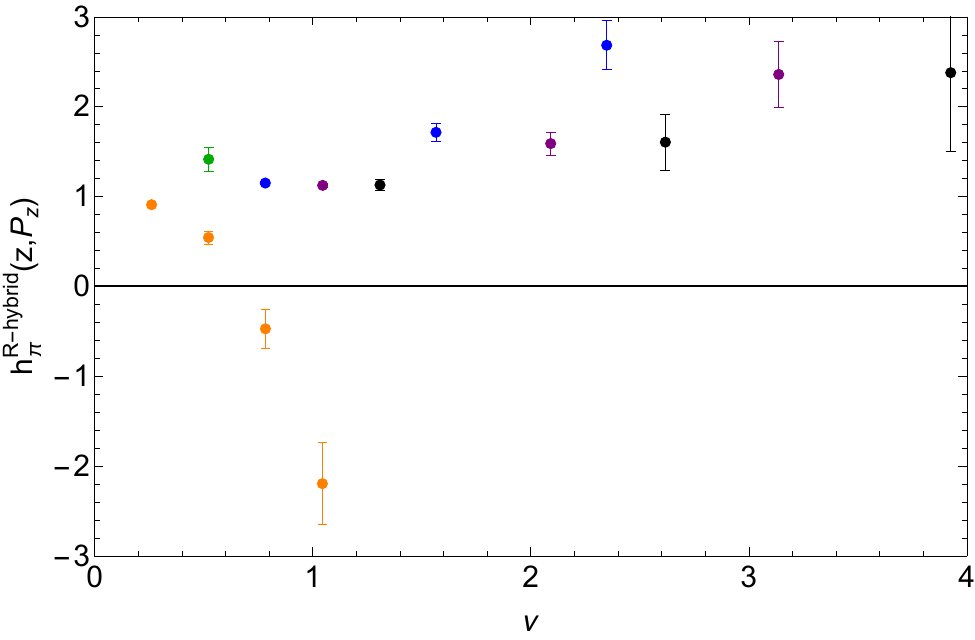}
\centering
        \includegraphics[width=.45\textwidth]{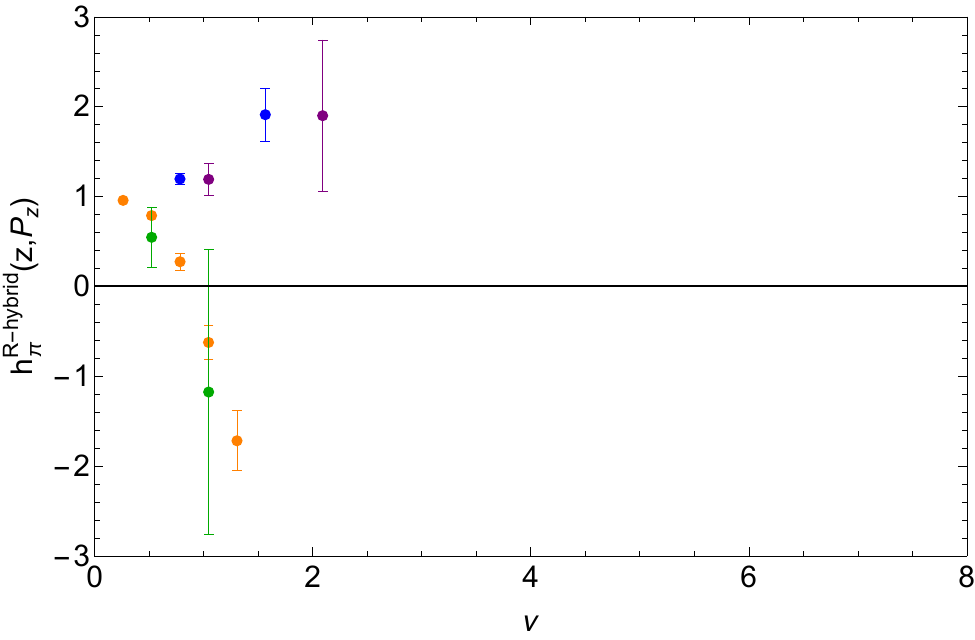}

    \caption{Hybrid-ratio renormalized matrix elements for the HYP5 smearing data for the strange nucleon, light nucleon, $\eta_s$, and $\pi$ (rows top to bottom) for each operator $O^{(1,2,3)}(z)$ (columns left to right). The renormalization is done with $z_s = 0.24$~fm.}
    \label{fig:hybrid-MEs_H5}
\end{figure*}

Now that $\delta m + m_0$ has been fit for $O^{(1)}$ and $O^{(2)}$ for each hadron and smearing, we can see if anything has changed in the behavior of the renormalized matrix elements.
We show the hybrid-ratio renormalized matrix elements with $z_s = 0.24$~fm in Fig.~\ref{fig:hybrid-MEs_W3} and \ref{fig:hybrid-MEs_H5} for Wilson-3 and HYP5 smearing respectively.
Again, we remove the many points with error over 200\% or means with magnitude greater than three for clarity and use the same plot ranges as for the ratio renormalized matrix elements.
The hybrid renormalized matrix elements exhibit similar behavior to the ratio renormalized ones, overall.
We see that $O^{(1)}$ (left column) has poor signal and gives inconsistent results at different momenta in nearly all cases.
We see again that the cleanest and most consistent signal comes from the strange nucleon and $O^{(2)}$ (right column) with more divergent behavior in the mesons (bottom two rows); however, the crossing below zero and overall divergent behavior at such short distances is still concerning for these operators.

At this level, it would not appear that the hybrid renormalization has made up for the ratio renormalized matrix elements quick decay, but it is worth exploring the phenomenological matrix elements obtained using the hybrid-ratio matching kernels instead.
We plot the $O^{(1)}$ and $O^{(2)}$ operator results for the strange and light nucleons compared the the phenomenological matrix elements in Figs.~\ref{fig:W3_ratio_pheno_lat_comp} and \ref{fig:H5_ratio_pheno_lat_comp} for the Wilson-3 and HYP5 smearing results respectively.
We again use the asymmetrical error formula for the phenomenological error bars, which clearly affects the results just after $\nu_s$.
We also quickly see that there is an interesting bump in the phenomenological matrix elements at $\nu_s = z_sP_z$ and also that the lattice matrix elements do not capture this bump at all.
In almost every case, the lattice matrix elements diverge quickly and cross zero in a way that is also not captured by the phenomenological matrix elements.
This suggests that the $O^{(3)}$ operator needs to be considered more closely.

\begin{figure*}\label{fig:W3_hybrid_pheno_lat_comp}
\centering
    \includegraphics[width=.45\textwidth]{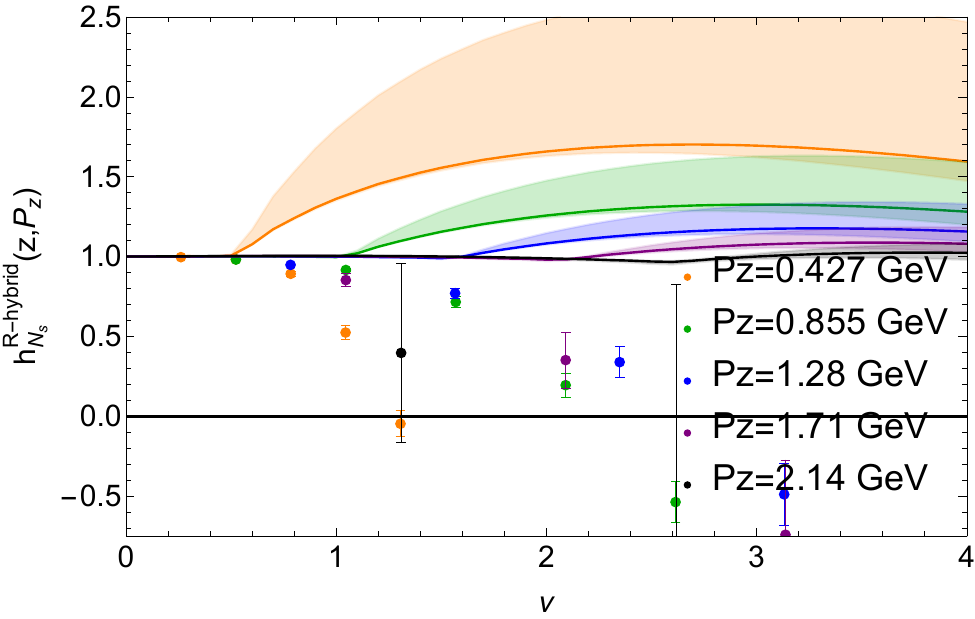}
\centering
    \includegraphics[width=.45\textwidth]{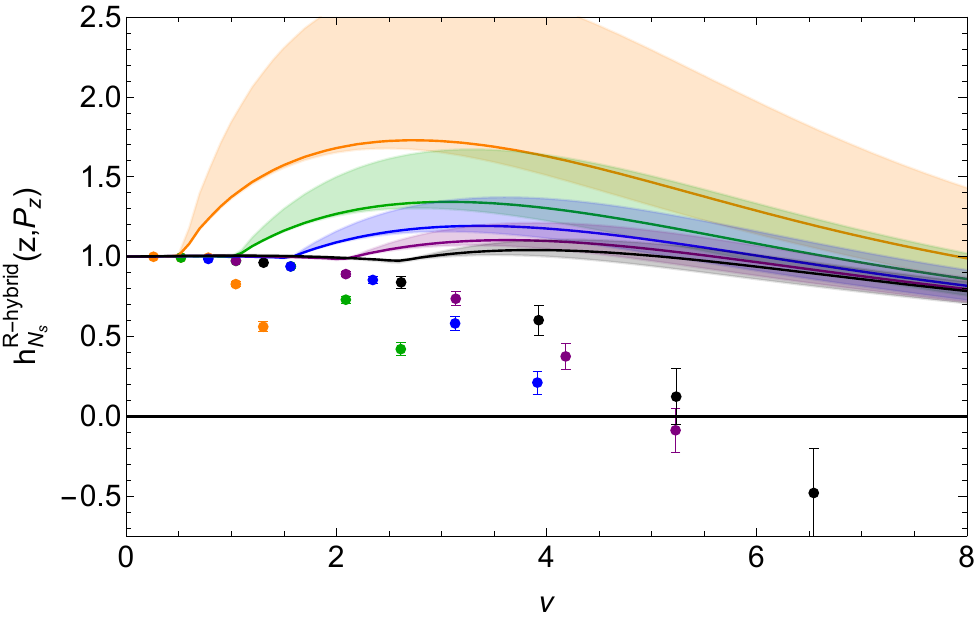}

\centering
    \includegraphics[width=.45\textwidth]{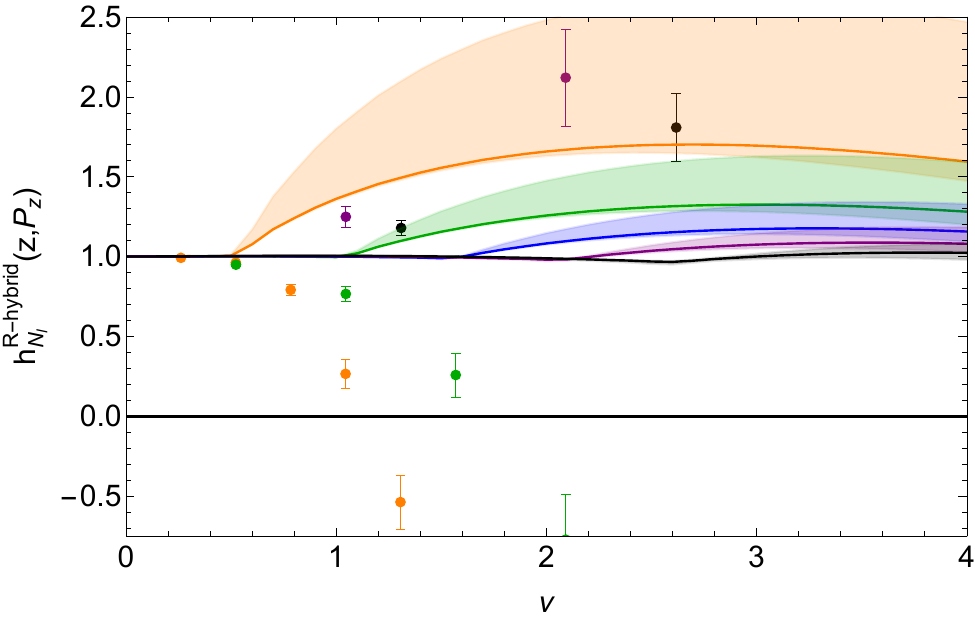}
\centering
    \includegraphics[width=.45\textwidth]{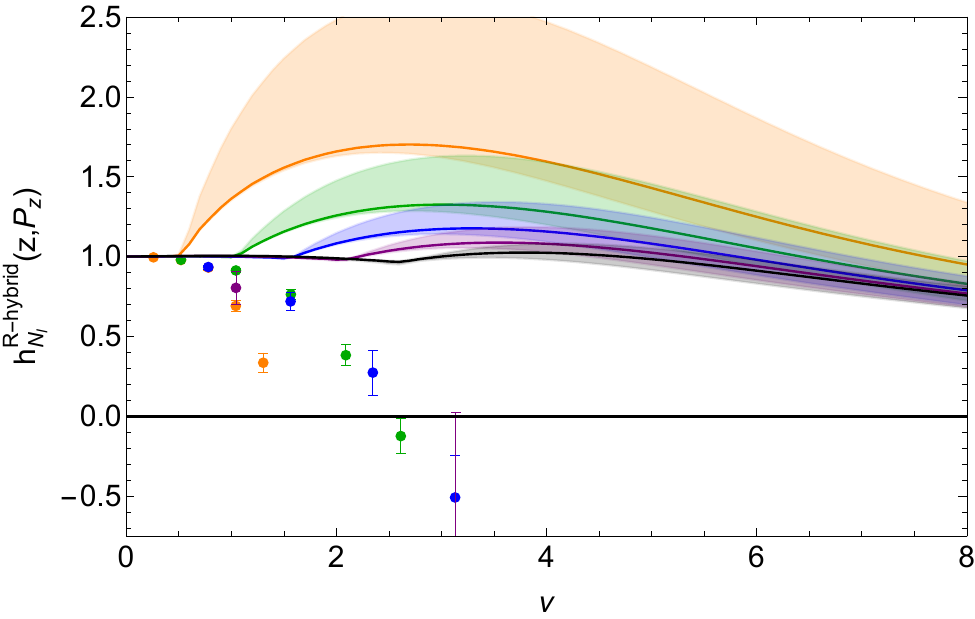}
    \caption{The lattice hybrid-ratio renormalized matrix elements for the strange (top) and light (bottom) nucleon for operator $O^{(1)}$ (left) and $O^{(2)}$ (right) with Wilson-3 smearing compared to the respective, reconstructed phenomenological matrix elements from the CT18 nucleon gluon PDF~\cite{Hou:2019efy}. The renormalization is done with $z_s = 0.24$~fm.}
\end{figure*}

\begin{figure*}\label{fig:H5_hybrid_pheno_lat_comp}
\centering
    \includegraphics[width=.45\textwidth]{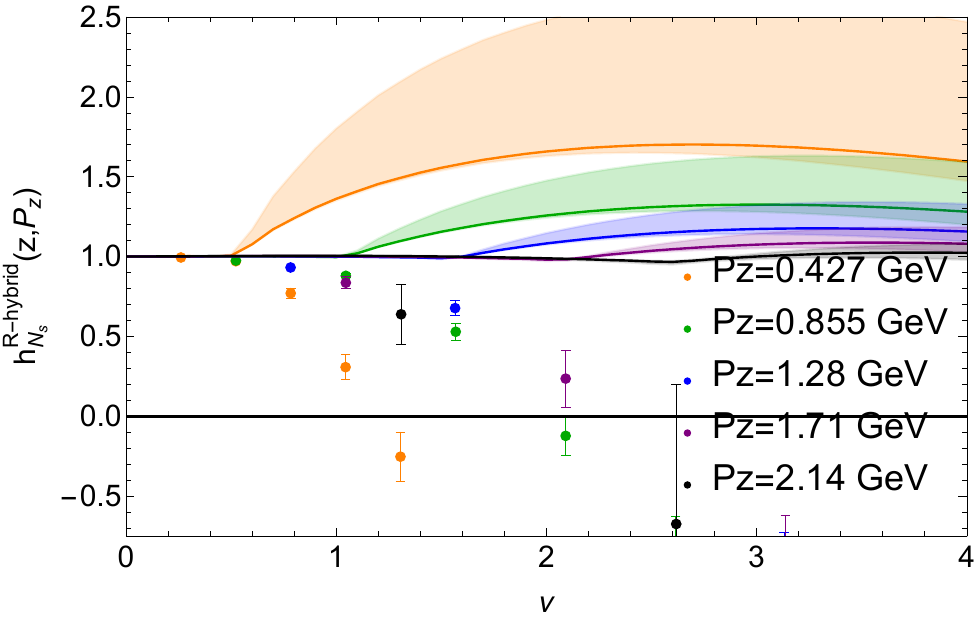}
\centering
    \includegraphics[width=.45\textwidth]{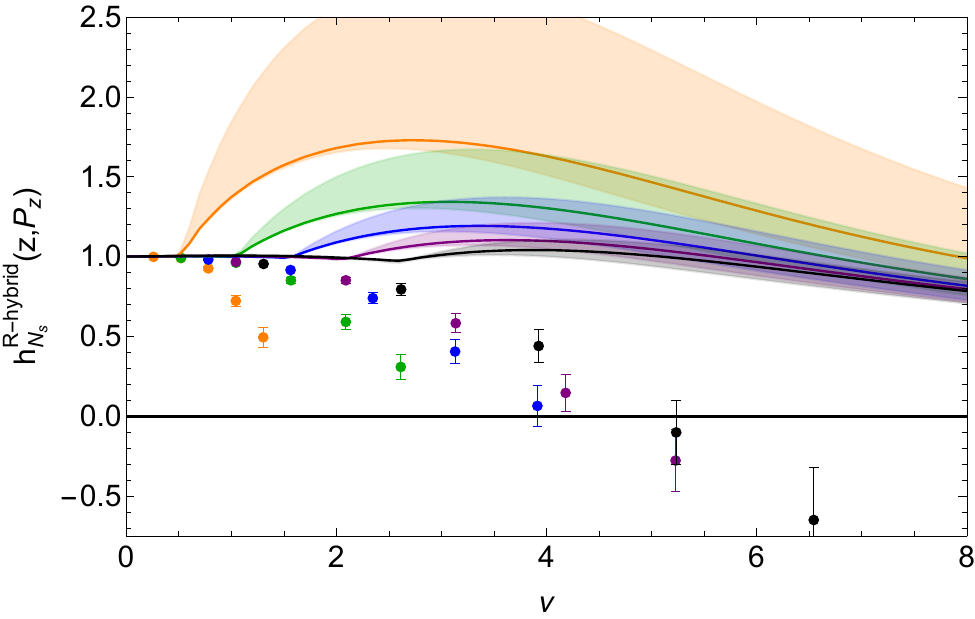}

\centering
    \includegraphics[width=.45\textwidth]{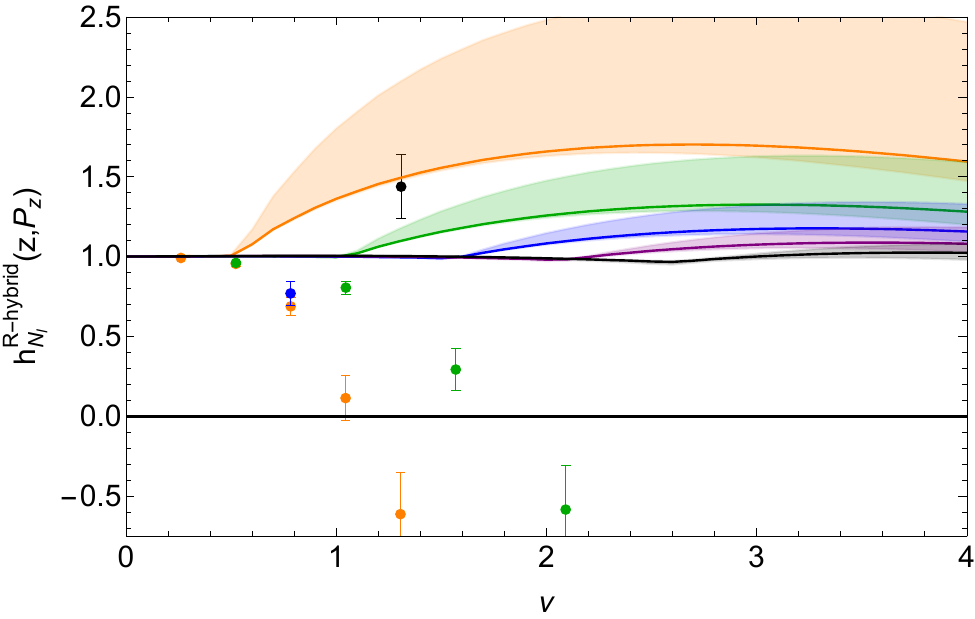}
\centering
    \includegraphics[width=.45\textwidth]{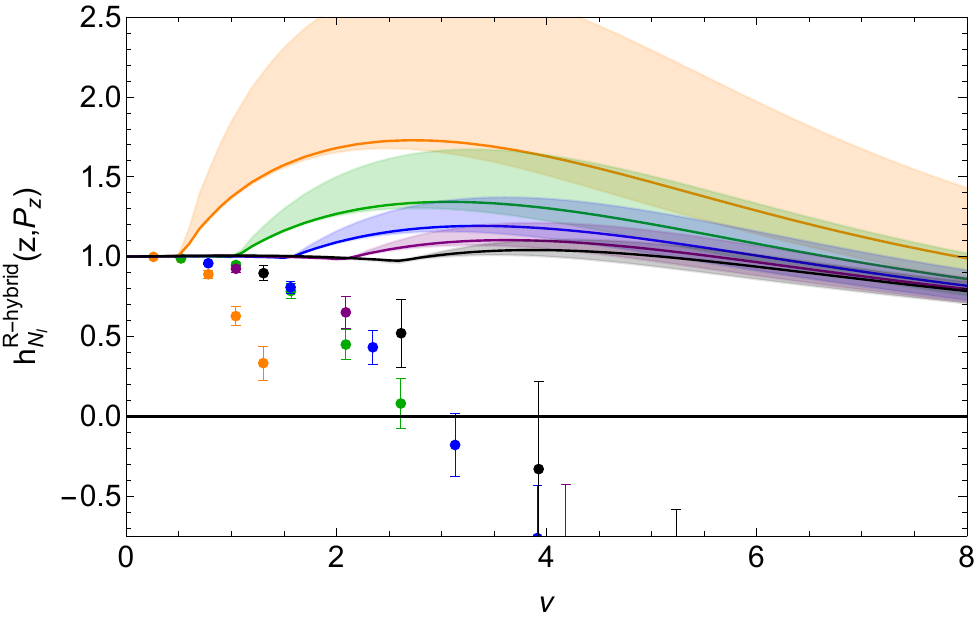}

    \caption{The lattice hybrid-ratio renormalized matrix elements for the strange (top) and light (bottom) nucleon for operator $O^{(1)}$ (left) and $O^{(2)}$ (right) with HYP5 smearing compared to the respective, reconstructed phenomenological matrix elements from the CT18 nucleon gluon PDF~\cite{Hou:2019efy}. The renormalization is done with $z_s = 0.24$~fm.}
\end{figure*}

\subsubsection{Operator $O^{(3)}$}

Although we do not have the Wilson coefficients for $O^{(3)}$ to fit $\delta m + m_0$ directly, we want to take a guess at what a hybrid-ratio renormalized matrix elements may look like from this operator.
We emphasize that this and the next section are hypothetical and rely on an educated, but still subjective, guess at a value of $\delta m + m_0$, along with using data that is likely smeared too much and has a much heavier than physical pion mass.
We considered our cleanest data, the strange nucleon with Wilson-3 smearing, and measured further to $z=23a$.
We only consider data up to $z=13a$, as the data becomes too noisy and likely contaminated by finite-volume effects beyond this point.

\begin{figure}
    \centering
    \includegraphics[width=0.95\linewidth]{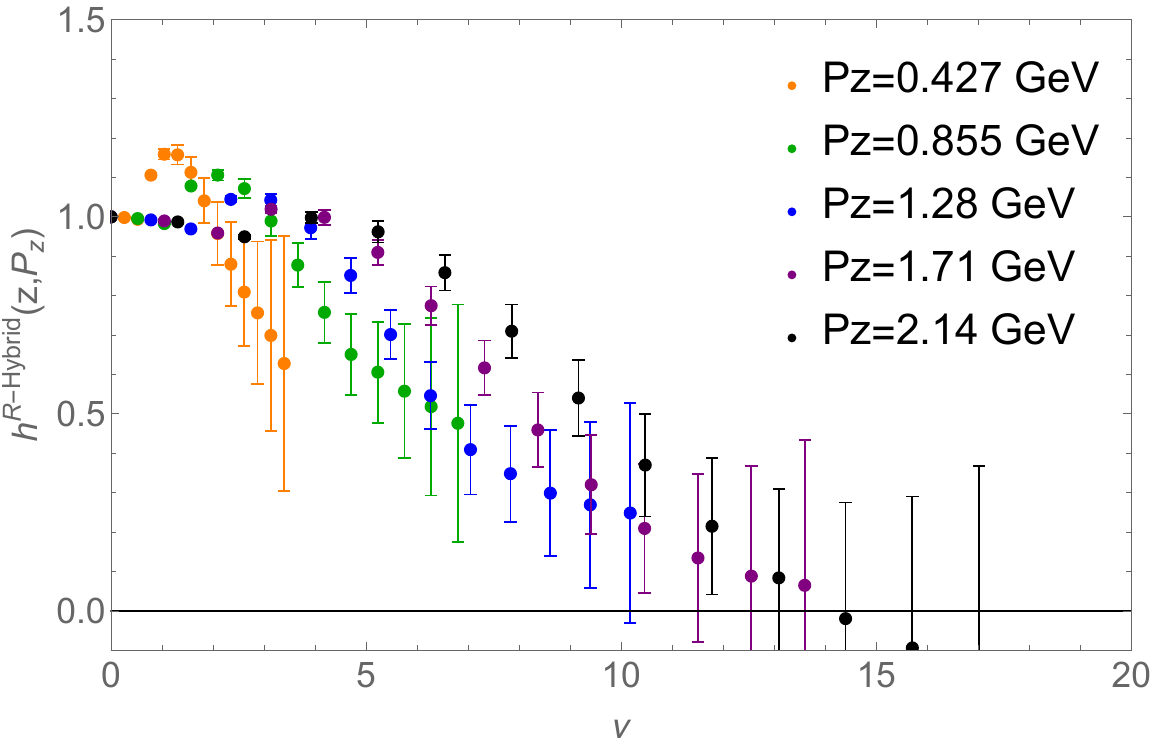}
    \caption{An hypothetical reconstruction of the hybrid-ratio renormalized $O^{(3)}$ matrix elements for the strange nucleon with Wilson-3 smearing using a qualitative guess of $\delta m + m_0 = 0.5$~GeV. The renormalization is done with $z_s = 0.24$~fm.}
    \label{fig:O0IB_hybrid_MEs}
\end{figure}

To guess $\delta m + m_0$, we fit the zero momentum bare matrix elements between $z = 7a$-$13a$ to a fit form $h^\text{fit}(z,0) = Ae^{-\delta mz}$, resulting in $\delta m = 0.65(94)$~GeV.
Though, not used in the final methodology here, some preliminary tests found for $O^{(1)}$ and $O^{(2)}$  that fitting $\delta m$ like this before fitting $m_0$ resulted consistently in negative values of $m_0$ with magnitude $O(100)$~MeV.
Starting from this information and attempting to get reasonable ``bump'' behavior seen in the phenomenological results for the other two operators, we decided $\delta m + m_0= 0.5$ GeV was a reasonable guess.
Above this point, the bump becomes unreasonably large, below this point, the matrix elements seem to decay too fast.
Again, we emphasize that the choice is subjective and that an objective result calls for the $O^{(3)}$ Wilson coefficients.
We plot our guess at hybrid renormalized $O^{(3)}$ matrix elements in with $z_s = 0.24$~fm in Fig.~\ref{fig:O0IB_hybrid_MEs}.
We see that with this choice, we recreate the bump after $\nu_s$, which is largest for the smallest $P_z$ and is gradually smoothed out at larger momentum.
After the bump, the different $P_z$ matrix elements start to become compatible again, as with the the phenomenological results seen before.
We move forward with the $P_z = 1.71$~GeV data, as it seems to be on a convergent path while the largest momentum likely displays more finite-volume effects.

\subsection{Quasi-PDF}

\begin{figure}
    \centering
    \includegraphics[width=0.95\linewidth]{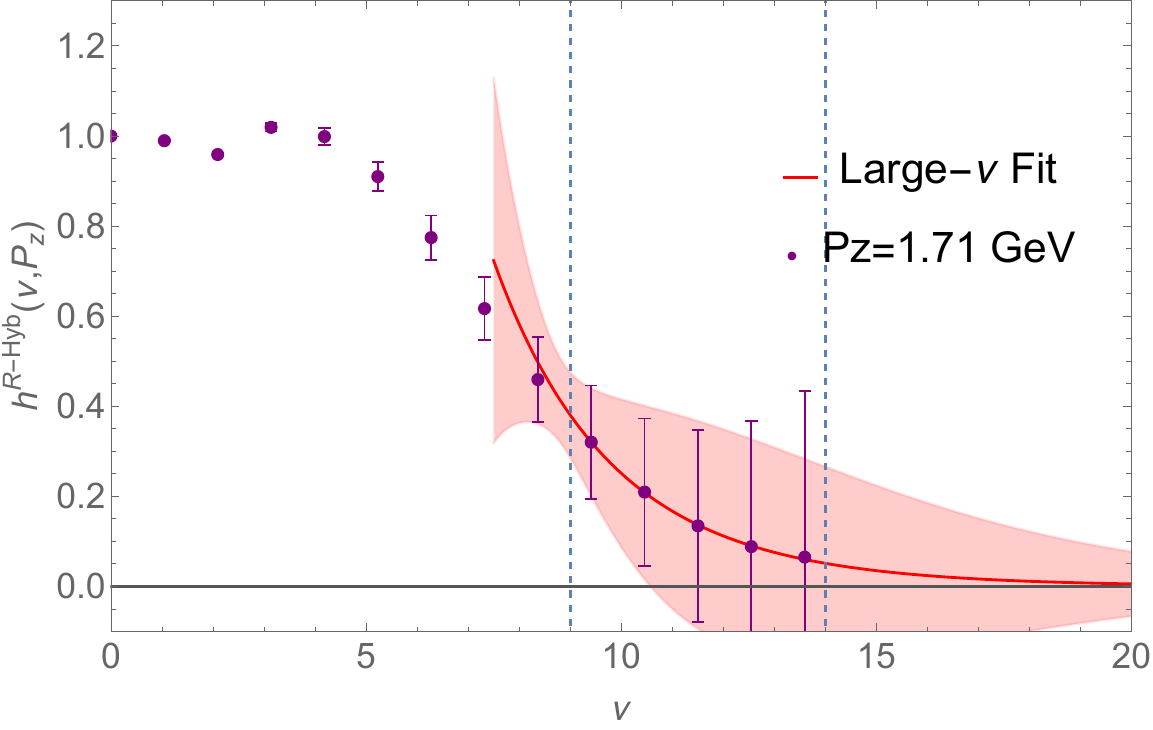}
    \caption{A large-$\nu$ extrapolation (red line and band) of the reconstructed hybrid-ratio renormalized $O^{(3)}$ matrix elements for the strange nucleon with Wilson-3 smearing (purple data points).
    The fit for the extrapolation uses the data between the light-blue lines.}
    \label{fig:O0IB_large-nu}
\end{figure}

Under the assumption that the small-$x$ behavior of the lightcone PDF trends like $x^{-\alpha}$, one may use an ansatz of the form~\cite{Ji:2020brr}
\begin{equation}\label{eq:large-nu-form}
    h^\text{R}(z,P_z) \approx A\frac{e^{-m\nu}}{|\nu|^d}
\end{equation}
to fit the large-$\nu$ data, where $A$, $m$ and $d$ are fitted parameters.
We use this form to fit our $P_z = 1.71$~GeV data from $z=9a$ to $13a$.
At this level of statistical precision and with only five data points, it is hard to separate the algebraic and exponential decay, causing a large amount of instability in the fitted parameters.
Nonetheless, we plot the results of this fit in Fig.~\ref{fig:O0IB_large-nu}.
We see qualitatively that the fit agrees well with the data.
At the largest $\nu$, the error becomes smaller than the data, and the mean decays quickly, suggesting that we can get a good Fourier transform.

\begin{figure}
    \centering{
    \includegraphics[width=0.95\linewidth]{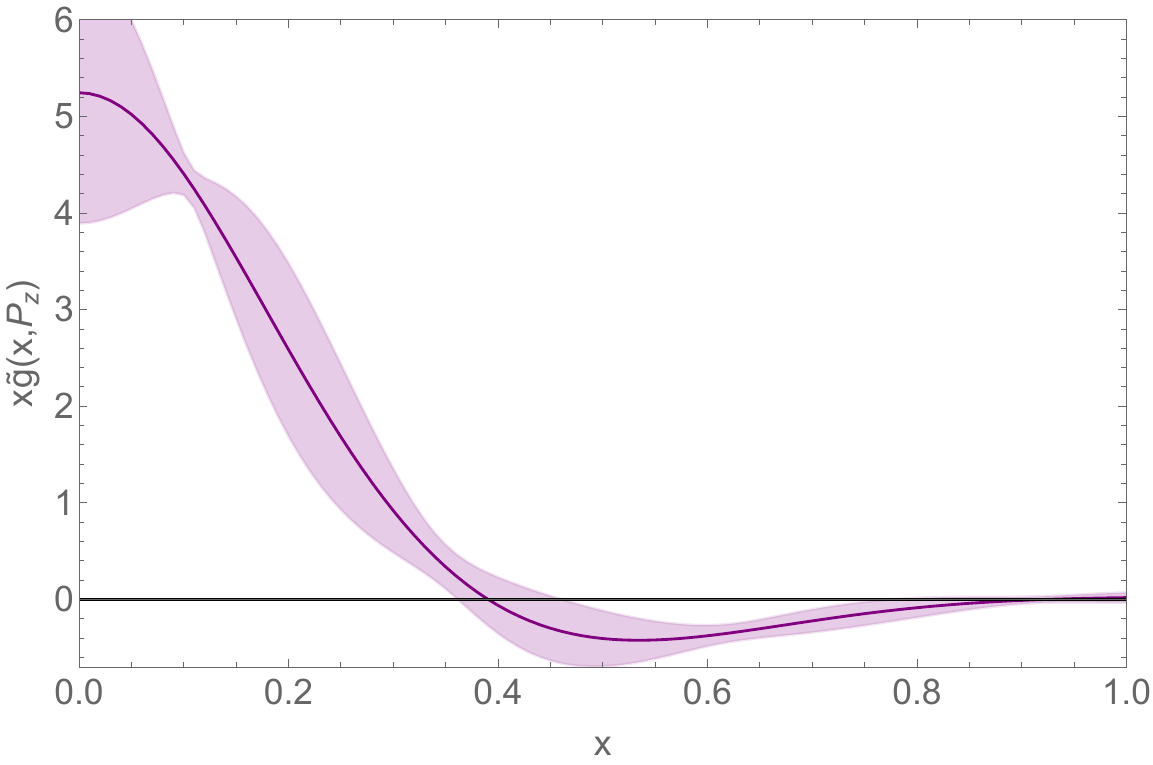}}
    \caption{The quasi-PDF for the strange nucleon with Wilson-3 smearing for our ``ansatz'' hybrid-ratio renormalization of $O^{(3)}$ at $P_z = 1.71$~GeV.}
    \label{fig:O0IB_qPDF}
\end{figure}

We perform an interpolation of the smaller-$\nu$ data and then use the extrapolated data beyond around $\nu=10$ to get a Fourier transform of the matrix elements, corresponding to $x\tilde{g}(x,P_z)$.
We show these results in Fig.~\ref{fig:O0IB_qPDF}.
We see that the uncertainty in the large-$\nu$ data seems to mostly affect the small-$x$ region.
Because the data and extrapolation do not go below zero, we see a finite value of $x\tilde{g}(x=0,P_z)$.
Interestingly, the quasi-PDF is negative in a range around $x \in [0.4,0.85]$ with some ranges being quite statistically significantly below zero.
It would be illuminating to see whether this effect is washed out by either a proper fit of $\delta m + m_0$ with the Wilson coefficients or whether this is something that is taken care of by the lightcone matching.
Overall, we are able to get the first gluon quasi-PDF from lattice data, but this required guess work for the hybrid renormalization due to the missing Wilson coefficient and a much heavier than physical pion mass and a large amount gauge-link smearing, both which likely affect the physics.
Nonetheless, this shows that we are very close to being able to extract a gluon PDF through LaMET.
Further signal improvements will be necessary to make a more confident extrapolation of the matrix elements and improve the error bars.

\subsection{Exploration of Coulomb Gauge Fixing}

It has been recently suggested and shown for quark PDFs and transverse momentum distributions (TMDs) that fixing to the Coulomb gauge and removing the Wilson line from the operator definitions reduces noise in the calculation, results in consistent lightcone PDFs, and sees minimal systematic uncertainty from Gribov copies~\cite{Gao:2023lny,Zhao:2023ptv,Gao:2024fbh}.
We wish to explore this for the gluon, naively following the methodology of Ref.~\cite{Gao:2023lny}.
We consider only Wilson-3 smeared data for the light nucleon for this preliminary study.
After applying the smearing, we fixed to the Coulomb gauge to an accuracy of $10^{-7}$ and measured each operator defined in Eqs.~\ref{eq:Op1},~\ref{eq:Op2}, and~\ref{eq:Op3} with the Wilson lines removed.

\begin{figure}
\centering
        \includegraphics[width=.45\textwidth]{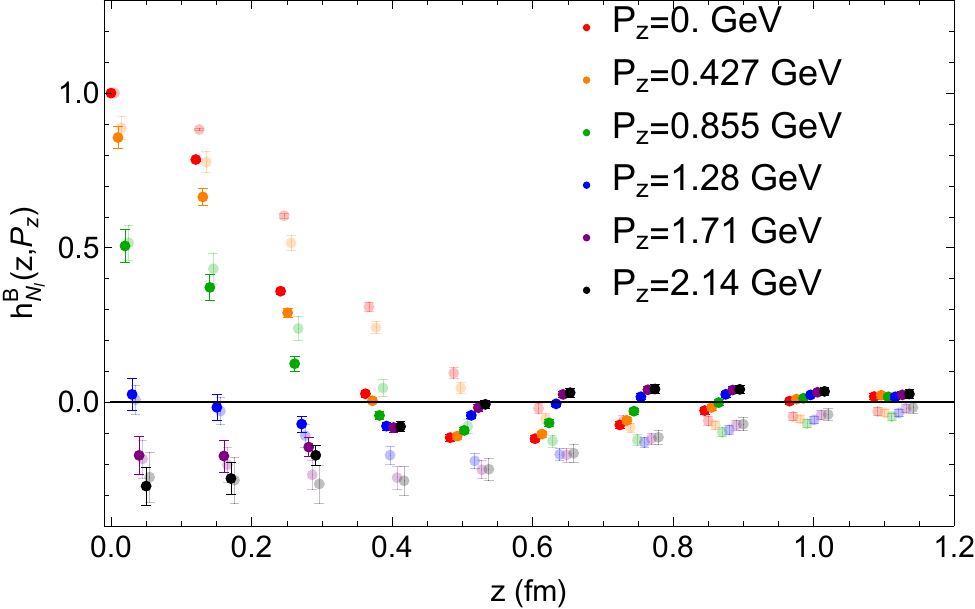}
\centering
        \includegraphics[width=.45\textwidth]{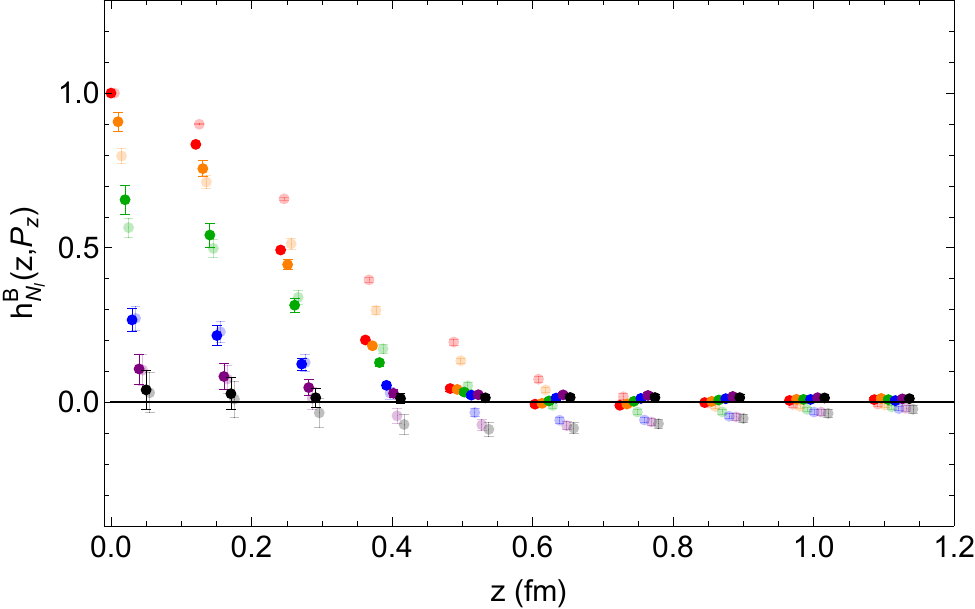}
\centering
        \includegraphics[width=.45\textwidth]{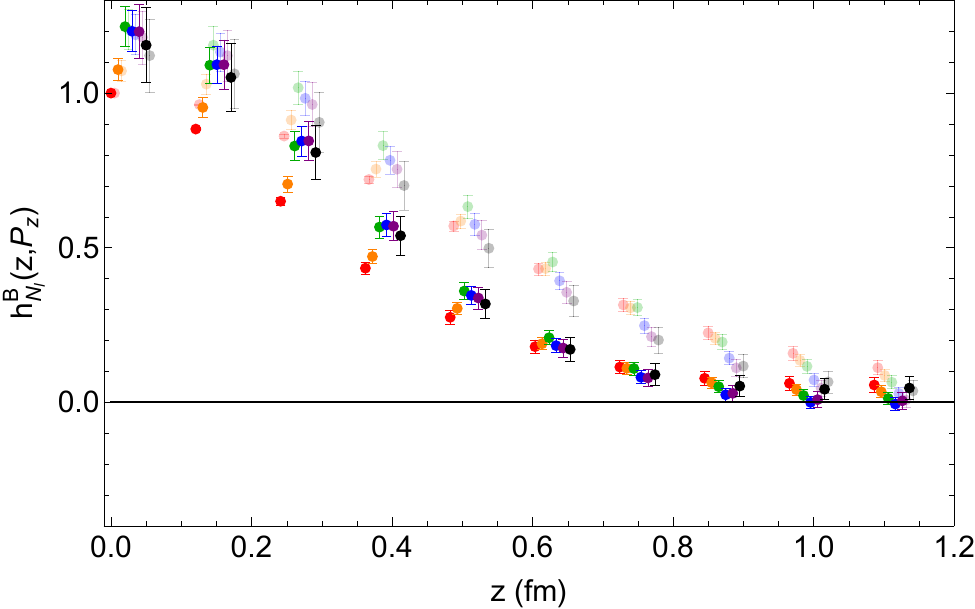}

    \caption{Bare matrix elements for the GI (filled circles) and CG (open circles) for the strange nucleon for each operator $O^{(1,2,3)}(z)$ (top to bottom) with Wilson-3 smearing.
    The data is normalized such that $h^{B}(0,0)=1$.
    The CG results are shifted slightly to the right for clarity}
    \label{fig:CG_Bare_MEs}
\end{figure}

We plot the bare matrix elements for Coulomb gauge (CG) (opaque markers) and gauge invariant (GI) (lighter markers) operators in Fig.~\ref{fig:CG_Bare_MEs} for each operator.
We see, as expected, the $z=0$ GI and CG matrix elements all agree well within statistical errors, except for the smallest two, nonzero, momenta for $O^{(2)}$.
Whatever, the cause of this disagreement, it is reduced at larger momenta.
Interestingly, the $P_z = 0$ and $0.427$~GeV CG matrix elements begin to disagree significantly with the GI matrix elements at $z=a$, while the larger momenta data are in better agreement until about $z=2a$-$3a$.
The CG data decays much faster than the GI results, as one would expect from the highly smeared gauge links in the GI results.
Overall, these observations suggest that the Coulomb gauge fixing is working as expected for these operators but large momenta may be more desirable to achieve the most consistent results at short distances.

\begin{figure}
\centering
        \includegraphics[width=.45\textwidth]{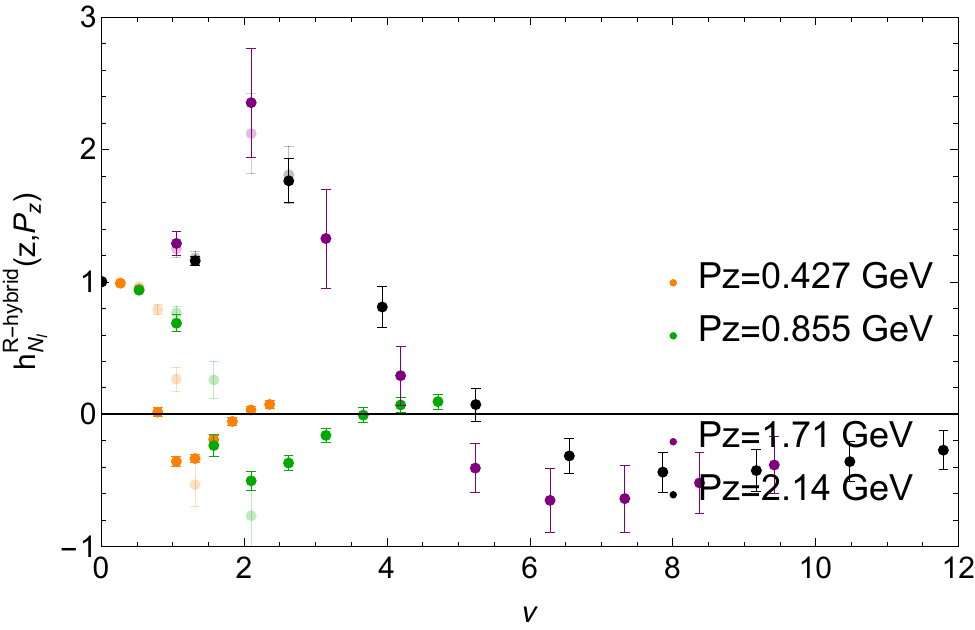}
\centering
        \includegraphics[width=.45\textwidth]{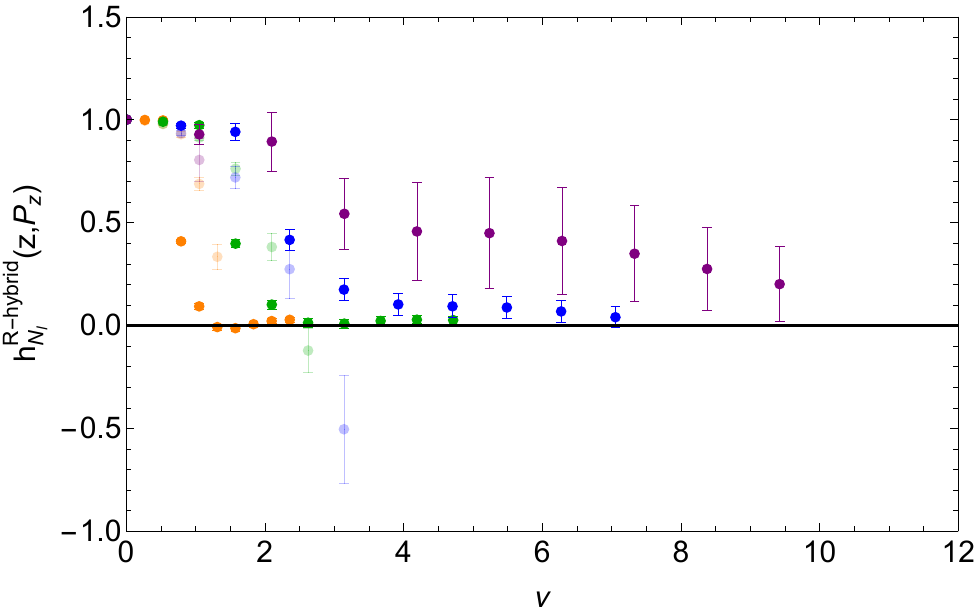}
\centering
        \includegraphics[width=.45\textwidth]{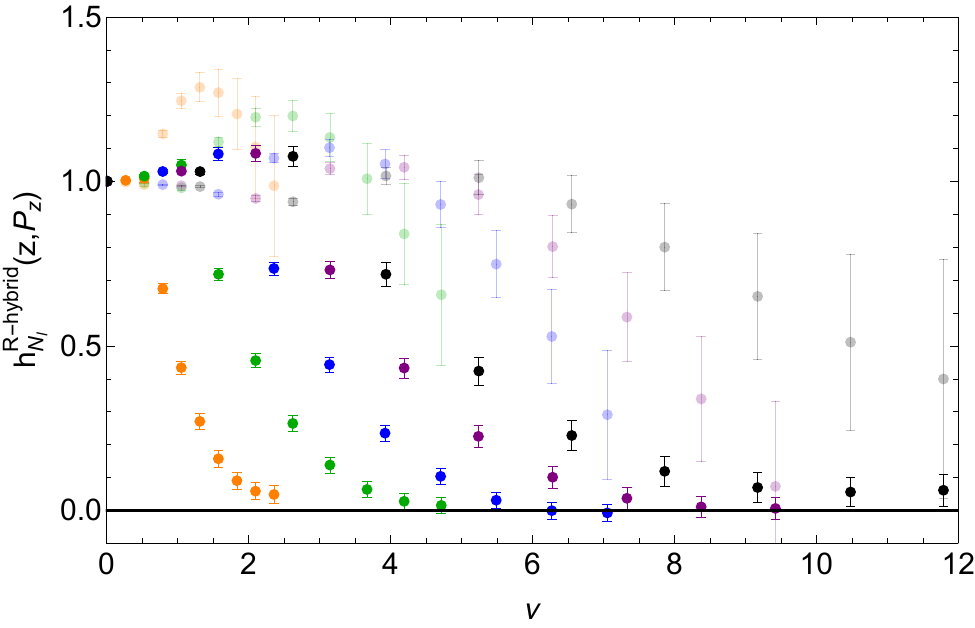}

    \caption{Hybrid renormalized matrix elements for the GI (filled circles) and CG (open circles) for the strange nucleon for each operator $O^{(1,2,3)}(z)$ (top to bottom) with Wilson-3 smearing.
    $\delta m + m_0$ is fixed to $0.55$~GeV for $O^{(3)}(z)$.
    The CG results are shifted slightly to the right for clarity}
    \label{fig:CG_hybrid_MEs}
\end{figure}

Following what was done for the quarks, we implement hybrid renormalization of the CG gluon matrix elements using Eq.\ref{eq:hyb_def}, setting $\delta m = m_0 = 0$, .
We present the hybrid renormalized matrix elements in Fig.~\ref{fig:CG_hybrid_MEs} for each operator.
It should be noted that the plots for $O^{(1)}$ and $O^{(2)}$ are missing the $P_z = 1.71$ and $2.14$~GeV data respectively because $h^\text{B}(0,P_z)$ in each case overlaps with zero, so the normalization term has well over 100\% error.
The same thing occurs to a lesser extent for $P_z = 1.71$ in $O^{(2)}$ as well, resulting in poor convergence.
We see that $O^{(1)}$ has particularly good agreement with the GI results at short distances, while the agreement for the other operators is not as good.
This may suggest different contaminations occur in the Coulomb gauge for these operators.
The gauge fixing still does not seem to fix the inconsistent behavior at different momenta for $O^{(1)}$.
However, every operator seem to now converge to zero much more quickly with far improved signal.
It is temping to be wary about the sharp behavior at $\nu_s$, especially at low momentum, but it seems smoothed out at larger momentum just like the sharp behavior seen in the phenomenological results for the GI matrix elements.
If this sharper decay behavior is behavior is confirmed to be reasonable by a calculation of the matching kernel for the Coulomb gauge operators used on phenomenological results, Coulomb gauge fixing could be a major step forward in gluon PDFs from the lattice.
More numerical study must be done here, too.
Smaller lattice spacings, larger volumes, less smearing, for example, should be considered.
It could be useful to consider more operators, as well.

\section{Conclusion and Outlook}\label{sec:conclusion}

We have presented our progress towards obtaining the first gluon PDF through LaMET with hybrid-ratio renormalization.
We consider three operators through which the quasi-PDF can be studied: $O^{(1)}$ and $O^{(2)}$ (Eqs.~\ref{eq:Op1} and \ref{eq:Op2}) which have recently had their Wilson coefficients and hybrid-renormalization matching kernels derived~\cite{Yao:2022vtp} and $O^{(3)}$ (Eq.~\ref{eq:Op3}) that was used in pseudo-PDF studies~\cite{Fan:2020cpa,Fan:2021bcr,HadStruc:2021wmh,Salas-Chavira:2021wui,Fan:2022kcb,Delmar:2023agv,Good:2023ecp}.
We found that operators $O^{(1)}$ and $O^{(2)}$ have consistently across hadrons and smearing techniques poorer signal than $O^{(3)}$.
We suggest that the $O^{(1)}$ and $O^{(2)}$ bare matrix elements crossing zero causes their ratio and hybrid-ratio renormalized matrix elements to have poor consistency between different momenta and to diverge towards $\pm \infty$.
We confirm that the behaviors in the renormalized matrix elements $O^{(1)}$ and $O^{(2)}$ do not reproduce the behavior of the matrix elements reconstructed from the CT18 nucleon gluon PDF global fit~\cite{Hou:2019efy}.
We found that $O^{(3)}$ for the nucleon with $M_\pi \approx 690$~MeV and Wilson-3 smearing has the best signal and used it to get a tentative first look at the hybrid renormalization for this operator by fitting $\delta m$ and making an estimate of $m_0$.
We found a balance of high momentum and good signal with the $P_z = 1.71$~GeV matrix elements, allowing us to fit a long-distance extrapolation and produce a quasi-PDF from this tentative data.
Overall, we suggest that operator $O^{(3)}$ is likely the best for studying the gluon PDF through LaMET;
we can obtain a quasi-PDF from this operator, but only in the case of heavy pion mass and a large amount of smearing, which may change the physics.
We conclude that numerical improvements are still needed to obtain a reliable long-range extrapolation with data that is closer to physical.

Finally, we explored the recent idea of Coulomb gauge fixing to improve signal of the matrix elements for the quark quasi-PDF and TMD~\cite{Gao:2023lny,Zhao:2023ptv,Gao:2024fbh}.
We naively follow the methodology for the quark, presenting a first limited study of gluon matrix elements from the lattice with Coulomb gauge fixing.
We found from our limited exploration, that the behavior of $O^{(2)}$ and $O^{(3)}$ show slightly different short-distance behavior between the Coulomb gauge and gauge-invariant results, possibly suggesting different contamination in the Coulomb gauge for these operators.
High momentum will be needed to smooth out the sharp behavior at $\nu_s$, but overall, Coulomb gauge fixing greatly improved the signal.

We have made major progress towards a gluon PDF from LaMET and identified more work to be done.
Once the the Wilson coefficients and the hybrid-ratio matching kernel for $O^{(3)}$ are derived explicitly, we can confirm our estimate for $\delta m + m_0$.
Further numerical improvements will allow us to go to higher momentum and obtain more reliable long distance extrapolations for our matrix elements.
Further details about the Coulomb gauge fixed gluon operators from the perturbative QCD side and the numerical side to fully utilize its power to improve the signal.

\section*{Acknowledgments}

We thank Jian-Hui Zhang for clarifying details for the $O^{(2)}$ operator Wilson coefficients and matching kernels.
We thank Yong Zhao, Xiangdong Ji, and many others who attending the LaMET2024 workshop for useful comments on this project.
We thank MILC Collaboration for sharing the lattices used to perform this study.
The LQCD calculations were performed using the Chroma software suite~\cite{Edwards:2004sx}.
This research used resources of the National Energy Research Scientific Computing Center, a DOE Office of Science User Facility supported by the Office of Science of the U.S. Department of Energy under Contract No. DE-AC02-05CH11231 through ERCAP;
facilities of the USQCD Collaboration, which are funded by the Office of Science of the U.S. Department of Energy,
and supported in part by Michigan State University through computational resources provided by the Institute for Cyber-Enabled Research (iCER).
The work of WG is supported by partially by MSU University Distinguished Fellowship and by U.S. Department of Energy, Office of Science, under grant DE-SC0024053 ``High Energy Physics Computing Traineeship for Lattice Gauge Theory''.
The work of KH is partially supported by the Professional Assistant program at Honors College at MSU and by the US National Science Foundation under grant PHY 2209424.
The work of HL is partially supported by the US National Science Foundation under grant PHY 2209424, and by the Research Corporation for Science Advancement through the Cottrell Scholar Award.

\bibliographystyle{unsrt}
\bibliography{refs}
\end{document}